\documentclass[longauth]{aa}
\usepackage{txfonts}
\usepackage{natbib}
\usepackage{graphicx}
\usepackage{hyperref}
\usepackage{listings}
\newcommand{\dibspec}{DIB-Spec}
\newcommand{\gspspec}{GSP-Spec}
\newcommand{\gspphot}{GSP-Phot}

\newcommand{\orcit}[1]{\protect\href{https://orcid.org/#1}{\protect\includegraphics[width=8pt]{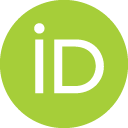}}}

\providecommand{\gaia}{\textit{Gaia}\xspace}
\providecommand{\gdr}[1]{\textit{Gaia}~DR{#1}\xspace}
\newcommand{\teff}{T_{\rm eff}}
\newcommand{\logg}{\log{g}}

\newcommand{\meta}{{\rm [M/H]}}

\newcommand{\vrad}{V_{\rm rad}}
\newcommand{\vraderr}{\sigma_{V_{\rm rad}}}

\def\kms{\,{\rm km\,s^{-1}}}
\def\kpc{\,{\rm kpc}}

\def\url#1{{\tt#1}}

\newcommand{\EBV}{E(B\,{-}\,V)}
\newcommand{\EBPRP}{\rm E(BP\,{-}\,RP)}

\newcommand{\dibdepth}{\mathcal{D}}
\newcommand{\diblambda}{\lambda_{\rm DIB}}
\newcommand{\dibwidth}{\sigma_{\rm DIB}}

\defcitealias{Zhao2022}{Z22}
\defcitealias{PVP}{S23}

\makeatletter
\renewcommand*\maketitle{%
  \thispagestyle{firstpage}
\begingroup
    \if@wideboxfn
    \setlength\bibindent{1.4\parindent}
    \else
    \setlength\bibindent{\parindent}
    \fi
    \renewcommand*\thefootnote{\@fnsymbol\c@footnote}%
    \renewcommand\@makefntext[1]{%
    \ifaa@longfn\hsize\textwidth\fi
    \noindent
    \hb@xt@\bibindent{\hss\@makefnmark\enspace}##1}
  \ifaa@twocolumn
  \begingroup
    \begin{aa@strip}
          \aa@maketitle
    \end{aa@strip}
    \@thanks            
  \endgroup
  \else
    \begingroup
      \let\thanks\footnote
      \aa@maketitle
    \endgroup
  \fi
\endgroup
  \setcounter{footnote}{0}%
}
\makeatother
\makeatletter
\renewcommand*\aa@pageof{, page \thepage{} of \pageref*{LastPage}}
\makeatother

\begin{document}

\title{\gaia Focused Product Release: Spatial distribution of two diffuse interstellar bands}

\author{
{\it Gaia} Collaboration
\and M.        ~Schultheis                    \orcit{0000-0002-6590-1657}\inst{\ref{inst:0001}}
\and H.        ~Zhao                          \orcit{0000-0003-2645-6869}\inst{\ref{inst:0001},\ref{inst:0003}}
\and T.        ~Zwitter                       \orcit{0000-0002-2325-8763}\inst{\ref{inst:0004}}
\and C.A.L.    ~Bailer-Jones                  \inst{\ref{inst:0005}}
\and R.        ~Carballo                      \orcit{0000-0001-7412-2498}\inst{\ref{inst:0006}}
\and R.        ~Sordo                         \orcit{0000-0003-4979-0659}\inst{\ref{inst:0007}}
\and R.        ~Drimmel                       \orcit{0000-0002-1777-5502}\inst{\ref{inst:0008}}
\and C.        ~Ordenovic                     \inst{\ref{inst:0001}}
\and F.        ~Pailler                       \orcit{0000-0002-4834-481X}\inst{\ref{inst:0010}}
\and M.        ~Fouesneau                     \orcit{0000-0001-9256-5516}\inst{\ref{inst:0005}}
\and O.L.      ~Creevey                       \orcit{0000-0003-1853-6631}\inst{\ref{inst:0001}}
\and U.        ~Heiter                        \orcit{0000-0001-6825-1066}\inst{\ref{inst:0013}}
\and A.        ~Recio-Blanco                  \orcit{0000-0002-6550-7377}\inst{\ref{inst:0001}}
\and G.        ~Kordopatis                    \orcit{0000-0002-9035-3920}\inst{\ref{inst:0001}}
\and P.        ~de Laverny                    \orcit{0000-0002-2817-4104}\inst{\ref{inst:0001}}
\and D.J.      ~Marshall                      \orcit{0000-0003-3956-3524}\inst{\ref{inst:0017}}
\and T.E.      ~Dharmawardena                 \orcit{0000-0002-9583-5216}\inst{\ref{inst:0005}}
\and A.G.A.    ~Brown                         \orcit{0000-0002-7419-9679}\inst{\ref{inst:0019}}
\and A.        ~Vallenari                     \orcit{0000-0003-0014-519X}\inst{\ref{inst:0007}}
\and T.        ~Prusti                        \orcit{0000-0003-3120-7867}\inst{\ref{inst:0021}}
\and J.H.J.    ~de Bruijne                    \orcit{0000-0001-6459-8599}\inst{\ref{inst:0021}}
\and F.        ~Arenou                        \orcit{0000-0003-2837-3899}\inst{\ref{inst:0023}}
\and C.        ~Babusiaux                     \orcit{0000-0002-7631-348X}\inst{\ref{inst:0024}}
\and A.        ~Barbier                       \orcit{0009-0004-0983-931X}\inst{\ref{inst:0010}}
\and M.        ~Biermann                      \orcit{0000-0002-5791-9056}\inst{\ref{inst:0026}}
\and C.        ~Ducourant                     \orcit{0000-0003-4843-8979}\inst{\ref{inst:0027}}
\and D.W.      ~Evans                         \orcit{0000-0002-6685-5998}\inst{\ref{inst:0028}}
\and L.        ~Eyer                          \orcit{0000-0002-0182-8040}\inst{\ref{inst:0029}}
\and R.        ~Guerra                        \orcit{0000-0002-9850-8982}\inst{\ref{inst:0030}}
\and A.        ~Hutton                        \inst{\ref{inst:0031}}
\and C.        ~Jordi                         \orcit{0000-0001-5495-9602}\inst{\ref{inst:0032},\ref{inst:0033},\ref{inst:0034}}
\and S.A.      ~Klioner                       \orcit{0000-0003-4682-7831}\inst{\ref{inst:0035}}
\and U.        ~Lammers                       \orcit{0000-0001-8309-3801}\inst{\ref{inst:0030}}
\and L.        ~Lindegren                     \orcit{0000-0002-5443-3026}\inst{\ref{inst:0037}}
\and X.        ~Luri                          \orcit{0000-0001-5428-9397}\inst{\ref{inst:0032},\ref{inst:0033},\ref{inst:0034}}
\and F.        ~Mignard                       \inst{\ref{inst:0001}}
\and S.        ~Randich                       \orcit{0000-0003-2438-0899}\inst{\ref{inst:0042}}
\and P.        ~Sartoretti                    \orcit{0000-0002-6574-7565}\inst{\ref{inst:0023}}
\and R.        ~Smiljanic                     \orcit{0000-0003-0942-7855}\inst{\ref{inst:0044}}
\and P.        ~Tanga                         \orcit{0000-0002-2718-997X}\inst{\ref{inst:0001}}
\and N.A.      ~Walton                        \orcit{0000-0003-3983-8778}\inst{\ref{inst:0028}}
\and U.        ~Bastian                       \orcit{0000-0002-8667-1715}\inst{\ref{inst:0026}}
\and M.        ~Cropper                       \orcit{0000-0003-4571-9468}\inst{\ref{inst:0048}}
\and D.        ~Katz                          \orcit{0000-0001-7986-3164}\inst{\ref{inst:0023}}
\and C.        ~Soubiran                      \orcit{0000-0003-3304-8134}\inst{\ref{inst:0027}}
\and F.        ~van Leeuwen                   \orcit{0000-0003-1781-4441}\inst{\ref{inst:0028}}
\and R.        ~Andrae                        \orcit{0000-0001-8006-6365}\inst{\ref{inst:0005}}
\and M.        ~Audard                        \orcit{0000-0003-4721-034X}\inst{\ref{inst:0029},\ref{inst:0054}}
\and J.        ~Bakker                        \inst{\ref{inst:0030}}
\and R.        ~Blomme                        \orcit{0000-0002-2526-346X}\inst{\ref{inst:0056}}
\and J.        ~Casta\~{n}eda                 \orcit{0000-0001-7820-946X}\inst{\ref{inst:0057},\ref{inst:0032},\ref{inst:0034}}
\and F.        ~De Angeli                     \orcit{0000-0003-1879-0488}\inst{\ref{inst:0028}}
\and C.        ~Fabricius                     \orcit{0000-0003-2639-1372}\inst{\ref{inst:0034},\ref{inst:0032},\ref{inst:0033}}
\and Y.        ~Fr\'{e}mat                    \orcit{0000-0002-4645-6017}\inst{\ref{inst:0056}}
\and L.        ~Galluccio                     \orcit{0000-0002-8541-0476}\inst{\ref{inst:0001}}
\and A.        ~Guerrier                      \inst{\ref{inst:0010}}
\and E.        ~Masana                        \orcit{0000-0002-4819-329X}\inst{\ref{inst:0034},\ref{inst:0032},\ref{inst:0033}}
\and R.        ~Messineo                      \inst{\ref{inst:0070}}
\and C.        ~Nicolas                       \inst{\ref{inst:0010}}
\and K.        ~Nienartowicz                  \orcit{0000-0001-5415-0547}\inst{\ref{inst:0072},\ref{inst:0054}}
\and P.        ~Panuzzo                       \orcit{0000-0002-0016-8271}\inst{\ref{inst:0023}}
\and F.        ~Riclet                        \inst{\ref{inst:0010}}
\and W.        ~Roux                          \orcit{0000-0002-7816-1950}\inst{\ref{inst:0010}}
\and G.M.      ~Seabroke                      \orcit{0000-0003-4072-9536}\inst{\ref{inst:0048}}
\and F.        ~Th\'{e}venin                  \orcit{0000-0002-5032-2476}\inst{\ref{inst:0001}}
\and G.        ~Gracia-Abril                  \inst{\ref{inst:0079},\ref{inst:0026}}
\and J.        ~Portell                       \orcit{0000-0002-8886-8925}\inst{\ref{inst:0032},\ref{inst:0033},\ref{inst:0034}}
\and D.        ~Teyssier                      \orcit{0000-0002-6261-5292}\inst{\ref{inst:0084}}
\and M.        ~Altmann                       \orcit{0000-0002-0530-0913}\inst{\ref{inst:0026},\ref{inst:0086}}
\and K.        ~Benson                        \inst{\ref{inst:0048}}
\and J.        ~Berthier                      \orcit{0000-0003-1846-6485}\inst{\ref{inst:0088}}
\and P.W.      ~Burgess                       \orcit{0009-0002-6668-4559}\inst{\ref{inst:0028}}
\and D.        ~Busonero                      \orcit{0000-0002-3903-7076}\inst{\ref{inst:0008}}
\and G.        ~Busso                         \orcit{0000-0003-0937-9849}\inst{\ref{inst:0028}}
\and H.        ~C\'{a}novas                   \orcit{0000-0001-7668-8022}\inst{\ref{inst:0084}}
\and B.        ~Carry                         \orcit{0000-0001-5242-3089}\inst{\ref{inst:0001}}
\and N.        ~Cheek                         \inst{\ref{inst:0094}}
\and G.        ~Clementini                    \orcit{0000-0001-9206-9723}\inst{\ref{inst:0095}}
\and Y.        ~Damerdji                      \orcit{0000-0002-3107-4024}\inst{\ref{inst:0096},\ref{inst:0097}}
\and M.        ~Davidson                      \orcit{0000-0001-9271-4411}\inst{\ref{inst:0098}}
\and P.        ~de Teodoro                    \inst{\ref{inst:0030}}
\and L.        ~Delchambre                    \orcit{0000-0003-2559-408X}\inst{\ref{inst:0096}}
\and A.        ~Dell'Oro                      \orcit{0000-0003-1561-9685}\inst{\ref{inst:0042}}
\and E.        ~Fraile Garcia                 \orcit{0000-0001-7742-9663}\inst{\ref{inst:0102}}
\and D.        ~Garabato                      \orcit{0000-0002-7133-6623}\inst{\ref{inst:0103}}
\and P.        ~Garc\'{i}a-Lario              \orcit{0000-0003-4039-8212}\inst{\ref{inst:0030}}
\and N.        ~Garralda Torres               \inst{\ref{inst:0105}}
\and P.        ~Gavras                        \orcit{0000-0002-4383-4836}\inst{\ref{inst:0102}}
\and R.        ~Haigron                       \inst{\ref{inst:0023}}
\and N.C.      ~Hambly                        \orcit{0000-0002-9901-9064}\inst{\ref{inst:0098}}
\and D.L.      ~Harrison                      \orcit{0000-0001-8687-6588}\inst{\ref{inst:0028},\ref{inst:0110}}
\and D.        ~Hatzidimitriou                \orcit{0000-0002-5415-0464}\inst{\ref{inst:0111}}
\and J.        ~Hern\'{a}ndez                 \orcit{0000-0002-0361-4994}\inst{\ref{inst:0030}}
\and S.T.      ~Hodgkin                       \orcit{0000-0002-5470-3962}\inst{\ref{inst:0028}}
\and B.        ~Holl                          \orcit{0000-0001-6220-3266}\inst{\ref{inst:0029},\ref{inst:0054}}
\and S.        ~Jamal                         \orcit{0000-0002-3929-6668}\inst{\ref{inst:0005}}
\and S.        ~Jordan                        \orcit{0000-0001-6316-6831}\inst{\ref{inst:0026}}
\and A.        ~Krone-Martins                 \orcit{0000-0002-2308-6623}\inst{\ref{inst:0118},\ref{inst:0119}}
\and A.C.      ~Lanzafame                     \orcit{0000-0002-2697-3607}\inst{\ref{inst:0120},\ref{inst:0121}}
\and W.        ~L\"{ o}ffler                  \inst{\ref{inst:0026}}
\and A.        ~Lorca                         \orcit{0000-0002-7985-250X}\inst{\ref{inst:0031}}
\and O.        ~Marchal                       \orcit{ 0000-0001-7461-892}\inst{\ref{inst:0124}}
\and P.M.      ~Marrese                       \orcit{0000-0002-8162-3810}\inst{\ref{inst:0125},\ref{inst:0126}}
\and A.        ~Moitinho                      \orcit{0000-0003-0822-5995}\inst{\ref{inst:0119}}
\and K.        ~Muinonen                      \orcit{0000-0001-8058-2642}\inst{\ref{inst:0128},\ref{inst:0129}}
\and M.        ~Nu\~{n}ez Campos              \inst{\ref{inst:0031}}
\and I.        ~Oreshina-Slezak               \inst{\ref{inst:0001}}
\and P.        ~Osborne                       \orcit{0000-0003-4482-3538}\inst{\ref{inst:0028}}
\and E.        ~Pancino                       \orcit{0000-0003-0788-5879}\inst{\ref{inst:0042},\ref{inst:0126}}
\and T.        ~Pauwels                       \inst{\ref{inst:0056}}
\and M.        ~Riello                        \orcit{0000-0002-3134-0935}\inst{\ref{inst:0028}}
\and L.        ~Rimoldini                     \orcit{0000-0002-0306-585X}\inst{\ref{inst:0054}}
\and A.C.      ~Robin                         \orcit{0000-0001-8654-9499}\inst{\ref{inst:0138}}
\and T.        ~Roegiers                      \orcit{0000-0002-1231-4440}\inst{\ref{inst:0139}}
\and L.M.      ~Sarro                         \orcit{0000-0002-5622-5191}\inst{\ref{inst:0140}}
\and C.        ~Siopis                        \orcit{0000-0002-6267-2924}\inst{\ref{inst:0141}}
\and M.        ~Smith                         \inst{\ref{inst:0048}}
\and A.        ~Sozzetti                      \orcit{0000-0002-7504-365X}\inst{\ref{inst:0008}}
\and E.        ~Utrilla                       \inst{\ref{inst:0031}}
\and M.        ~van Leeuwen                   \orcit{0000-0001-9698-2392}\inst{\ref{inst:0028}}
\and K.        ~Weingrill                     \orcit{0000-0002-8163-2493}\inst{\ref{inst:0146}}
\and U.        ~Abbas                         \orcit{0000-0002-5076-766X}\inst{\ref{inst:0008}}
\and P.        ~\'{A}brah\'{a}m               \orcit{0000-0001-6015-646X}\inst{\ref{inst:0148},\ref{inst:0149}}
\and A.        ~Abreu Aramburu                \orcit{0000-0003-3959-0856}\inst{\ref{inst:0105}}
\and C.        ~Aerts                         \orcit{0000-0003-1822-7126}\inst{\ref{inst:0151},\ref{inst:0152},\ref{inst:0005}}
\and G.        ~Altavilla                     \orcit{0000-0002-9934-1352}\inst{\ref{inst:0125},\ref{inst:0126}}
\and M.A.      ~\'{A}lvarez                   \orcit{0000-0002-6786-2620}\inst{\ref{inst:0103}}
\and J.        ~Alves                         \orcit{0000-0002-4355-0921}\inst{\ref{inst:0157}}
\and F.        ~Anders                        \inst{\ref{inst:0032},\ref{inst:0033},\ref{inst:0034}}
\and R.I.      ~Anderson                      \orcit{0000-0001-8089-4419}\inst{\ref{inst:0161}}
\and T.        ~Antoja                        \orcit{0000-0003-2595-5148}\inst{\ref{inst:0032},\ref{inst:0033},\ref{inst:0034}}
\and D.        ~Baines                        \orcit{0000-0002-6923-3756}\inst{\ref{inst:0165}}
\and S.G.      ~Baker                         \orcit{0000-0002-6436-1257}\inst{\ref{inst:0048}}
\and Z.        ~Balog                         \orcit{0000-0003-1748-2926}\inst{\ref{inst:0026},\ref{inst:0005}}
\and C.        ~Barache                       \inst{\ref{inst:0086}}
\and D.        ~Barbato                       \inst{\ref{inst:0029},\ref{inst:0008}}
\and M.        ~Barros                        \orcit{0000-0002-9728-9618}\inst{\ref{inst:0172}}
\and M.A.      ~Barstow                       \orcit{0000-0002-7116-3259}\inst{\ref{inst:0173}}
\and S.        ~Bartolom\'{e}                 \orcit{0000-0002-6290-6030}\inst{\ref{inst:0034},\ref{inst:0032},\ref{inst:0033}}
\and D.        ~Bashi                         \orcit{0000-0002-9035-2645}\inst{\ref{inst:0177},\ref{inst:0178}}
\and N.        ~Bauchet                       \orcit{0000-0002-2307-8973}\inst{\ref{inst:0023}}
\and N.        ~Baudeau                       \inst{\ref{inst:0180}}
\and U.        ~Becciani                      \orcit{0000-0002-4389-8688}\inst{\ref{inst:0120}}
\and L.R.      ~Bedin                         \inst{\ref{inst:0007}}
\and I.        ~Bellas-Velidis                \inst{\ref{inst:0183}}
\and M.        ~Bellazzini                    \orcit{0000-0001-8200-810X}\inst{\ref{inst:0095}}
\and W.        ~Beordo                        \orcit{0000-0002-5094-1306}\inst{\ref{inst:0008},\ref{inst:0186}}
\and A.        ~Berihuete                     \orcit{0000-0002-8589-4423}\inst{\ref{inst:0187}}
\and M.        ~Bernet                        \orcit{0000-0001-7503-1010}\inst{\ref{inst:0032},\ref{inst:0033},\ref{inst:0034}}
\and C.        ~Bertolotto                    \inst{\ref{inst:0070}}
\and S.        ~Bertone                       \orcit{0000-0001-9885-8440}\inst{\ref{inst:0008}}
\and L.        ~Bianchi                       \orcit{0000-0002-7999-4372}\inst{\ref{inst:0193}}
\and A.        ~Binnenfeld                    \orcit{0000-0002-9319-3838}\inst{\ref{inst:0194}}
\and A.        ~Blazere                       \inst{\ref{inst:0195}}
\and T.        ~Boch                          \orcit{0000-0001-5818-2781}\inst{\ref{inst:0124}}
\and A.        ~Bombrun                       \inst{\ref{inst:0197}}
\and S.        ~Bouquillon                    \inst{\ref{inst:0086},\ref{inst:0199}}
\and A.        ~Bragaglia                     \orcit{0000-0002-0338-7883}\inst{\ref{inst:0095}}
\and J.        ~Braine                        \orcit{0000-0003-1740-1284}\inst{\ref{inst:0027}}
\and L.        ~Bramante                      \inst{\ref{inst:0070}}
\and E.        ~Breedt                        \orcit{0000-0001-6180-3438}\inst{\ref{inst:0028}}
\and A.        ~Bressan                       \orcit{0000-0002-7922-8440}\inst{\ref{inst:0204}}
\and N.        ~Brouillet                     \orcit{0000-0002-3274-7024}\inst{\ref{inst:0027}}
\and E.        ~Brugaletta                    \orcit{0000-0003-2598-6737}\inst{\ref{inst:0120}}
\and B.        ~Bucciarelli                   \orcit{0000-0002-5303-0268}\inst{\ref{inst:0008},\ref{inst:0186}}
\and A.G.      ~Butkevich                     \orcit{0000-0002-4098-3588}\inst{\ref{inst:0008}}
\and R.        ~Buzzi                         \orcit{0000-0001-9389-5701}\inst{\ref{inst:0008}}
\and E.        ~Caffau                        \orcit{0000-0001-6011-6134}\inst{\ref{inst:0023}}
\and R.        ~Cancelliere                   \orcit{0000-0002-9120-3799}\inst{\ref{inst:0212}}
\and S.        ~Cannizzo                      \inst{\ref{inst:0213}}
\and T.        ~Carlucci                      \inst{\ref{inst:0086}}
\and M.I.      ~Carnerero                     \orcit{0000-0001-5843-5515}\inst{\ref{inst:0008}}
\and J.M.      ~Carrasco                      \orcit{0000-0002-3029-5853}\inst{\ref{inst:0034},\ref{inst:0032},\ref{inst:0033}}
\and J.        ~Carretero                     \orcit{0000-0002-3130-0204}\inst{\ref{inst:0220},\ref{inst:0221}}
\and S.        ~Carton                        \inst{\ref{inst:0213}}
\and L.        ~Casamiquela                   \orcit{0000-0001-5238-8674}\inst{\ref{inst:0027},\ref{inst:0023}}
\and M.        ~Castellani                    \orcit{0000-0002-7650-7428}\inst{\ref{inst:0125}}
\and A.        ~Castro-Ginard                 \orcit{0000-0002-9419-3725}\inst{\ref{inst:0019}}
\and V.        ~Cesare                        \orcit{0000-0003-1119-4237}\inst{\ref{inst:0120}}
\and P.        ~Charlot                       \orcit{0000-0002-9142-716X}\inst{\ref{inst:0027}}
\and L.        ~Chemin                        \orcit{0000-0002-3834-7937}\inst{\ref{inst:0229}}
\and V.        ~Chiaramida                    \inst{\ref{inst:0070}}
\and A.        ~Chiavassa                     \orcit{0000-0003-3891-7554}\inst{\ref{inst:0001}}
\and N.        ~Chornay                       \orcit{0000-0002-8767-3907}\inst{\ref{inst:0028},\ref{inst:0054}}
\and R.        ~Collins                       \orcit{0000-0001-8437-1703}\inst{\ref{inst:0098}}
\and G.        ~Contursi                      \orcit{0000-0001-5370-1511}\inst{\ref{inst:0001}}
\and W.J.      ~Cooper                        \orcit{0000-0003-3501-8967}\inst{\ref{inst:0236},\ref{inst:0008}}
\and T.        ~Cornez                        \inst{\ref{inst:0213}}
\and M.        ~Crosta                        \orcit{0000-0003-4369-3786}\inst{\ref{inst:0008},\ref{inst:0240}}
\and C.        ~Crowley                       \orcit{0000-0002-9391-9360}\inst{\ref{inst:0197}}
\and C.        ~Dafonte                       \orcit{0000-0003-4693-7555}\inst{\ref{inst:0103}}
\and F.        ~De Luise                      \orcit{0000-0002-6570-8208}\inst{\ref{inst:0244}}
\and R.        ~De March                      \orcit{0000-0003-0567-842X}\inst{\ref{inst:0070}}
\and R.        ~de Souza                      \orcit{0009-0007-7669-0254}\inst{\ref{inst:0246}}
\and A.        ~de Torres                     \inst{\ref{inst:0197}}
\and E.F.      ~del Peloso                    \inst{\ref{inst:0026}}
\and M.        ~Delbo                         \orcit{0000-0002-8963-2404}\inst{\ref{inst:0001}}
\and A.        ~Delgado                       \inst{\ref{inst:0102}}
\and S.        ~Diakite                       \inst{\ref{inst:0251}}
\and C.        ~Diener                        \inst{\ref{inst:0028}}
\and E.        ~Distefano                     \orcit{0000-0002-2448-2513}\inst{\ref{inst:0120}}
\and C.        ~Dolding                       \inst{\ref{inst:0048}}
\and K.        ~Dsilva                        \orcit{0000-0002-1476-9772}\inst{\ref{inst:0141}}
\and J.        ~Dur\'{a}n                     \inst{\ref{inst:0102}}
\and H.        ~Enke                          \orcit{0000-0002-2366-8316}\inst{\ref{inst:0146}}
\and P.        ~Esquej                        \orcit{0000-0001-8195-628X}\inst{\ref{inst:0102}}
\and C.        ~Fabre                         \inst{\ref{inst:0195}}
\and M.        ~Fabrizio                      \orcit{0000-0001-5829-111X}\inst{\ref{inst:0125},\ref{inst:0126}}
\and S.        ~Faigler                       \orcit{0000-0002-8368-5724}\inst{\ref{inst:0177}}
\and M.        ~Fatovi\'{c}                   \orcit{0000-0003-1911-4326}\inst{\ref{inst:0263}}
\and G.        ~Fedorets                      \orcit{0000-0002-8418-4809}\inst{\ref{inst:0128},\ref{inst:0265}}
\and J.        ~Fern\'{a}ndez-Hern\'{a}ndez   \inst{\ref{inst:0102}}
\and P.        ~Fernique                      \orcit{0000-0002-3304-2923}\inst{\ref{inst:0124}}
\and F.        ~Figueras                      \orcit{0000-0002-3393-0007}\inst{\ref{inst:0032},\ref{inst:0033},\ref{inst:0034}}
\and Y.        ~Fournier                      \orcit{0000-0002-6633-9088}\inst{\ref{inst:0146}}
\and C.        ~Fouron                        \inst{\ref{inst:0180}}
\and M.        ~Gai                           \orcit{0000-0001-9008-134X}\inst{\ref{inst:0008}}
\and M.        ~Galinier                      \orcit{0000-0001-7920-0133}\inst{\ref{inst:0001}}
\and A.        ~Garcia-Gutierrez              \inst{\ref{inst:0034},\ref{inst:0032},\ref{inst:0033}}
\and M.        ~Garc\'{i}a-Torres             \orcit{0000-0002-6867-7080}\inst{\ref{inst:0278}}
\and A.        ~Garofalo                      \orcit{0000-0002-5907-0375}\inst{\ref{inst:0095}}
\and E.        ~Gerlach                       \orcit{0000-0002-9533-2168}\inst{\ref{inst:0035}}
\and R.        ~Geyer                         \orcit{0000-0001-6967-8707}\inst{\ref{inst:0035}}
\and P.        ~Giacobbe                      \orcit{0000-0001-7034-7024}\inst{\ref{inst:0008}}
\and G.        ~Gilmore                       \orcit{0000-0003-4632-0213}\inst{\ref{inst:0028},\ref{inst:0284}}
\and S.        ~Girona                        \orcit{0000-0002-1975-1918}\inst{\ref{inst:0285}}
\and G.        ~Giuffrida                     \orcit{0000-0002-8979-4614}\inst{\ref{inst:0125}}
\and R.        ~Gomel                         \inst{\ref{inst:0177}}
\and A.        ~Gomez                         \orcit{0000-0002-3796-3690}\inst{\ref{inst:0103}}
\and J.        ~Gonz\'{a}lez-N\'{u}\~{n}ez    \orcit{0000-0001-5311-5555}\inst{\ref{inst:0289}}
\and I.        ~Gonz\'{a}lez-Santamar\'{i}a   \orcit{0000-0002-8537-9384}\inst{\ref{inst:0103}}
\and E.        ~Gosset                        \inst{\ref{inst:0096},\ref{inst:0292}}
\and M.        ~Granvik                       \orcit{0000-0002-5624-1888}\inst{\ref{inst:0128},\ref{inst:0294}}
\and V.        ~Gregori Barrera               \inst{\ref{inst:0034},\ref{inst:0032},\ref{inst:0033}}
\and R.        ~Guti\'{e}rrez-S\'{a}nchez     \orcit{0009-0003-1500-4733}\inst{\ref{inst:0084}}
\and M.        ~Haywood                       \orcit{0000-0003-0434-0400}\inst{\ref{inst:0023}}
\and A.        ~Helmer                        \inst{\ref{inst:0213}}
\and A.        ~Helmi                         \orcit{0000-0003-3937-7641}\inst{\ref{inst:0301}}
\and K.        ~Henares                       \inst{\ref{inst:0165}}
\and S.L.      ~Hidalgo                       \orcit{0000-0002-0002-9298}\inst{\ref{inst:0303},\ref{inst:0304}}
\and T.        ~Hilger                        \orcit{0000-0003-1646-0063}\inst{\ref{inst:0035}}
\and D.        ~Hobbs                         \orcit{0000-0002-2696-1366}\inst{\ref{inst:0037}}
\and C.        ~Hottier                       \orcit{0000-0002-3498-3944}\inst{\ref{inst:0023}}
\and H.E.      ~Huckle                        \inst{\ref{inst:0048}}
\and M.        ~Jab\l{}o\'{n}ska              \orcit{0000-0001-6962-4979}\inst{\ref{inst:0309},\ref{inst:0310}}
\and F.        ~Jansen                        \inst{\ref{inst:0311}}
\and \'{O}.    ~Jim\'{e}nez-Arranz            \orcit{0000-0001-7434-5165}\inst{\ref{inst:0032},\ref{inst:0033},\ref{inst:0034}}
\and J.        ~Juaristi Campillo             \inst{\ref{inst:0026}}
\and S.        ~Khanna                        \orcit{0000-0002-2604-4277}\inst{\ref{inst:0008},\ref{inst:0301}}
\and A.J.      ~Korn                          \orcit{0000-0002-3881-6756}\inst{\ref{inst:0013}}
\and \'{A}     ~K\'{o}sp\'{a}l                \orcit{0000-0001-7157-6275}\inst{\ref{inst:0148},\ref{inst:0005},\ref{inst:0149}}
\and Z.        ~Kostrzewa-Rutkowska           \inst{\ref{inst:0019}}
\and M.        ~Kun                           \orcit{0000-0002-7538-5166}\inst{\ref{inst:0148}}
\and S.        ~Lambert                       \orcit{0000-0001-6759-5502}\inst{\ref{inst:0086}}
\and A.F.      ~Lanza                         \orcit{0000-0001-5928-7251}\inst{\ref{inst:0120}}
\and J.-F.     ~Le Campion                    \inst{\ref{inst:0027}}
\and Y.        ~Lebreton                      \orcit{0000-0002-4834-2144}\inst{\ref{inst:0327},\ref{inst:0328}}
\and T.        ~Lebzelter                     \orcit{0000-0002-0702-7551}\inst{\ref{inst:0157}}
\and S.        ~Leccia                        \orcit{0000-0001-5685-6930}\inst{\ref{inst:0330}}
\and I.        ~Lecoeur-Taibi                 \orcit{0000-0003-0029-8575}\inst{\ref{inst:0054}}
\and G.        ~Lecoutre                      \inst{\ref{inst:0138}}
\and S.        ~Liao                          \orcit{0000-0002-9346-0211}\inst{\ref{inst:0333},\ref{inst:0008},\ref{inst:0335}}
\and L.        ~Liberato                      \orcit{0000-0003-3433-6269}\inst{\ref{inst:0001},\ref{inst:0337}}
\and E.        ~Licata                        \orcit{0000-0002-5203-0135}\inst{\ref{inst:0008}}
\and H.E.P.    ~Lindstr{\o}m                  \orcit{0009-0004-8864-5459}\inst{\ref{inst:0008},\ref{inst:0340},\ref{inst:0341}}
\and T.A.      ~Lister                        \orcit{0000-0002-3818-7769}\inst{\ref{inst:0342}}
\and E.        ~Livanou                       \orcit{0000-0003-0628-2347}\inst{\ref{inst:0111}}
\and A.        ~Lobel                         \orcit{0000-0001-5030-019X}\inst{\ref{inst:0056}}
\and C.        ~Loup                          \inst{\ref{inst:0124}}
\and L.        ~Mahy                          \orcit{0000-0003-0688-7987}\inst{\ref{inst:0056}}
\and R.G.      ~Mann                          \orcit{0000-0002-0194-325X}\inst{\ref{inst:0098}}
\and M.        ~Manteiga                      \orcit{0000-0002-7711-5581}\inst{\ref{inst:0348}}
\and J.M.      ~Marchant                      \orcit{0000-0002-3678-3145}\inst{\ref{inst:0349}}
\and M.        ~Marconi                       \orcit{0000-0002-1330-2927}\inst{\ref{inst:0330}}
\and D.        ~Mar\'{i}n Pina                \orcit{0000-0001-6482-1842}\inst{\ref{inst:0032},\ref{inst:0033},\ref{inst:0034}}
\and S.        ~Marinoni                      \orcit{0000-0001-7990-6849}\inst{\ref{inst:0125},\ref{inst:0126}}
\and J.        ~Mart\'{i}n Lozano             \orcit{0009-0001-2435-6680}\inst{\ref{inst:0094}}
\and J.M.      ~Mart\'{i}n-Fleitas            \orcit{0000-0002-8594-569X}\inst{\ref{inst:0031}}
\and G.        ~Marton                        \orcit{0000-0002-1326-1686}\inst{\ref{inst:0148}}
\and N.        ~Mary                          \inst{\ref{inst:0213}}
\and A.        ~Masip                         \orcit{0000-0003-1419-0020}\inst{\ref{inst:0034},\ref{inst:0032},\ref{inst:0033}}
\and D.        ~Massari                       \orcit{0000-0001-8892-4301}\inst{\ref{inst:0095}}
\and A.        ~Mastrobuono-Battisti          \orcit{0000-0002-2386-9142}\inst{\ref{inst:0023}}
\and T.        ~Mazeh                         \orcit{0000-0002-3569-3391}\inst{\ref{inst:0177}}
\and P.J.      ~McMillan                      \orcit{0000-0002-8861-2620}\inst{\ref{inst:0037}}
\and J.        ~Meichsner                     \orcit{0000-0002-9900-7864}\inst{\ref{inst:0035}}
\and S.        ~Messina                       \orcit{0000-0002-2851-2468}\inst{\ref{inst:0120}}
\and D.        ~Michalik                      \orcit{0000-0002-7618-6556}\inst{\ref{inst:0021}}
\and N.R.      ~Millar                        \inst{\ref{inst:0028}}
\and A.        ~Mints                         \orcit{0000-0002-8440-1455}\inst{\ref{inst:0146}}
\and D.        ~Molina                        \orcit{0000-0003-4814-0275}\inst{\ref{inst:0033},\ref{inst:0032},\ref{inst:0034}}
\and R.        ~Molinaro                      \orcit{0000-0003-3055-6002}\inst{\ref{inst:0330}}
\and L.        ~Moln\'{a}r                    \orcit{0000-0002-8159-1599}\inst{\ref{inst:0148},\ref{inst:0377},\ref{inst:0149}}
\and G.        ~Monari                        \orcit{0000-0002-6863-0661}\inst{\ref{inst:0124}}
\and M.        ~Mongui\'{o}                   \orcit{0000-0002-4519-6700}\inst{\ref{inst:0032},\ref{inst:0033},\ref{inst:0034}}
\and P.        ~Montegriffo                   \orcit{0000-0001-5013-5948}\inst{\ref{inst:0095}}
\and A.        ~Montero                       \inst{\ref{inst:0094}}
\and R.        ~Mor                           \orcit{0000-0002-8179-6527}\inst{\ref{inst:0385},\ref{inst:0033},\ref{inst:0034}}
\and A.        ~Mora                          \inst{\ref{inst:0031}}
\and R.        ~Morbidelli                    \orcit{0000-0001-7627-4946}\inst{\ref{inst:0008}}
\and T.        ~Morel                         \orcit{0000-0002-8176-4816}\inst{\ref{inst:0096}}
\and D.        ~Morris                        \orcit{0000-0002-1952-6251}\inst{\ref{inst:0098}}
\and N.        ~Mowlavi                       \orcit{0000-0003-1578-6993}\inst{\ref{inst:0029}}
\and D.        ~Munoz                         \inst{\ref{inst:0213}}
\and T.        ~Muraveva                      \orcit{0000-0002-0969-1915}\inst{\ref{inst:0095}}
\and C.P.      ~Murphy                        \inst{\ref{inst:0030}}
\and I.        ~Musella                       \orcit{0000-0001-5909-6615}\inst{\ref{inst:0330}}
\and Z.        ~Nagy                          \orcit{0000-0002-3632-1194}\inst{\ref{inst:0148}}
\and S.        ~Nieto                         \inst{\ref{inst:0102}}
\and L.        ~Noval                         \inst{\ref{inst:0213}}
\and A.        ~Ogden                         \inst{\ref{inst:0028}}
\and C.        ~Pagani                        \orcit{0000-0001-5477-4720}\inst{\ref{inst:0401}}
\and I.        ~Pagano                        \orcit{0000-0001-9573-4928}\inst{\ref{inst:0120}}
\and L.        ~Palaversa                     \orcit{0000-0003-3710-0331}\inst{\ref{inst:0263}}
\and P.A.      ~Palicio                       \orcit{0000-0002-7432-8709}\inst{\ref{inst:0001}}
\and L.        ~Pallas-Quintela               \orcit{0000-0001-9296-3100}\inst{\ref{inst:0103}}
\and A.        ~Panahi                        \orcit{0000-0001-5850-4373}\inst{\ref{inst:0177}}
\and C.        ~Panem                         \inst{\ref{inst:0010}}
\and S.        ~Payne-Wardenaar               \inst{\ref{inst:0026}}
\and L.        ~Pegoraro                      \inst{\ref{inst:0010}}
\and A.        ~Penttil\"{ a}                 \orcit{0000-0001-7403-1721}\inst{\ref{inst:0128}}
\and P.        ~Pesciullesi                   \inst{\ref{inst:0102}}
\and A.M.      ~Piersimoni                    \orcit{0000-0002-8019-3708}\inst{\ref{inst:0244}}
\and M.        ~Pinamonti                     \orcit{0000-0002-4445-1845}\inst{\ref{inst:0008}}
\and F.-X.     ~Pineau                        \orcit{0000-0002-2335-4499}\inst{\ref{inst:0124}}
\and E.        ~Plachy                        \orcit{0000-0002-5481-3352}\inst{\ref{inst:0148},\ref{inst:0377},\ref{inst:0149}}
\and G.        ~Plum                          \inst{\ref{inst:0023}}
\and E.        ~Poggio                        \orcit{0000-0003-3793-8505}\inst{\ref{inst:0001},\ref{inst:0008}}
\and D.        ~Pourbaix$^\dagger$            \orcit{0000-0002-3020-1837}\inst{\ref{inst:0141},\ref{inst:0292}}
\and A.        ~Pr\v{s}a                      \orcit{0000-0002-1913-0281}\inst{\ref{inst:0423}}
\and L.        ~Pulone                        \orcit{0000-0002-5285-998X}\inst{\ref{inst:0125}}
\and E.        ~Racero                        \orcit{0000-0002-6101-9050}\inst{\ref{inst:0094},\ref{inst:0426}}
\and M.        ~Rainer                        \orcit{0000-0002-8786-2572}\inst{\ref{inst:0042},\ref{inst:0428}}
\and C.M.      ~Raiteri                       \orcit{0000-0003-1784-2784}\inst{\ref{inst:0008}}
\and P.        ~Ramos                         \orcit{0000-0002-5080-7027}\inst{\ref{inst:0430},\ref{inst:0032},\ref{inst:0034}}
\and M.        ~Ramos-Lerate                  \orcit{0009-0005-4677-8031}\inst{\ref{inst:0084}}
\and M.        ~Ratajczak                     \orcit{0000-0002-3218-2684}\inst{\ref{inst:0309}}
\and P.        ~Re Fiorentin                  \orcit{0000-0002-4995-0475}\inst{\ref{inst:0008}}
\and S.        ~Regibo                        \orcit{0000-0001-7227-9563}\inst{\ref{inst:0151}}
\and C.        ~Reyl\'{e}                     \orcit{0000-0003-2258-2403}\inst{\ref{inst:0138}}
\and V.        ~Ripepi                        \orcit{0000-0003-1801-426X}\inst{\ref{inst:0330}}
\and A.        ~Riva                          \orcit{0000-0002-6928-8589}\inst{\ref{inst:0008}}
\and H.-W.     ~Rix                           \orcit{0000-0003-4996-9069}\inst{\ref{inst:0005}}
\and G.        ~Rixon                         \orcit{0000-0003-4399-6568}\inst{\ref{inst:0028}}
\and N.        ~Robichon                      \orcit{0000-0003-4545-7517}\inst{\ref{inst:0023}}
\and C.        ~Robin                         \inst{\ref{inst:0213}}
\and M.        ~Romero-G\'{o}mez              \orcit{0000-0003-3936-1025}\inst{\ref{inst:0032},\ref{inst:0033},\ref{inst:0034}}
\and N.        ~Rowell                        \orcit{0000-0003-3809-1895}\inst{\ref{inst:0098}}
\and F.        ~Royer                         \orcit{0000-0002-9374-8645}\inst{\ref{inst:0023}}
\and D.        ~Ruz Mieres                    \orcit{0000-0002-9455-157X}\inst{\ref{inst:0028}}
\and K.A.      ~Rybicki                       \orcit{0000-0002-9326-9329}\inst{\ref{inst:0450}}
\and G.        ~Sadowski                      \orcit{0000-0002-3411-1003}\inst{\ref{inst:0141}}
\and A.        ~S\'{a}ez N\'{u}\~{n}ez        \orcit{0009-0001-6078-0868}\inst{\ref{inst:0034},\ref{inst:0032},\ref{inst:0033}}
\and A.        ~Sagrist\`{a} Sell\'{e}s       \orcit{0000-0001-6191-2028}\inst{\ref{inst:0026}}
\and J.        ~Sahlmann                      \orcit{0000-0001-9525-3673}\inst{\ref{inst:0102}}
\and V.        ~Sanchez Gimenez               \orcit{0000-0003-1797-3557}\inst{\ref{inst:0034},\ref{inst:0032},\ref{inst:0033}}
\and N.        ~Sanna                         \orcit{0000-0001-9275-9492}\inst{\ref{inst:0042}}
\and R.        ~Santove\~{n}a                 \orcit{0000-0002-9257-2131}\inst{\ref{inst:0103}}
\and M.        ~Sarasso                       \orcit{0000-0001-5121-0727}\inst{\ref{inst:0008}}
\and C.        ~Sarrate Riera                 \inst{\ref{inst:0057},\ref{inst:0032},\ref{inst:0034}}
\and E.        ~Sciacca                       \orcit{0000-0002-5574-2787}\inst{\ref{inst:0120}}
\and J.C.      ~Segovia                       \inst{\ref{inst:0094}}
\and D.        ~S\'{e}gransan                 \orcit{0000-0003-2355-8034}\inst{\ref{inst:0029}}
\and S.        ~Shahaf                        \orcit{0000-0001-9298-8068}\inst{\ref{inst:0450}}
\and A.        ~Siebert                       \orcit{0000-0001-8059-2840}\inst{\ref{inst:0124},\ref{inst:0471}}
\and L.        ~Siltala                       \orcit{0000-0002-6938-794X}\inst{\ref{inst:0128}}
\and E.        ~Slezak                        \inst{\ref{inst:0001}}
\and R.L.      ~Smart                         \orcit{0000-0002-4424-4766}\inst{\ref{inst:0008},\ref{inst:0236}}
\and O.N.      ~Snaith                        \orcit{0000-0003-1414-1296}\inst{\ref{inst:0023},\ref{inst:0477}}
\and E.        ~Solano                        \orcit{0000-0003-1885-5130}\inst{\ref{inst:0478}}
\and F.        ~Solitro                       \inst{\ref{inst:0070}}
\and D.        ~Souami                        \orcit{0000-0003-4058-0815}\inst{\ref{inst:0327},\ref{inst:0481}}
\and J.        ~Souchay                       \inst{\ref{inst:0086}}
\and L.        ~Spina                         \orcit{0000-0002-9760-6249}\inst{\ref{inst:0007}}
\and E.        ~Spitoni                       \orcit{0000-0001-9715-5727}\inst{\ref{inst:0001},\ref{inst:0485}}
\and F.        ~Spoto                         \orcit{0000-0001-7319-5847}\inst{\ref{inst:0486}}
\and L.A.      ~Squillante                    \inst{\ref{inst:0070}}
\and I.A.      ~Steele                        \orcit{0000-0001-8397-5759}\inst{\ref{inst:0349}}
\and H.        ~Steidelm\"{ u}ller            \inst{\ref{inst:0035}}
\and J.        ~Surdej                        \orcit{0000-0002-7005-1976}\inst{\ref{inst:0096}}
\and L.        ~Szabados                      \orcit{0000-0002-2046-4131}\inst{\ref{inst:0148}}
\and F.        ~Taris                         \inst{\ref{inst:0086}}
\and M.B.      ~Taylor                        \orcit{0000-0002-4209-1479}\inst{\ref{inst:0493}}
\and R.        ~Teixeira                      \orcit{0000-0002-6806-6626}\inst{\ref{inst:0246}}
\and K.        ~Tisani\'{c}                   \orcit{0000-0001-6382-4937}\inst{\ref{inst:0263}}
\and L.        ~Tolomei                       \orcit{0000-0002-3541-3230}\inst{\ref{inst:0070}}
\and F.        ~Torra                         \orcit{0000-0002-8429-299X}\inst{\ref{inst:0057},\ref{inst:0032},\ref{inst:0034}}
\and G.        ~Torralba Elipe                \orcit{0000-0001-8738-194X}\inst{\ref{inst:0103},\ref{inst:0501},\ref{inst:0502}}
\and M.        ~Trabucchi                     \orcit{0000-0002-1429-2388}\inst{\ref{inst:0503},\ref{inst:0029}}
\and M.        ~Tsantaki                      \orcit{0000-0002-0552-2313}\inst{\ref{inst:0042}}
\and A.        ~Ulla                          \orcit{0000-0001-6424-5005}\inst{\ref{inst:0506},\ref{inst:0507}}
\and N.        ~Unger                         \orcit{0000-0003-3993-7127}\inst{\ref{inst:0029}}
\and O.        ~Vanel                         \orcit{0000-0002-7898-0454}\inst{\ref{inst:0023}}
\and A.        ~Vecchiato                     \orcit{0000-0003-1399-5556}\inst{\ref{inst:0008}}
\and D.        ~Vicente                       \orcit{0000-0002-1584-1182}\inst{\ref{inst:0285}}
\and S.        ~Voutsinas                     \inst{\ref{inst:0098}}
\and M.        ~Weiler                        \inst{\ref{inst:0034},\ref{inst:0032},\ref{inst:0033}}
\and \L{}.     ~Wyrzykowski                   \orcit{0000-0002-9658-6151}\inst{\ref{inst:0309}}
\and J.        ~Zorec                         \orcit{0000-0003-1257-6915}\inst{\ref{inst:0517}}
\and L.        ~Balaguer-N\'{u}\~{n}ez      \orcit{0000-0001-9789-7069}\inst{\ref{inst:0034},\ref{inst:0032},\ref{inst:0033}}
\and N.        ~Leclerc                      \orcit{0009-0001-5569-6098}\inst{\ref{inst:0023}}
\and S.        ~Morgenthaler               \orcit{0009-0005-6349-3716}\inst{\ref{inst:0555}}
\and G.        ~Robert                         \inst{\ref{inst:0213}}
\and S.        ~Zucker                    \orcit{0000-0003-3173-3138}\inst{\ref{inst:0194}}
}
\institute{
     Universit\'{e} C\^{o}te d'Azur, Observatoire de la C\^{o}te d'Azur, CNRS, Laboratoire Lagrange, Bd de l'Observatoire, CS 34229, 06304 Nice Cedex 4, France\relax                                                                                                                                                                                                                                                \label{inst:0001}
\and Purple Mountain Observatory, Chinese Academy of Sciences, Nanjing 210023, China\relax                                                                                                                                                                                                                                                                                                                           \label{inst:0003}
\and Faculty of Mathematics and Physics, University of Ljubljana, Jadranska ulica 19, 1000 Ljubljana, Slovenia\relax                                                                                                                                                                                                                                                                                                 \label{inst:0004}
\and Max Planck Institute for Astronomy, K\"{ o}nigstuhl 17, 69117 Heidelberg, Germany\relax                                                                                                                                                                                                                                                                                                                         \label{inst:0005}
\and Dpto. de Matem\'{a}tica Aplicada y Ciencias de la Computaci\'{o}n, Univ. de Cantabria, ETS Ingenieros de Caminos, Canales y Puertos, Avda. de los Castros s/n, 39005 Santander, Spain\relax                                                                                                                                                                                                                     \label{inst:0006}
\and INAF - Osservatorio astronomico di Padova, Vicolo Osservatorio 5, 35122 Padova, Italy\relax                                                                                                                                                                                                                                                                                                                     \label{inst:0007}
\and INAF - Osservatorio Astrofisico di Torino, via Osservatorio 20, 10025 Pino Torinese (TO), Italy\relax                                                                                                                                                                                                                                                                                                           \label{inst:0008}
\and CNES Centre Spatial de Toulouse, 18 avenue Edouard Belin, 31401 Toulouse Cedex 9, France\relax                                                                                                                                                                                                                                                                                                                  \label{inst:0010}
\and Observational Astrophysics, Division of Astronomy and Space Physics, Department of Physics and Astronomy, Uppsala University, Box 516, 751 20 Uppsala, Sweden\relax                                                                                                                                                                                                                                             \label{inst:0013}
\and IRAP, Universit\'{e} de Toulouse, CNRS, UPS, CNES, 9 Av. colonel Roche, BP 44346, 31028 Toulouse Cedex 4, France\relax                                                                                                                                                                                                                                                                                          \label{inst:0017}
\and Leiden Observatory, Leiden University, Niels Bohrweg 2, 2333 CA Leiden, The Netherlands\relax                                                                                                                                                                                                                                                                                                                   \label{inst:0019}
\and European Space Agency (ESA), European Space Research and Technology Centre (ESTEC), Keplerlaan 1, 2201AZ, Noordwijk, The Netherlands\relax                                                                                                                                                                                                                                                                      \label{inst:0021}
\and GEPI, Observatoire de Paris, Universit\'{e} PSL, CNRS, 5 Place Jules Janssen, 92190 Meudon, France\relax                                                                                                                                                                                                                                                                                                        \label{inst:0023}
\and Univ. Grenoble Alpes, CNRS, IPAG, 38000 Grenoble, France\relax                                                                                                                                                                                                                                                                                                                                                  \label{inst:0024}
\and Astronomisches Rechen-Institut, Zentrum f\"{ u}r Astronomie der Universit\"{ a}t Heidelberg, M\"{ o}nchhofstr. 12-14, 69120 Heidelberg, Germany\relax                                                                                                                                                                                                                                                           \label{inst:0026}
\and Laboratoire d'astrophysique de Bordeaux, Univ. Bordeaux, CNRS, B18N, all{\'e}e Geoffroy Saint-Hilaire, 33615 Pessac, France\relax                                                                                                                                                                                                                                                                               \label{inst:0027}
\and Institute of Astronomy, University of Cambridge, Madingley Road, Cambridge CB3 0HA, United Kingdom\relax                                                                                                                                                                                                                                                                                                        \label{inst:0028}
\and Department of Astronomy, University of Geneva, Chemin Pegasi 51, 1290 Versoix, Switzerland\relax                                                                                                                                                                                                                                                                                                                \label{inst:0029}
\and European Space Agency (ESA), European Space Astronomy Centre (ESAC), Camino bajo del Castillo, s/n, Urbanizaci\'{o}n Villafranca del Castillo, Villanueva de la Ca\~{n}ada, 28692 Madrid, Spain\relax                                                                                                                                                                                                           \label{inst:0030}
\and Aurora Technology for European Space Agency (ESA), Camino bajo del Castillo, s/n, Urbanizaci\'{o}n Villafranca del Castillo, Villanueva de la Ca\~{n}ada, 28692 Madrid, Spain\relax                                                                                                                                                                                                                             \label{inst:0031}
\and Institut de Ci\`{e}ncies del Cosmos (ICCUB), Universitat  de  Barcelona  (UB), Mart\'{i} i  Franqu\`{e}s  1, 08028 Barcelona, Spain\relax                                                                                                                                                                                                                                                                       \label{inst:0032}
\and Departament de F\'{i}sica Qu\`{a}ntica i Astrof\'{i}sica (FQA), Universitat de Barcelona (UB), c. Mart\'{i} i Franqu\`{e}s 1, 08028 Barcelona, Spain\relax                                                                                                                                                                                                                                                      \label{inst:0033}
\and Institut d'Estudis Espacials de Catalunya (IEEC), c. Gran Capit\`{a}, 2-4, 08034 Barcelona, Spain\relax                                                                                                                                                                                                                                                                                                         \label{inst:0034}
\and Lohrmann Observatory, Technische Universit\"{ a}t Dresden, Mommsenstra{\ss}e 13, 01062 Dresden, Germany\relax                                                                                                                                                                                                                                                                                                   \label{inst:0035}
\and Lund Observatory, Division of Astrophysics, Department of Physics, Lund University, Box 43, 22100 Lund, Sweden\relax                                                                                                                                                                                                                                                                                            \label{inst:0037}
\and INAF - Osservatorio Astrofisico di Arcetri, Largo Enrico Fermi 5, 50125 Firenze, Italy\relax                                                                                                                                                                                                                                                                                                                    \label{inst:0042}
\and Nicolaus Copernicus Astronomical Center, Polish Academy of Sciences, ul. Bartycka 18, 00-716 Warsaw, Poland\relax                                                                                                                                                                                                                                                                                               \label{inst:0044}
\and Mullard Space Science Laboratory, University College London, Holmbury St Mary, Dorking, Surrey RH5 6NT, United Kingdom\relax                                                                                                                                                                                                                                                                                    \label{inst:0048}
\and Department of Astronomy, University of Geneva, Chemin d'Ecogia 16, 1290 Versoix, Switzerland\relax                                                                                                                                                                                                                                                                                                              \label{inst:0054}
\and Royal Observatory of Belgium, Ringlaan 3, 1180 Brussels, Belgium\relax                                                                                                                                                                                                                                                                                                                                          \label{inst:0056}
\and DAPCOM Data Services, c. dels Vilabella, 5-7, 80500 Vic, Barcelona, Spain\relax                                                                                                                                                                                                                                                                                                                                 \label{inst:0057}
\and ALTEC S.p.a, Corso Marche, 79,10146 Torino, Italy\relax                                                                                                                                                                                                                                                                                                                                                         \label{inst:0070}
\and Sednai S\`{a}rl, Geneva, Switzerland\relax                                                                                                                                                                                                                                                                                                                                                                      \label{inst:0072}
\and Gaia DPAC Project Office, ESAC, Camino bajo del Castillo, s/n, Urbanizaci\'{o}n Villafranca del Castillo, Villanueva de la Ca\~{n}ada, 28692 Madrid, Spain\relax                                                                                                                                                                                                                                                \label{inst:0079}
\and Telespazio UK S.L. for European Space Agency (ESA), Camino bajo del Castillo, s/n, Urbanizaci\'{o}n Villafranca del Castillo, Villanueva de la Ca\~{n}ada, 28692 Madrid, Spain\relax                                                                                                                                                                                                                            \label{inst:0084}
\and SYRTE, Observatoire de Paris, Universit\'{e} PSL, CNRS, Sorbonne Universit\'{e}, LNE, 61 avenue de l'Observatoire 75014 Paris, France\relax                                                                                                                                                                                                                                                                     \label{inst:0086}
\and IMCCE, Observatoire de Paris, Universit\'{e} PSL, CNRS, Sorbonne Universit{\'e}, Univ. Lille, 77 av. Denfert-Rochereau, 75014 Paris, France\relax                                                                                                                                                                                                                                                               \label{inst:0088}
\and Serco Gesti\'{o}n de Negocios for European Space Agency (ESA), Camino bajo del Castillo, s/n, Urbanizaci\'{o}n Villafranca del Castillo, Villanueva de la Ca\~{n}ada, 28692 Madrid, Spain\relax                                                                                                                                                                                                                 \label{inst:0094}
\and INAF - Osservatorio di Astrofisica e Scienza dello Spazio di Bologna, via Piero Gobetti 93/3, 40129 Bologna, Italy\relax                                                                                                                                                                                                                                                                                        \label{inst:0095}
\and Institut d'Astrophysique et de G\'{e}ophysique, Universit\'{e} de Li\`{e}ge, 19c, All\'{e}e du 6 Ao\^{u}t, B-4000 Li\`{e}ge, Belgium\relax                                                                                                                                                                                                                                                                      \label{inst:0096}
\and CRAAG - Centre de Recherche en Astronomie, Astrophysique et G\'{e}ophysique, Route de l'Observatoire Bp 63 Bouzareah 16340 Algiers, Algeria\relax                                                                                                                                                                                                                                                               \label{inst:0097}
\and Institute for Astronomy, University of Edinburgh, Royal Observatory, Blackford Hill, Edinburgh EH9 3HJ, United Kingdom\relax                                                                                                                                                                                                                                                                                    \label{inst:0098}
\and RHEA for European Space Agency (ESA), Camino bajo del Castillo, s/n, Urbanizaci\'{o}n Villafranca del Castillo, Villanueva de la Ca\~{n}ada, 28692 Madrid, Spain\relax                                                                                                                                                                                                                                          \label{inst:0102}
\and CIGUS CITIC - Department of Computer Science and Information Technologies, University of A Coru\~{n}a, Campus de Elvi\~{n}a s/n, A Coru\~{n}a, 15071, Spain\relax                                                                                                                                                                                                                                               \label{inst:0103}
\and ATG Europe for European Space Agency (ESA), Camino bajo del Castillo, s/n, Urbanizaci\'{o}n Villafranca del Castillo, Villanueva de la Ca\~{n}ada, 28692 Madrid, Spain\relax                                                                                                                                                                                                                                    \label{inst:0105}
\and Kavli Institute for Cosmology Cambridge, Institute of Astronomy, Madingley Road, Cambridge, CB3 0HA\relax                                                                                                                                                                                                                                                                                                       \label{inst:0110}
\and Department of Astrophysics, Astronomy and Mechanics, National and Kapodistrian University of Athens, Panepistimiopolis, Zografos, 15783 Athens, Greece\relax                                                                                                                                                                                                                                                    \label{inst:0111}
\and Donald Bren School of Information and Computer Sciences, University of California, Irvine, CA 92697, USA\relax                                                                                                                                                                                                                                                                                                  \label{inst:0118}
\and CENTRA, Faculdade de Ci\^{e}ncias, Universidade de Lisboa, Edif. C8, Campo Grande, 1749-016 Lisboa, Portugal\relax                                                                                                                                                                                                                                                                                              \label{inst:0119}
\and INAF - Osservatorio Astrofisico di Catania, via S. Sofia 78, 95123 Catania, Italy\relax                                                                                                                                                                                                                                                                                                                         \label{inst:0120}
\and Dipartimento di Fisica e Astronomia ""Ettore Majorana"", Universit\`{a} di Catania, Via S. Sofia 64, 95123 Catania, Italy\relax                                                                                                                                                                                                                                                                                 \label{inst:0121}
\and Universit\'{e} de Strasbourg, CNRS, Observatoire astronomique de Strasbourg, UMR 7550,  11 rue de l'Universit\'{e}, 67000 Strasbourg, France\relax                                                                                                                                                                                                                                                              \label{inst:0124}
\and INAF - Osservatorio Astronomico di Roma, Via Frascati 33, 00078 Monte Porzio Catone (Roma), Italy\relax                                                                                                                                                                                                                                                                                                         \label{inst:0125}
\and Space Science Data Center - ASI, Via del Politecnico SNC, 00133 Roma, Italy\relax                                                                                                                                                                                                                                                                                                                               \label{inst:0126}
\and Department of Physics, University of Helsinki, P.O. Box 64, 00014 Helsinki, Finland\relax                                                                                                                                                                                                                                                                                                                       \label{inst:0128}
\and Finnish Geospatial Research Institute FGI, Vuorimiehentie 5, 02150 Espoo, Finland\relax                                                                                                                                                                                                                                                                                                                         \label{inst:0129}
\and Institut UTINAM CNRS UMR6213, Universit\'{e} de Franche-Comt\'{e}, OSU THETA Franche-Comt\'{e} Bourgogne, Observatoire de Besan\c{c}on, BP1615, 25010 Besan\c{c}on Cedex, France\relax                                                                                                                                                                                                                          \label{inst:0138}
\and HE Space Operations BV for European Space Agency (ESA), Keplerlaan 1, 2201AZ, Noordwijk, The Netherlands\relax                                                                                                                                                                                                                                                                                                  \label{inst:0139}
\and Dpto. de Inteligencia Artificial, UNED, c/ Juan del Rosal 16, 28040 Madrid, Spain\relax                                                                                                                                                                                                                                                                                                                         \label{inst:0140}
\and Institut d'Astronomie et d'Astrophysique, Universit\'{e} Libre de Bruxelles CP 226, Boulevard du Triomphe, 1050 Brussels, Belgium\relax                                                                                                                                                                                                                                                                         \label{inst:0141}
\and Leibniz Institute for Astrophysics Potsdam (AIP), An der Sternwarte 16, 14482 Potsdam, Germany\relax                                                                                                                                                                                                                                                                                                            \label{inst:0146}
\and Konkoly Observatory, Research Centre for Astronomy and Earth Sciences, E\"{ o}tv\"{ o}s Lor{\'a}nd Research Network (ELKH), MTA Centre of Excellence, Konkoly Thege Mikl\'{o}s \'{u}t 15-17, 1121 Budapest, Hungary\relax                                                                                                                                                                                       \label{inst:0148}
\and ELTE E\"{ o}tv\"{ o}s Lor\'{a}nd University, Institute of Physics, 1117, P\'{a}zm\'{a}ny P\'{e}ter s\'{e}t\'{a}ny 1A, Budapest, Hungary\relax                                                                                                                                                                                                                                                                   \label{inst:0149}
\and Instituut voor Sterrenkunde, KU Leuven, Celestijnenlaan 200D, 3001 Leuven, Belgium\relax                                                                                                                                                                                                                                                                                                                        \label{inst:0151}
\and Department of Astrophysics/IMAPP, Radboud University, P.O.Box 9010, 6500 GL Nijmegen, The Netherlands\relax                                                                                                                                                                                                                                                                                                     \label{inst:0152}
\and University of Vienna, Department of Astrophysics, T\"{ u}rkenschanzstra{\ss}e 17, A1180 Vienna, Austria\relax                                                                                                                                                                                                                                                                                                   \label{inst:0157}
\and Institute of Physics, Ecole Polytechnique F\'ed\'erale de Lausanne (EPFL), Observatoire de Sauverny, 1290 Versoix, Switzerland\relax                                                                                                                                                                                                                                                                            \label{inst:0161}
\and Quasar Science Resources for European Space Agency (ESA), Camino bajo del Castillo, s/n, Urbanizaci\'{o}n Villafranca del Castillo, Villanueva de la Ca\~{n}ada, 28692 Madrid, Spain\relax                                                                                                                                                                                                                      \label{inst:0165}
\and LASIGE, Faculdade de Ci\^{e}ncias, Universidade de Lisboa, Edif. C6, Campo Grande, 1749-016 Lisboa, Portugal\relax                                                                                                                                                                                                                                                                                              \label{inst:0172}
\and School of Physics and Astronomy , University of Leicester, University Road, Leicester LE1 7RH, United Kingdom\relax                                                                                                                                                                                                                                                                                             \label{inst:0173}
\and School of Physics and Astronomy, Tel Aviv University, Tel Aviv 6997801, Israel\relax                                                                                                                                                                                                                                                                                                                            \label{inst:0177}
\and Cavendish Laboratory, JJ Thomson Avenue, Cambridge CB3 0HE, United Kingdom\relax                                                                                                                                                                                                                                                                                                                                \label{inst:0178}
\and Telespazio for CNES Centre Spatial de Toulouse, 18 avenue Edouard Belin, 31401 Toulouse Cedex 9, France\relax                                                                                                                                                                                                                                                                                                   \label{inst:0180}
\and National Observatory of Athens, I. Metaxa and Vas. Pavlou, Palaia Penteli, 15236 Athens, Greece\relax                                                                                                                                                                                                                                                                                                           \label{inst:0183}
\and University of Turin, Department of Physics, Via Pietro Giuria 1, 10125 Torino, Italy\relax                                                                                                                                                                                                                                                                                                                      \label{inst:0186}
\and Depto. Estad\'istica e Investigaci\'on Operativa. Universidad de C\'adiz, Avda. Rep\'ublica Saharaui s/n, 11510 Puerto Real, C\'adiz, Spain\relax                                                                                                                                                                                                                                                               \label{inst:0187}
\and EURIX S.r.l., Corso Vittorio Emanuele II 61, 10128, Torino, Italy\relax                                                                                                                                                                                                                                                                                                                                         \label{inst:0193}
\and Porter School of the Environment and Earth Sciences, Tel Aviv University, Tel Aviv 6997801, Israel\relax                                                                                                                                                                                                                                                                                                        \label{inst:0194}
\and ATOS for CNES Centre Spatial de Toulouse, 18 avenue Edouard Belin, 31401 Toulouse Cedex 9, France\relax                                                                                                                                                                                                                                                                                                         \label{inst:0195}
\and HE Space Operations BV for European Space Agency (ESA), Camino bajo del Castillo, s/n, Urbanizaci\'{o}n Villafranca del Castillo, Villanueva de la Ca\~{n}ada, 28692 Madrid, Spain\relax                                                                                                                                                                                                                        \label{inst:0197}
\and LFCA/DAS,Universidad de Chile,CNRS,Casilla 36-D, Santiago, Chile\relax                                                                                                                                                                                                                                                                                                                                          \label{inst:0199}
\and SISSA - Scuola Internazionale Superiore di Studi Avanzati, via Bonomea 265, 34136 Trieste, Italy\relax                                                                                                                                                                                                                                                                                                          \label{inst:0204}
\and University of Turin, Department of Computer Sciences, Corso Svizzera 185, 10149 Torino, Italy\relax                                                                                                                                                                                                                                                                                                             \label{inst:0212}
\and Thales Services for CNES Centre Spatial de Toulouse, 18 avenue Edouard Belin, 31401 Toulouse Cedex 9, France\relax                                                                                                                                                                                                                                                                                              \label{inst:0213}
\and Institut de F\'{i}sica d'Altes Energies (IFAE), The Barcelona Institute of Science and Technology, Campus UAB, 08193 Bellaterra (Barcelona), Spain\relax                                                                                                                                                                                                                                                        \label{inst:0220}
\and Port d'Informaci\'{o} Cient\'{i}fica (PIC), Campus UAB, C. Albareda s/n, 08193 Bellaterra (Barcelona), Spain\relax                                                                                                                                                                                                                                                                                              \label{inst:0221}
\and Instituto de Astrof\'{i}sica, Universidad Andres Bello, Fernandez Concha 700, Las Condes, Santiago RM, Chile\relax                                                                                                                                                                                                                                                                                              \label{inst:0229}
\and Centre for Astrophysics Research, University of Hertfordshire, College Lane, AL10 9AB, Hatfield, United Kingdom\relax                                                                                                                                                                                                                                                                                           \label{inst:0236}
\and University of Turin, Mathematical Department ""G.Peano"", Via Carlo Alberto 10, 10123 Torino, Italy\relax                                                                                                                                                                                                                                                                                                       \label{inst:0240}
\and INAF - Osservatorio Astronomico d'Abruzzo, Via Mentore Maggini, 64100 Teramo, Italy\relax                                                                                                                                                                                                                                                                                                                       \label{inst:0244}
\and Instituto de Astronomia, Geof\`{i}sica e Ci\^{e}ncias Atmosf\'{e}ricas, Universidade de S\~{a}o Paulo, Rua do Mat\~{a}o, 1226, Cidade Universitaria, 05508-900 S\~{a}o Paulo, SP, Brazil\relax                                                                                                                                                                                                                  \label{inst:0246}
\and M\'{e}socentre de calcul de Franche-Comt\'{e}, Universit\'{e} de Franche-Comt\'{e}, 16 route de Gray, 25030 Besan\c{c}on Cedex, France\relax                                                                                                                                                                                                                                                                    \label{inst:0251}
\and Ru{\dj}er Bo\v{s}kovi\'{c} Institute, Bijeni\v{c}ka cesta 54, 10000 Zagreb, Croatia\relax                                                                                                                                                                                                                                                                                                                       \label{inst:0263}
\and Astrophysics Research Centre, School of Mathematics and Physics, Queen's University Belfast, Belfast BT7 1NN, UK\relax                                                                                                                                                                                                                                                                                          \label{inst:0265}
\and Data Science and Big Data Lab, Pablo de Olavide University, 41013, Seville, Spain\relax                                                                                                                                                                                                                                                                                                                         \label{inst:0278}
\and Institute of Astrophysics, FORTH, Crete, Greece\relax                                                                                                                                                                                                                                                                                                                                                           \label{inst:0284}
\and Barcelona Supercomputing Center (BSC), Pla\c{c}a Eusebi G\"{ u}ell 1-3, 08034-Barcelona, Spain\relax                                                                                                                                                                                                                                                                                                            \label{inst:0285}
\and ETSE Telecomunicaci\'{o}n, Universidade de Vigo, Campus Lagoas-Marcosende, 36310 Vigo, Galicia, Spain\relax                                                                                                                                                                                                                                                                                                     \label{inst:0289}
\and F.R.S.-FNRS, Rue d'Egmont 5, 1000 Brussels, Belgium\relax                                                                                                                                                                                                                                                                                                                                                       \label{inst:0292}
\and Asteroid Engineering Laboratory, Lule\aa{} University of Technology, Box 848, S-981 28 Kiruna, Sweden\relax                                                                                                                                                                                                                                                                                                     \label{inst:0294}
\and Kapteyn Astronomical Institute, University of Groningen, Landleven 12, 9747 AD Groningen, The Netherlands\relax                                                                                                                                                                                                                                                                                                 \label{inst:0301}
\and IAC - Instituto de Astrofisica de Canarias, Via L\'{a}ctea s/n, 38200 La Laguna S.C., Tenerife, Spain\relax                                                                                                                                                                                                                                                                                                     \label{inst:0303}
\and Department of Astrophysics, University of La Laguna, Via L\'{a}ctea s/n, 38200 La Laguna S.C., Tenerife, Spain\relax                                                                                                                                                                                                                                                                                            \label{inst:0304}
\and Astronomical Observatory, University of Warsaw,  Al. Ujazdowskie 4, 00-478 Warszawa, Poland\relax                                                                                                                                                                                                                                                                                                               \label{inst:0309}
\and Research School of Astronomy \& Astrophysics, Australian National University, Cotter Road, Weston, ACT 2611, Australia\relax                                                                                                                                                                                                                                                                                     \label{inst:0310}
\and European Space Agency (ESA, retired), European Space Research and Technology Centre (ESTEC), Keplerlaan 1, 2201AZ, Noordwijk, The Netherlands\relax                                                                                                                                                                                                                                                             \label{inst:0311}
\and LESIA, Observatoire de Paris, Universit\'{e} PSL, CNRS, Sorbonne Universit\'{e}, Universit\'{e} de Paris, 5 Place Jules Janssen, 92190 Meudon, France\relax                                                                                                                                                                                                                                                     \label{inst:0327}
\and Universit\'{e} Rennes, CNRS, IPR (Institut de Physique de Rennes) - UMR 6251, 35000 Rennes, France\relax                                                                                                                                                                                                                                                                                                        \label{inst:0328}
\and INAF - Osservatorio Astronomico di Capodimonte, Via Moiariello 16, 80131, Napoli, Italy\relax                                                                                                                                                                                                                                                                                                                   \label{inst:0330}
\and Shanghai Astronomical Observatory, Chinese Academy of Sciences, 80 Nandan Road, Shanghai 200030, People's Republic of China\relax                                                                                                                                                                                                                                                                               \label{inst:0333}
\and University of Chinese Academy of Sciences, No.19(A) Yuquan Road, Shijingshan District, Beijing 100049, People's Republic of China\relax                                                                                                                                                                                                                                                                         \label{inst:0335}
\and S\~{a}o Paulo State University, Grupo de Din\^{a}mica Orbital e Planetologia, CEP 12516-410, Guaratinguet\'{a}, SP, Brazil\relax                                                                                                                                                                                                                                                                                \label{inst:0337}
\and Niels Bohr Institute, University of Copenhagen, Juliane Maries Vej 30, 2100 Copenhagen {\O}, Denmark\relax                                                                                                                                                                                                                                                                                                      \label{inst:0340}
\and DXC Technology, Retortvej 8, 2500 Valby, Denmark\relax                                                                                                                                                                                                                                                                                                                                                          \label{inst:0341}
\and Las Cumbres Observatory, 6740 Cortona Drive Suite 102, Goleta, CA 93117, USA\relax                                                                                                                                                                                                                                                                                                                              \label{inst:0342}
\and CIGUS CITIC, Department of Nautical Sciences and Marine Engineering, University of A Coru\~{n}a, Paseo de Ronda 51, 15071, A Coru\~{n}a, Spain\relax                                                                                                                                                                                                                                                            \label{inst:0348}
\and Astrophysics Research Institute, Liverpool John Moores University, 146 Brownlow Hill, Liverpool L3 5RF, United Kingdom\relax                                                                                                                                                                                                                                                                                    \label{inst:0349}
\and MTA CSFK Lend\"{ u}let Near-Field Cosmology Research Group, Konkoly Observatory, MTA Research Centre for Astronomy and Earth Sciences, Konkoly Thege Mikl\'{o}s \'{u}t 15-17, 1121 Budapest, Hungary\relax                                                                                                                                                                                                      \label{inst:0377}
\and Pervasive Technologies s.l., c. Saragossa 118, 08006 Barcelona, Spain\relax                                                                                                                                                                                                                                                                                                                                     \label{inst:0385}
\and School of Physics and Astronomy, University of Leicester, University Road, Leicester LE1 7RH, United Kingdom\relax                                                                                                                                                                                                                                                                                              \label{inst:0401}
\and Villanova University, Department of Astrophysics and Planetary Science, 800 E Lancaster Avenue, Villanova PA 19085, USA\relax                                                                                                                                                                                                                                                                                   \label{inst:0423}
\and Departmento de F\'{i}sica de la Tierra y Astrof\'{i}sica, Universidad Complutense de Madrid, 28040 Madrid, Spain\relax                                                                                                                                                                                                                                                                                          \label{inst:0426}
\and INAF - Osservatorio Astronomico di Brera, via E. Bianchi, 46, 23807 Merate (LC), Italy\relax                                                                                                                                                                                                                                                                                                                    \label{inst:0428}
\and National Astronomical Observatory of Japan, 2-21-1 Osawa, Mitaka, Tokyo 181-8588, Japan\relax                                                                                                                                                                                                                                                                                                                   \label{inst:0430}
\and Department of Particle Physics and Astrophysics, Weizmann Institute of Science, Rehovot 7610001, Israel\relax                                                                                                                                                                                                                                                                                                   \label{inst:0450}
\and Centre de Donn\'{e}es Astronomique de Strasbourg, Strasbourg, France\relax                                                                                                                                                                                                                                                                                                                                      \label{inst:0471}
\and University of Exeter, School of Physics and Astronomy, Stocker Road, Exeter, EX2 7SJ, United Kingdom\relax                                                                                                                                                                                                                                                                                                      \label{inst:0477}
\and Departamento de Astrof\'{i}sica, Centro de Astrobiolog\'{i}a (CSIC-INTA), ESA-ESAC. Camino Bajo del Castillo s/n. 28692 Villanueva de la Ca\~{n}ada, Madrid, Spain\relax                                                                                                                                                                                                                                        \label{inst:0478}
\and naXys, Department of Mathematics, University of Namur, Rue de Bruxelles 61, 5000 Namur, Belgium\relax                                                                                                                                                                                                                                                                                                           \label{inst:0481}
\and INAF. Osservatorio Astronomico di Trieste, via G.B. Tiepolo 11, 34131, Trieste, Italy\relax                                                                                                                                                                                                                                                                                                                     \label{inst:0485}
\and Harvard-Smithsonian Center for Astrophysics, 60 Garden St., MS 15, Cambridge, MA 02138, USA\relax                                                                                                                                                                                                                                                                                                               \label{inst:0486}
\and H H Wills Physics Laboratory, University of Bristol, Tyndall Avenue, Bristol BS8 1TL, United Kingdom\relax                                                                                                                                                                                                                                                                                                      \label{inst:0493}
\and Escuela de Arquitectura y Polit\'{e}cnica - Universidad Europea de Valencia, Spain\relax                                                                                                                                                                                                                                                                                                                        \label{inst:0501}
\and Escuela Superior de Ingenier\'{i}a y Tecnolog\'{i}a - Universidad Internacional de la Rioja, Spain\relax                                                                                                                                                                                                                                                                                                        \label{inst:0502}
\and Department of Physics and Astronomy G. Galilei, University of Padova, Vicolo dell'Osservatorio 3, 35122, Padova, Italy\relax                                                                                                                                                                                                                                                                                    \label{inst:0503}
\and Applied Physics Department, Universidade de Vigo, 36310 Vigo, Spain\relax                                                                                                                                                                                                                                                                                                                                       \label{inst:0506}
\and Instituto de F{'i}sica e Ciencias Aeroespaciais (IFCAE), Universidade de Vigo‚ \'{A} Campus de As Lagoas, 32004 Ourense, Spain\relax                                                                                                                                                                                                                                                                          \label{inst:0507}
\and Sorbonne Universit\'{e}, CNRS, UMR7095, Institut d'Astrophysique de Paris, 98bis bd. Arago, 75014 Paris, France\relax                                                                                                                                                                                                                                                                                           \label{inst:0517}
\and Institute of Mathematics, Ecole Polytechnique F\'ed\'erale de Lausanne (EPFL), Switzerland\relax
\label{inst:0555}
}

\date{Received ??; accepted ??}
\abstract{Diffuse interstellar bands (DIBs) are absorption features seen in optical and infrared spectra of stars and extragalactic objects 
that are probably caused by large and complex molecules in the galactic interstellar medium (ISM). Here we investigate the Galactic 
distribution and properties of two DIBs identified in almost six million stellar spectra collected by the \gaia Radial Velocity 
Spectrometer. These measurements constitute a part of the \gaia Focused Product Release to be made public between the \gaia DR3 and 
DR4 data releases. In order to isolate the DIB signal from the stellar features in each individual spectrum, we identified a set of 
160 000 spectra at high Galactic latitudes ($|b|\,{\geqslant}\,65^{\circ}$) covering a range of stellar parameters which we 
consider to be the DIB-free reference sample. Matching each target spectrum to its closest reference spectra in stellar parameter space 
allowed us to remove the stellar spectrum empirically, without reference to stellar models, leaving a set of six million ISM spectra. 
Using the star's parallax and sky coordinates, we then allocated each ISM spectrum to a voxel (VOlume piXEL) on a contiguous 
three-dimensional grid with an angular size of $1.8^{\circ}$ (level 5 HEALPix) and 29 unequally sized distance bins. Identifying the 
two DIBs at 862.1\,nm ($\lambda$862.1) and 864.8\,nm ($\lambda$864.8) in the stacked spectra, we modelled their shapes and report the 
depth, central wavelength, width, and equivalent width (EW) for each, along with confidence bounds on these measurements. We then 
explored the properties and distributions of these quantities and compared them with similar measurements from other surveys. Our main 
results are as follows: (1) the strength and spatial distribution of the DIB\,$\lambda$862.1 are very consistent with what was found in \gaia DR3, 
but for this work we attained a higher signal-to-noise ratio in the stacked spectra to larger distances, which allowed us to trace DIBs in the outer 
spiral arm and beyond the Scutum--Centaurus spiral arm; (2) we produced an all-sky map below ${\pm}65^{\circ}$ of Galactic latitude 
to $\sim$4000\,pc of both DIB features and their correlations; (3) we detected the signals of DIB\,$\lambda$862.1 inside the 
Local Bubble ($\lesssim$200\,pc); and (4) there is a reasonable correlation with the dust reddening found from stellar absorption and EWs 
of both DIBs with a correlation coefficient of 0.90 for $\lambda$862.1 and 0.77 for $\lambda$864.8.
}


\titlerunning{DIBspec}

\maketitle

\section{Introduction}

Diffuse interstellar bands (DIBs) are a set of ubiquitous interstellar absorption features that primarily exist in the optical 
and near-infrared wavelength range (about 0.4--2.4\,$\mu$m) of the spectra of stars (\citealt{Fan2019}; \citealt{Hamano2022}; 
\citealt{Ebenbichler2022}), galaxies \citep[e.g.][]{Monreal-Ibero2018}, and distant quasars \citep[e.g.][]{Monreal-Ibero2015a}. 
DIBs presumably originate from the electronic transitions of carbon-bearing molecules and are now well recognized as the signatures
of complex molecules \citep{Tielens2014iaus}, although the exact species of their carriers remain largely unidentified. Based on 
high-resolution spectrometry, DIBs can be used to probe the variation of the interstellar environments in clouds \citep{Cordiner2013},  
and reveal the physical and chemical process of the interstellar medium \citep[ISM;][]{Welty2014iaus} and the formation and 
development of chemical complexity in space \citep{Tielens2014iaus}.
%--------------------------------------------------------------------------------------------------------------------------------

While in the past DIB studies were mainly concentrated on small dedicated sample sizes, large Galactic spectroscopic surveys during 
the last decade such as \gaia--ESO, APOGEE, SDSS, RAVE, and GALAH have given rise to numerous DIB detections in our Milky Way 
\citep[e.g.][]{Puspitarini2015,Zasowski2015c,Elyajouri2019,Lan2015,Baron2015b,Kos2013,Vogrincic2023}, which has enabled the investigation 
of the kinematics \citep{Zasowski2015c,hz2021b} and three-dimensional (3D) distribution \citep{Kos2014} of DIB carriers that trace 
the large-scale structures of our Milky Way. The latest \gaia data release 3 (DR3) contains the largest catalogue so far for the DIB 
at 862.1\,nm in air ($\lambda$862.1). DIB\,$\lambda$862.1 was detected and measured in \gaia Radial Velocity Spectrometer 
\citep[RVS;][]{Seabroke2022} spectra of individual stars, by the General Stellar Parametrizer from spectroscopy ({\gspspec}) module 
\citep{GSPspecDR3} of the Astrophysical parameters inference system \citep[Apsis;][]{Bailer-Jones2013,Creevey2022}. Productive 
analysis and results of the DR3 DIB catalogue were performed and presented in \citet[][hereafter S23]{PVP} in which we  built a 
map of the median DIB\,$\lambda$862.1 strength, covering all the longitudinal directions, mainly within 3\,kpc from the Sun and 
1\,kpc above and below the Galactic plane. The rest-frame wavelength of $\lambda$862.1 was determined as $\lambda_0\,{=}\,8623.23\pm0.019$\,{\AA} 
in vacuum. An average scale height for the carrier of $\lambda$862.1 was estimated as $\rm 98.60_{-8.46}^{+11.10}$\,pc assuming 
a simple exponential distribution of the carrier perpendicular to the Galactic plane. The longitudinal variations of the radial 
velocity of the $\lambda$862.1 carrier were clearly shown as well.
%--------------------------------------------------------------------------------------------------------------------------------

% The latest \gdr3 revealed the largest number of individual DIBs 
% detected so far in our Galaxy i(\citealt{GSPspecDR3},\citealt[][hereafter S22]{PVP}) enabling for the first time to obtain a full-sky coverage of the 
% DIB at 862.1\,nm ($\lambda$862.1) with productive results, such as a median map of the DIB strength mainly within 3\,kpc from the
% Sun and 1\,kpc above and below the Galactic plane, a precisely determined rest-frame wavelength ($\lambda_0\,{=}\,8623.23\pm0.019$\,{\AA} 
% in vacuum), an average scale height for the carrier of DIB\,$\lambda$862.1 ($\rm 98.60_{-8.46}^{+11.10}$\,pc) assuming a simple
% exponential distribution, and a velocity curve of the carrier of $\lambda$862.1.
%--------------------------------------------------------------------------------------------------------------------------------

Nevertheless, due to the limitation of the signal-to-noise ratio (S/N) of the individual RVS spectra, DIB\,$\lambda$862.1 could 
be successfully measured in only $\sim$10\% of RVS objects. Restricting the DIB sample further to reliable high-quality measurement 
reduced the sample to only $\sim$140\,000 (see \citetalias{PVP} Sect. 3 for the definition of the high-quality sample). Another 
major limitation is the usage of synthetic spectra in the process of measuring the DIB \citep{GSPspecDR3}, assuming that they 
represent the stellar components in the observed spectra perfectly. The complexity of the description of the full stellar physics 
in stellar atmosphere models as well as uncertainties in the atomic line list could easily lead to inappropriate modelling of stellar 
lines around the DIB signal, which would introduce further uncertainties in the fitting of the DIB profile.
%--------------------------------------------------------------------------------------------------------------------------------

% The RVS SNR distribution in Gaia DR3 contained only a fraction of $\sim 10\%$ for which the DIB could be estimated on an individual 
% spectrum basis. Restricting further the DIB sample to reliable high-quality measurement reduces the sample to only $\sim$140\,000 
% DIB detections (see \citetalias{PVP} Sect. 3). Another major limitation is the usage of synthetic spectra in the process of measuring 
% the DIB (S22) assuming that they represent perfectly the observed spectrum. The complexity of the description of the full stellar 
% physics in stellar atmosphere models as well as uncertainties in the atomic line list could easily lead to mismatches between the 
% observed spectrum and synthetic spectra affecting the fit of the DIB profile.  
%--------------------------------------------------------------------------------------------------------------------------------

To overcome the disadvantage of using synthetic spectra and the constraint of the S/N of individual RVS spectra for DIB
measurement, we developed an Apsis module of the \gaia Data Processing and Analysis Consortium (DPAC), called \gaia DPAC/{\dibspec},
which processed 6.8 million RVS spectra and conducted new DIB measurements. To avoid using synthetic spectra, \citet{Kos2013} 
developed a data-driven method, called the  BNM, to detect DIB signals in the spectra of late-type stars 
using artificial stellar templates constructed from real spectra observed at high latitudes that have a similar morphology to the 
spectrum of the target star but are likely to be free of the signature of ISM, in analogy to the distribution of interstellar 
extinction. \citet[][hereafter Z22]{Zhao2022} applied BNM to the publicly available RVS spectra within \gdr3 where they confirmed 
the presence of the weak DIB at 864.8\,nm\footnote{The accurate rest-frame wavelength of the DIB $\lambda$864.8 has not been 
determined, and we therefore name it $\lambda$864.8 following previous suggestions.} in RVS spectra. On the other hand, stacking 
spectra in a given spatial volume would significantly increase the S/N of spectra and thus allows the detection of much
weaker DIB signals \citep[e.g.][]{Kos2013,Baron2015b,Lan2015,Zhao2022,Zhao2023a}. With these two techniques, {\dibspec}
could detect and measure DIB signals in more distant zones compared to the results in DR3 and reveal the large-scale spatial
distribution of the DIB carriers (stacking reduces the spatial resolution). Furthermore, an increased S/N of stacked spectra
enables {\dibspec} to measure DIB\,$\lambda$864.8 as well. The \gaia Focused Product Release (FPR) contains the parameters of two
DIBs, $\lambda$862.1 and $\lambda$864.8, fitted by {\dibspec} and the stacked ISM spectra (spectra only containing interstellar 
features) in each defined VOlume piXEL (voxel, or 3D display element). The aim of this paper is to introduce the {\dibspec} module and
present a preliminary analysis of the DIB measurements.

The paper is outlined as follows: A brief description of the input \gaia RVS spectra is provided in Sect. \ref{sect:input}. Section 
\ref{sect:pipeline} explains the pipeline of {the \dibspec} module, including the construction of target and reference samples, 
deriving ISM spectra, the stacking of ISM spectra, and the DIB measurement. Section \ref{sect:catalog} describes and discusses 
in detail the outputs of {\dibspec}, the fitted DIB parameters, and their stacked ISM spectra. The performed validation of {\dibspec} 
outputs are presented in Sect. \ref{sect:validation}. We discuss in Sect. \ref{localbubble} detections of DIB\,$\lambda$862.1 inside 
the Local Bubble. We finish with some caveats about the usage of the {\dibspec} outputs in Sect. \ref{sect:caveats} and our main 
conclusions in Sect. \ref{sect:conclusion}.

\section{Input \gaia RVS data} \label{sect:input}

The input data for {\dibspec} are based on the \gaia RVS spectra that were processed by the \gaia DPAC Coordination Unit 6 (CU6). 
%which provides wavelength-calibrated spectra, normalized to the continuum and resampled
The processing includes removal of cosmic rays, wavelength calibration, normalization to the continuum, and resampling
from 846 to 870\,nm with a spacing of 0.01\,nm (2400 wavelength bins, \citealt{Sartoretti2018A,2023A&A...674A...6S}). % Sartoretti et al.
The resolving power is $R\,{=}\,\lambda/\Delta \lambda\,{\sim}\,11500$ 
(\citealt{Cropper2018}). The RVS spectra were then processed by {\gspspec} to estimate their stellar atmospheric parameters (effective 
temperature $\teff$, surface gravity $\logg$, metallicity $\meta$, and $\rm [\alpha/Fe]$) without taking into account any post-processing 
steps. The RVS spectra were renormalized and rebinned by {\gspspec}, from 2400 to 800 wavelength bins, sampled every 0.03\,nm to 
increase their S/N. In total, there are 6\,862\,982 RVS spectra with $\rm S/N\,{\geqslant}\,20$ (S/N is provided by the CU6 analysis 
\citealt{Seabroke2022}). These RVS spectra, normalized by {\gspspec}, and their stellar atmospheric parameters, as well as three 
other parameters (parallax $\varpi$, stellar radial velocity $\vrad$, and the velocity uncertainty $\vraderr$), are used as the 
basic input for {\dibspec}. We want to stress that the stellar atmospheric parameters are only needed to speed up the procedure 
of deriving ISM spectra (see Sect.~\ref{subsect:ism-spec}) but are not indispensable in the data-driven method. We also note that 
not all the 6.8 million spectra have published stellar parameters in \gaia DR3 (about 1.2 million were filtered out), but their 
parameter distributions are shown in this paper, like Fig. \ref{fig:sp}.  
%--------------------------------------------------------------------------------------------------------------------------------

% The RVS spectra used in {\dibspec} are rebinned in the 
% same way as \citet{GSPspecDR3}, from 2400 to 800 wavelength bins, sampled every 0.03\,nm to increase their SNR. We refer to 
% \citet{Seabroke2022} for a detailed description of \gaia RVS spectra. In total, we have 6\,862\,982 RVS spectra with 
% $\rm SNR\,{\geqslant}\,20$ as input data for {\dibspec}. This number is the total number of RVS spectra in cycle 3 processed by GSPspec without considering any further postprocessing steps (\citealt{GSPspecDR3})
% The $\rm SNR$ is provided by the CU6 analysis \citep{Seabroke2022}. 
% %--------------------------------------------------------------------------------------------------------------------------------

% We use as an input the full sample of RVS spectra (6\,862\,982 spectra) processed by the General Stellar Parametriser-spectroscopy 
% module ({\gspspec}; \citealt{GSPspecDR3}) without taking into account any post-processing steps. We use the normalized spectra of 
% {\gspspec} as well as the stellar parameters from {\gspspec} as an input for {\dibspec}. We want to stress that the stellar parameters 
% are only needed to speed up the search of the best neighbour (see Sect.~\ref{subsect:ism-spec}) but are not at all required. 
% %--------------------------------------------------------------------------------------------------------------------------------

\begin{figure}
  \centering
  \includegraphics[width=8cm]{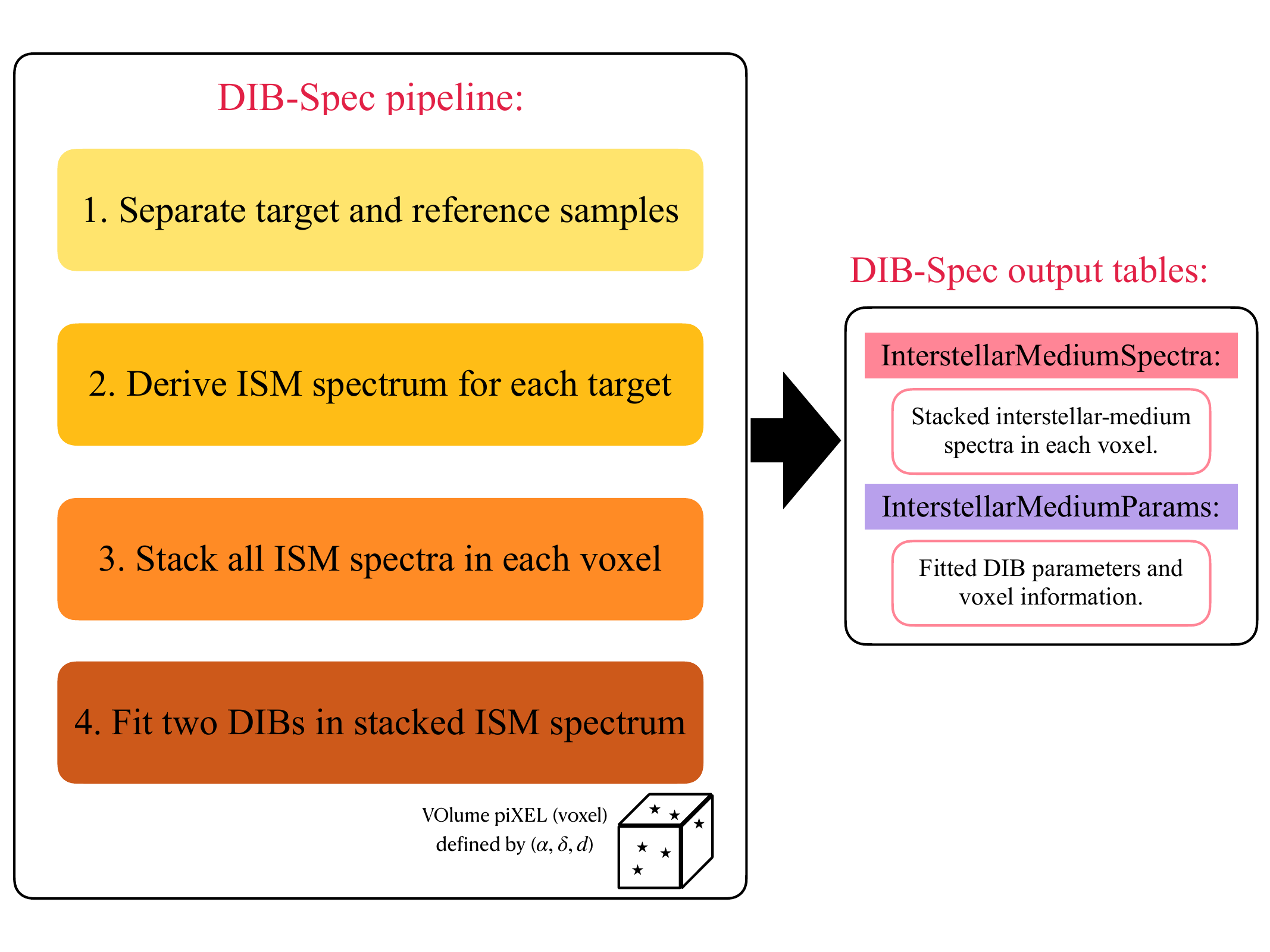}
  \caption{Schematic workflow of the {\dibspec} module.}
  \label{fig:dpcc}
\end{figure}
%--------------------------------------------------------------------------------------------------------------------------------

\section{The pipeline of the {\dibspec} module} \label{sect:pipeline}

Figure \ref{fig:dpcc} shows the general overview of the pipeline of {\dibspec} module, which was operated by CU8 at the 
Data Processing Center CNES (DPCC) in Toulouse, France. %\hzhao{Do we have the total duration of the DIB-Spec run?}
% which is part of the \gaia Coordination Unit 8 (CU8) Apsis pipeline (\citealt{Bailer-Jones2013}). An 
% updated description of this pipeline as used in the most recent \gaia DR3 is provided by \citet{Creevey2022}. 
%--------------------------------------------------------------------------------------------------------------------------------

{\dibspec} contains three main steps, which are explained in detail below. First, based on the quality of the RVS spectra and the 
results of {\gspspec}, {\dibspec} builds two samples: a sample of `target' stars whose spectra are expected to contain DIB signals, 
and a sample of `reference' stars at high latitudes whose spectra have presumably no DIB features. Second, the reference 
spectra are used to derive the ISM spectra for each target star. Finally, {\dibspec} stacks the ISM spectra of individual target 
stars in each voxel and fits the two DIBs in the stacked ISM spectra.

\subsection{Target and reference samples} \label{subsect:ref-spec}

\begin{figure}
  \centering
  \includegraphics[width=8.4cm]{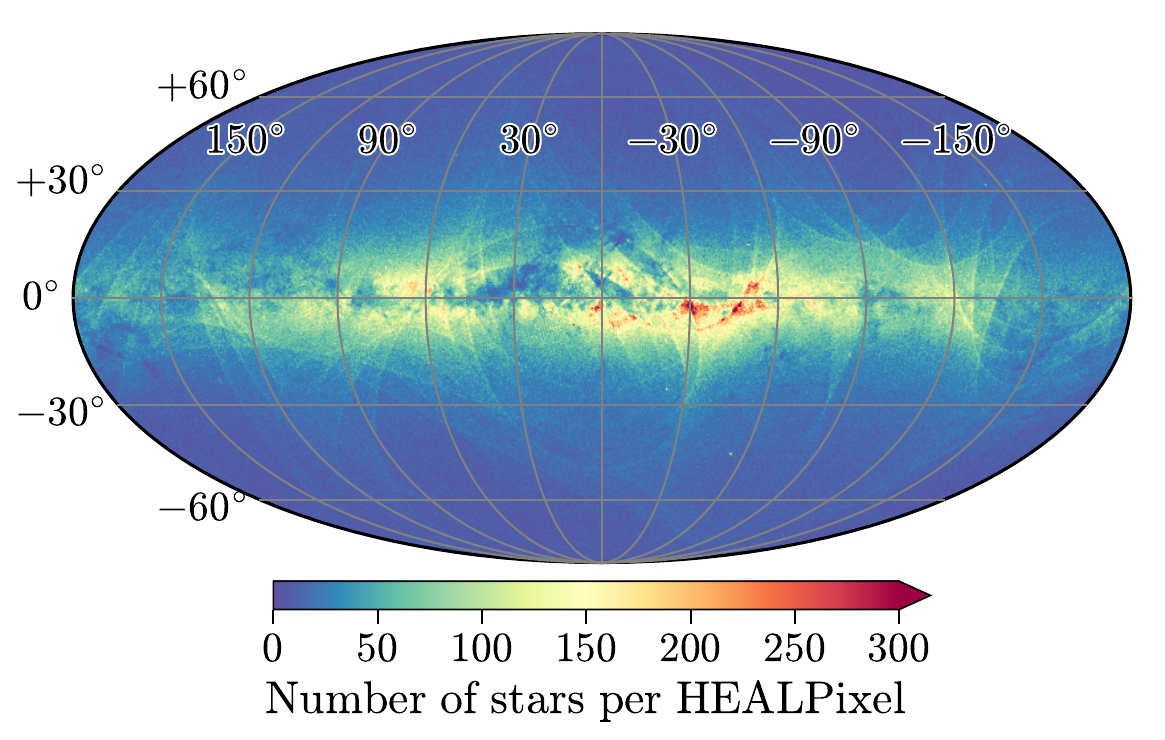}
  \caption{Galactic spatial density distribution of the 6\,143\,681 RVS spectra used in {\dibspec}. This HEALPix (\citealt{Gorski2005}) 
  map has a level of 7, corresponding to a spatial resolution of $0.46^{\circ}$.}
  \label{fig:rvs}
\end{figure}

\begin{figure}
  \centering
  \includegraphics[width=7cm]{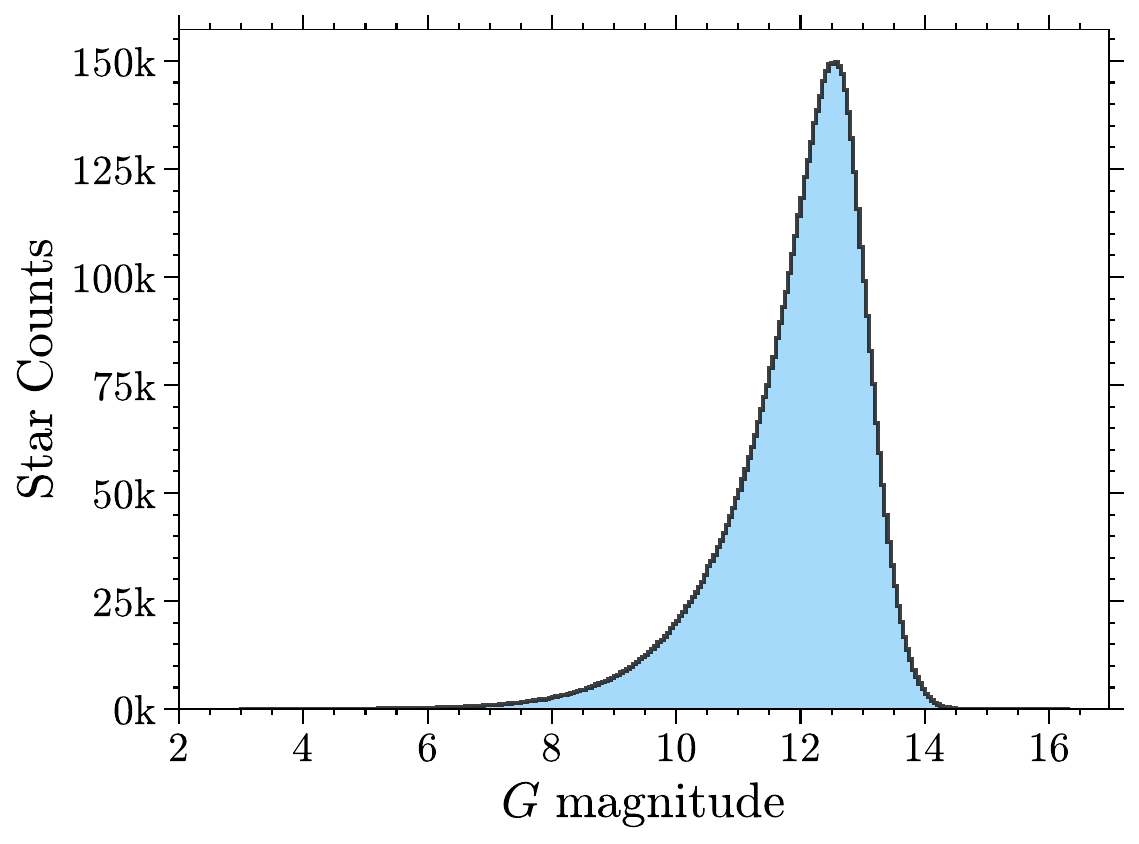}
  \caption{\gaia $G$--band magnitude distribution of the 6\,143\,681 RVS objects used in {\dibspec}.}
  \label{fig:Gmag}
\end{figure}

The RVS objects must fulfil a set of requirements in order to be retained in the {\dibspec} analysis. First, the RVS objects must 
have a measurement of $\vrad$ derived from CU6 with $\vraderr\,{\leqslant}\,5\,\kms$, because their ISM spectra have to be shifted from the 
stellar frame (as provided by CU6) back into the heliocentric frame before stacking. Then, the parallax must be above 0.1\,mas, 
which corresponds to objects within 10\,kpc if they have small parallax uncertainties. We expect to detect the DIB signals in very 
distant zones, but it should be noted that most of the RVS objects (91.5\%) are within 4\,kpc (see Fig. \ref{fig:sp}). These criteria 
result in a set of 6\,143\,681 objects to be used. Their number density distribution in Galactic coordinates and $G$--band magnitude 
distribution are shown in Figs. \ref{fig:rvs} and \ref{fig:Gmag}, respectively. Most of these RVS objects are located close to the 
Galactic plane and their $G$--band magnitudes are mainly within 8--14\,mag.
%--------------------------------------------------------------------------------------------------------------------------------

%--------------------------------------------------------------------------------------------------------------------------------

%--------------------------------------------------------------------------------------------------------------------------------

\begin{figure}
  \centering
  \includegraphics[width=7cm]{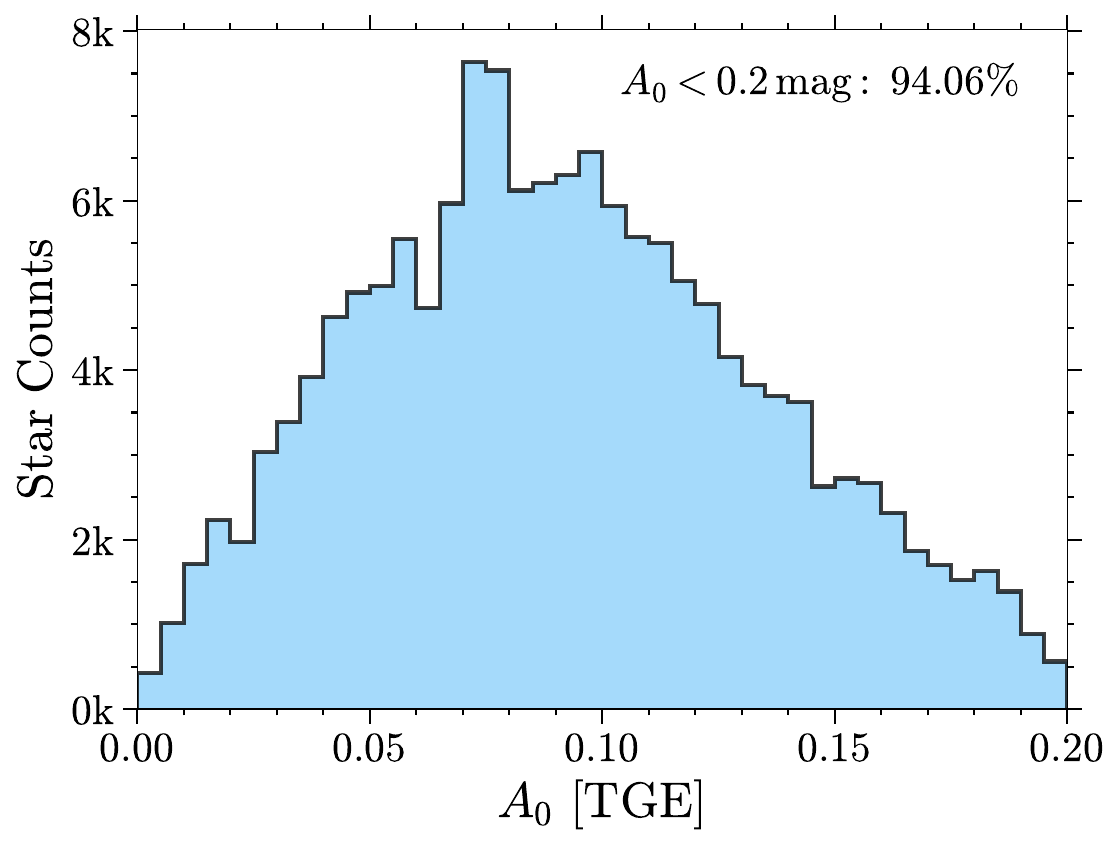}
  \caption{Distribution of $A_0$ from \gaia TGE map \citep{Delchambre2022} for 160\,392 reference stars.}
  \label{fig:ref-A0}
\end{figure}

\begin{figure*}
  \centering
  \includegraphics[width=16.8cm]{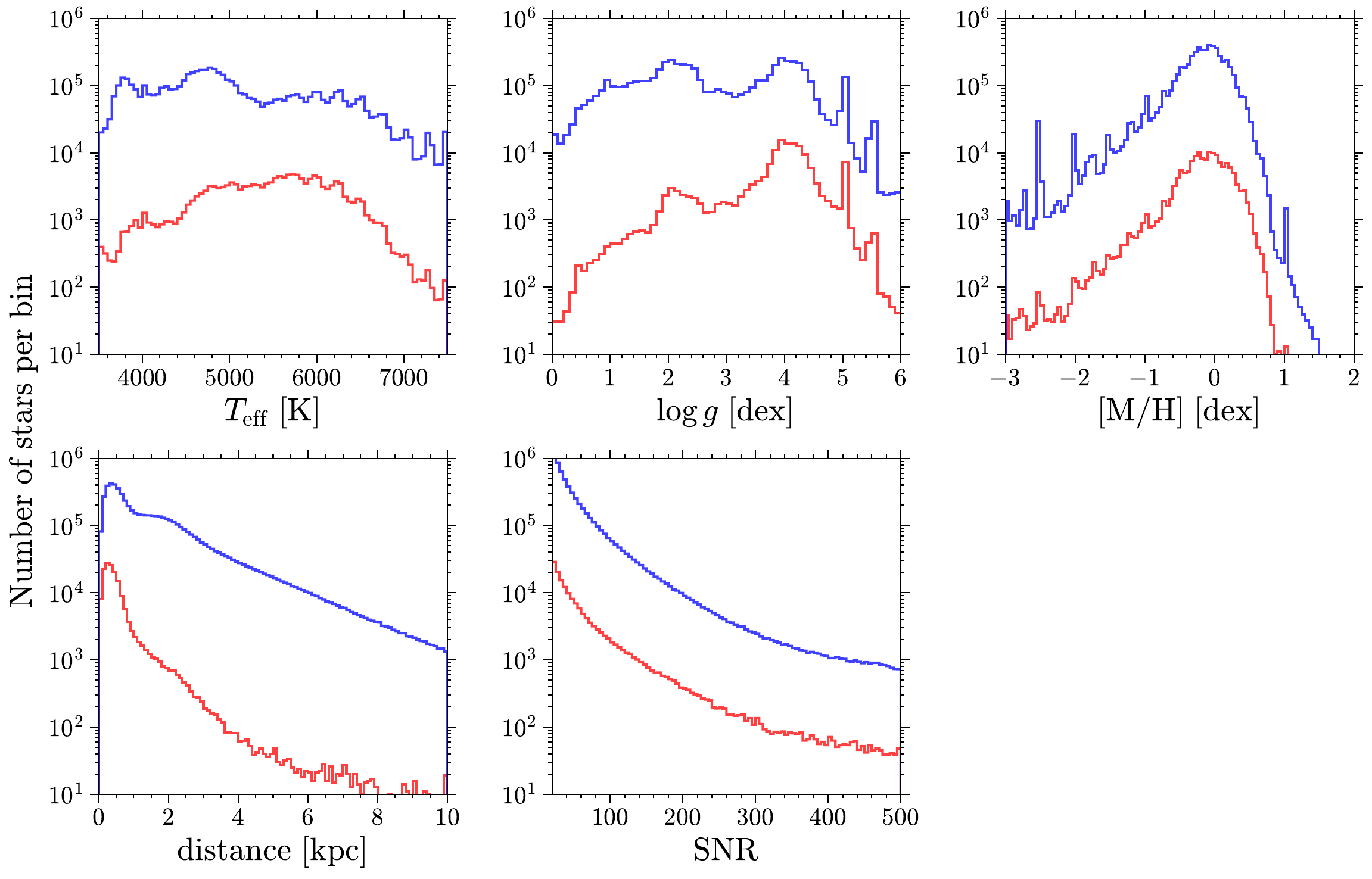}
  \caption{Distribution of $\teff$, $\logg$, $\meta$, the distance of stars, and spectral S/N for both reference stars (in 
  red) and target stars (in blue). These parameters have not been filtered out in the post-processing.}
  \label{fig:sp}
\end{figure*}

The selected stars are separated into two sets according to their Galactic latitudes: 5\,983\,289 target stars with $|b|\,{<}\,65^{\circ}$
and, 160\,392 reference stars with $|b|\,{\geqslant}\,65^{\circ}$. The DIB strength has a strong dependence on the Galactic latitude 
since the stars at high latitudes usually contain very weak DIB signals in their spectra (see the maps of DIB strength in \citealt{Lan2015}, 
\citealt{Baron2015a}, and \citetalias{PVP} for example). In \gaia DR3, we detected $\lambda$862.1 in 247 sightlines with 
$|b|\,{\geqslant}\,65^{\circ}$ in the high-quality sample ($\sim$140\,000 in total) of the DIB catalogue (see Sect. 3 in 
\citetalias{PVP}). Their mean depth is only 1.5\% of the continuum with a very small mean $A_0$ (monochromatic extinction at 541.4\,nm 
\citealt{Fitzpatrick1999}) of 0.104\,mag estimated by the Total Galactic extinction (TGE) map of \gaia DR3 \citep{Delchambre2022}. 
The mean $A_0$, on the other hand, for the reference stars is 0.103\,mag (see Fig. \ref{fig:ref-A0}).
Therefore, 1.5\% would be the maximum error we expect in the derived ISM spectra within the DIB region of target spectra introduced 
by the possible DIB signals at high latitudes when using the reference spectra. It should be noted that the real errors could be much 
lower because the TGE map would overestimate $A_0$ for nearby stars. Furthermore, as discussed in \citet{Kos2013}, any stars with  
unusually high extinctions or strong DIBs will be averaged out because the stellar template is produced by averaging several reference 
spectra. In summary, the reference spectra could be treated as DIB-free spectra and used to model stellar components in 
the DIB region of the target spectra. 
%--------------------------------------------------------------------------------------------------------------------------------
% colour excess $\EBV$ of 0.017\,mag from \citet{Planck2016dust}. 
% a mean EW of 0.029

% The mean $\EBV$ of the reference stars used in {\dibspec} is 0.016\,mag (see Fig. \ref{fig:ref-ebv} for the $\EBV$ distribution). 
% \Tomaz{One could do median instead of mean, which avoids outliers even more effectively. I propose we explain we use median or we 
% add a sentence that using median makes no difference, so outliers are not a problem.}
% \hzhao{We take the weighted mean of selected reference spectra (best neighbors) to build the stellar template for the target spectrum.
% And we take the median of spectra flux when stacking the ISM spectra in each voxel.}

%--------------------------------------------------------------------------------------------------------------------------------

% \begin{figure}
%   \centering
%   \includegraphics[width=7cm]{figures/ref-ebv.pdf}
%   \caption{Distribution of $\EBV$ from \citet{Planck2016dust} for the 160\,392 reference stars.}
%   \label{fig:ref-ebv}
% \end{figure}
%--------------------------------------------------------------------------------------------------------------------------------

The target and reference stars have a similar distribution in $\teff$, $\logg$, and $\meta$ (Fig. \ref{fig:sp}). Therefore, 
most of the target spectra should have enough neighbour candidates selected by comparing their atmospheric parameters (see Sect. 
\ref{subsect:ism-spec} for more details) to model their stellar components. The target and reference samples also have a similar 
distribution of spectral S/N, although the reference sample has a higher proportion of nearby stars (87.5\% within 1\,kpc) compared 
to the target sample (47.1\%).

%--------------------------------------------------------------------------------------------------------------------------------

\subsection{Derivation of ISM spectra for individual targets with the  BNM} \label{subsect:ism-spec}

To detect and measure the two DIBs, $\lambda$862.1 and $\lambda$864.8, in target spectra, the stellar components need to be 
subtracted by the BNM. We follow here the main principles in \citet{Kos2013}. For each target, BNM selects a set of reference 
spectra with similar spectral morphology to that of the target spectrum and combines them by averaging their normalized 
flux weighted by the spectral S/N to build the stellar template for the target spectrum. In practice, BNM first constrains the 
stellar atmospheric parameters (from {\gspspec}) of the reference stars, to be around the values of the target, in the ranges 
$\teff\,{\pm}\,20\%$, $\logg\,{\pm}\,0.6$\,dex, and $\meta\,{\pm}\,0.4$\,dex. The ranges follow those in \citet{Kos2013}. 
This step reduces the number of considered reference stars and consequently speeds up the procedure. 
%--------------------------------------------------------------------------------------------------------------------------------

%The similarity of the spectral morphology between the target and reference spectra is characterized as the inverse of the weighted 
%mean of the absolute difference of their normalized fluxes at all wavelength bins.

To compare the spectral morphology between the target and reference spectra, we first calculate the absolute difference and 
their normalized fluxes at each wavelength bin. Then we take the weighted mean of these flux differences, where the inverse of the 
weighted mean can be seen as the similarity of the spectral morphology. The weights are set to 0.0 in the range 860--868\,nm 
where the two DIBs are located, and to 1.0 elsewhere. An exception to this is the central regions of the \ion{Ca}{ii} lines (849.43--851.03\,nm 
and 853.73--855.73\,nm) where the weights are set to 0.7 because the \ion{Ca}{ii} lines are dominating the RVS spectra of cool stars 
but do not affect the DIB measurements which rely more heavily on the modelling of the \ion{Fe}{i} lines (\citealt{Kos2013}; 
\citetalias{Zhao2022}). Reference spectra with similarity values smaller than three times the S/N of the target spectrum are 
discarded. The remaining spectra are sorted by decreasing similarity, and the first 200 at most are selected as the best
neighbours for the given target spectrum, as best neighbours do not significantly increase the accuracy of the stellar template (\citetalias{Zhao2022}).
We note that the neighbourhood used in BNM refers to the parameter space rather than the physical space, that is the best
neighbours and target should have similar stellar parameters and spectral morphology, but not necessarily similar sky positions.
The best neighbours are averaged with their spectral S/N as weights in order to build a stellar template. The ISM spectrum is defined 
as the target spectrum divided by the stellar template. If the number of the best neighbours is less than 10, no stellar template 
or ISM spectrum is generated. {\dibspec} successfully generated stellar templates for 4\,595\,489 target spectra (76.8\% of the 
target sample) and consequently derived their ISM spectra. 
%--------------------------------------------------------------------------------------------------------------------------------
% \Tomaz{I find it somewhat confusing that a larger value of "similarity" corresponds to spectra with larger differences. I propose that either we inverse the definition of similarity or we call it differently.}
% \Tomaz{It may be worth to add a sentence explaining why we limit the list to just 200 closest neighbours (that you can't take much more anyway, as a maximum in a voxel is 386, may be a good point). In the final statistics it may be worth mentioning what is an average number of neighbours considered (the adopted mininum is 10 and maximum is 200).}
% \hzhao{Sorry Tomaz, the maximum number in the voxel is not the number of best neighbors of a target, but the number of individual
% ISM spectra in the voxel for stacking. We have lost the information about the number of best neighbors for each target star.}

\begin{figure*}[!h]
  \centering
  \includegraphics[width=16.8cm]{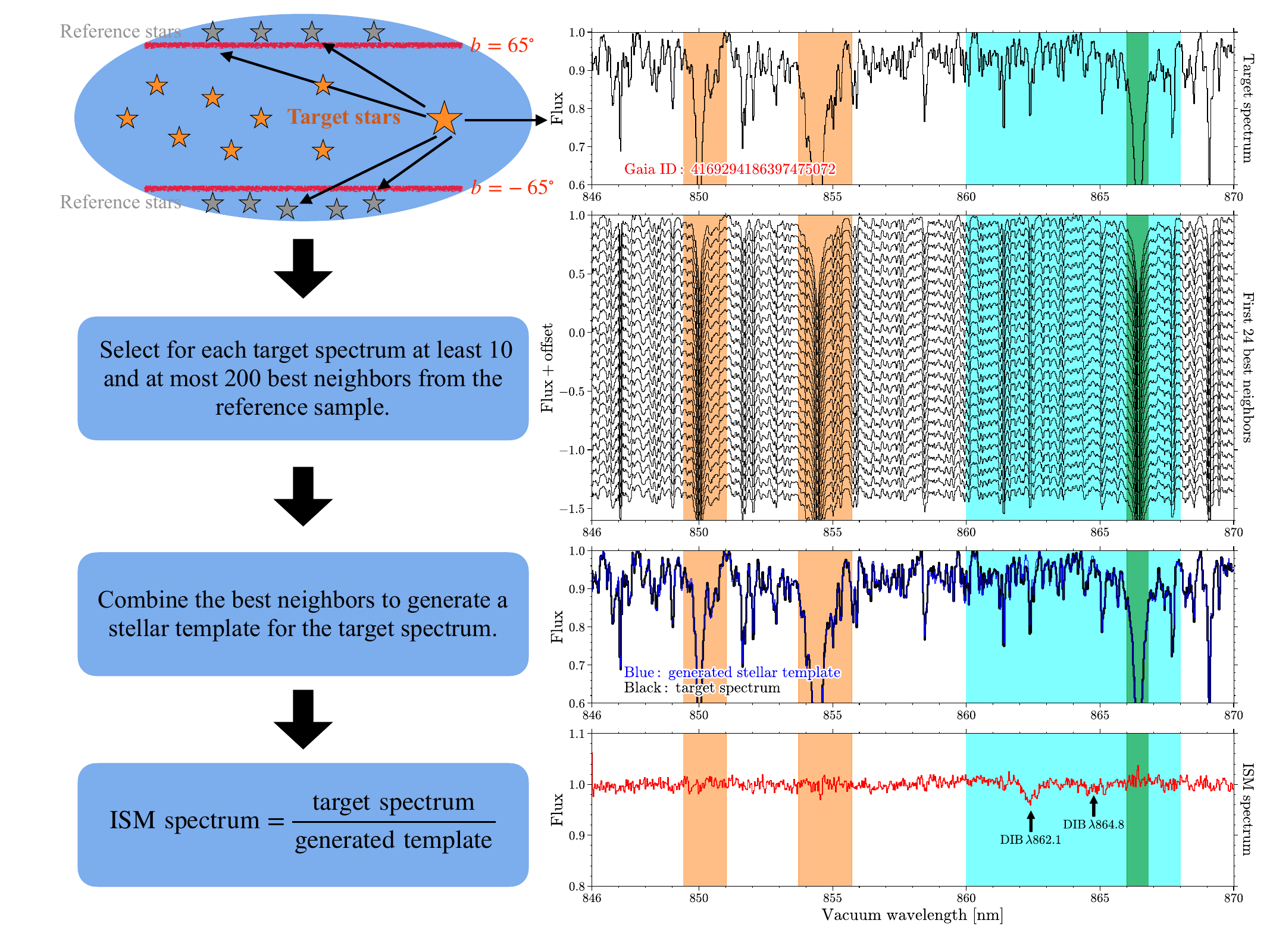}
  \caption{Illustration of deriving ISM spectra for individual targets. {\it Left:} Schematic view of each main step. 
  Detailed explanations are in Sect. \ref{subsect:ism-spec}. {\it Right:} Example for a target (\gaia ID: 4169294186397475072) 
  having $\teff\,{=}\,4445_{-11}^{+15}$\,K, $\logg\,{=}\,1.86\,{\pm}\,0.03$\,dex, $\meta\,{=}\,0.12_{-0.01}^{+0.02}$\,dex,
  and spectral S/N of 110.3. From top to bottom panels: 1) Observed RVS spectrum of this target; 2) First 24 best neighbours
  for this target; 3) The black line is the target spectrum and the blue line is its stellar template built by averaging 
  200 best neighbours; 4) Derived ISM spectrum for this target. The positions of the two DIBs are also indicated. The orange 
  shades indicate the regions of two \ion{Ca}{ii} lines where the weights are set to 0.7 when comparing the target 
  spectrum and reference spectra. The cyan region (860--868\,nm) indicates the spectral window used for fitting the two 
  DIBs, with a masked region (green, 866.0--866.8\,nm) during the fitting because of the residuals of the \ion{Ca}{ii} line.}
  \label{fig:ism-spec}
\end{figure*}

An illustration of BNM is shown in Fig. \ref{fig:ism-spec}: the main steps are summarized on the left side and an example is 
presented on the right side. We made a test of BNM with the reference sample (see Appendix \ref{app:ref-check} for details) and 
found that the performance depends strongly on the spectral S/N. For $\rm S/N\,{>}\,50$, the average flux residuals between 
the RVS spectra and the derived stellar templates are mainly within 0.02 in the DIB window (861.2--866.0\,nm). While the residuals 
also vary with the stellar atmospheric parameters, which is caused by the change in the number density of stars and the accuracy of 
{\gspspec} light for different types of stars. As BNM constrains the SED similarity by S/N and requires at least ten best neighbours 
for a given target spectrum, a target star that has an incorrect spectral type (e.g. an early-type star gains a very low $\teff$) 
or rare reference stars in its vicinity will not get a generated stellar template.
%--------------------------------------------------------------------------------------------------------------------------------

%--------------------------------------------------------------------------------------------------------------------------------

\subsection{Stacking and fitting in each voxel} \label{subsect:stack-fit}

The ISM spectra of individual targets are stacked in each voxel according to their equatorial coordinates ($\alpha,\delta$) and 
the distances ($1/\varpi$) to the target stars to get a much higher spectral S/N than individuals' for a reliable measurement of 
the two DIBs. As the DIB carriers should be located between the observer and the background star, we measure in each
voxel  the sum of the DIB materials from the observer up to the voxel. Considering the size of each voxel, the measured DIB 
feature would be averaged according to the spatial and distance resolution of the voxels. The pixelation in ($\alpha,\delta$) 
is done at level 5 in the HEALPix\footnote{\url{https://healpix.sourceforge.io}} scheme \citep{Gorski2005}, corresponding to 
12\,288 pixels with an equivalent spatial resolution  of $1.8^{\circ}$. For the binning in distance, we selected some adaptive steps instead of a 
uniform bin size because the distribution of the target stars is inhomogeneous in space. Furthermore, as the resolution of the 
sky ($\alpha,\delta$) is fixed as $1.8^{\circ}$, we would like to get as high a resolution in distance as possible, especially 
for nearby regions. 
%--------------------------------------------------------------------------------------------------------------------------------

The S/N of the stacked spectra is a good criterion for determining the size of the distance bins, but the main restriction is that
{\dibspec} runs with defined distance bins as input. Therefore, we cannot use the derived ISM spectra of individual
stars, which were intermediate products of {\dibspec}, to find the best solution for the adaptive
distance bins. Instead, $\rm S/N^{\prime}\,{=}\,\sqrt{\sum {\rm S/N}_i^2}$, where ${\rm S/N}_i$ is the S/N of the $i$th individual 
RVS spectra in a voxel, is used to characterize the stacked ISM spectra in each voxel. The size of the distance bins was determined
by the requirement that in each bin the number of HEALPixels having $\rm S/N^{\prime}\,{\geqslant}\,200$ is larger than 85\% of the
total when distance is smaller than 4.5\kpc. And a constant size of 0.5\,kpc was applied for bins between 4.5 and 10\,kpc. In this
way, there are finally 29 distance bins whose ranges are listed in Table \ref{tab:dist-bin}, together with the number of HEALPixels
in each bin with at least one target star. Certainly, $\rm S/N^{\prime}$ would be much higher than the true S/N of the stacked ISM 
spectra, as it does not account for the uncertainty caused by the subtraction of stellar components. In fact, only $\sim$50\% voxels 
within 4.5\,kpc have $\rm S/N\,{\geqslant}\,200$ (here is the truly calculated S/N of the stacked ISM spectra, see its definition
below). Table \ref{tab:dist-bin} lists the fraction of $\rm S/N\,{\geqslant}\,200$ in each distance bin. The fraction could exceed
85\% when we consider $\rm S/N\,{\geqslant}\,150$, but it will also significantly reduce beyond 1.67\,kpc.
%--------------------------------------------------------------------------------------------------------------------------------

% To produce 3D voxels from the 2D HEALPixels, we define 29 distance bins for each HEALPix
% pixel. The bin size was determined by the requirement that in each distance bin the number of HEALPixels having $\rm SNR^{\prime}\,{\geqslant}\,200$ 
% is larger than 85\% of total within 4.5\,kpc, with a constant bin size of 0.5\,kpc for bins between 4.5 and 10\,kpc. 
% $\rm SNR^{\prime}\,{=}\,\sqrt{\sum {\rm SNR}^2}$ is used to estimate the SNR of the stacked ISM spectra in each voxel. Nevertheless,
% the final calculated SNR of stacked ISM spectra (see its definition below) is lower than $\rm SNR^{\prime}$ because 1) additional
% uncertainty was introduced when deriving ISM spectra; 2) The number of target spectra in a voxel would reduce for the failure of
% the generation of the stellar template. Consequently, only $\sim$50\% voxels (only count voxels containing at least one target,
% $N_{\rm tar}\,{\geqslant}\,1$) within 4.5\,kpc have $\rm SNR\,{\geqslant}\,200$. Table \ref{tab:dist-bin} lists the range of the 
% 29 distance bins, the number of HEALPixels with $N_{\rm tar}\,{\geqslant}\,1$ in each distance bin, and the fraction of $\rm 
% SNR\,{\geqslant}\,150$.
%--------------------------------------------------------------------------------------------------------------------------------

\begin{table}
\centering
\caption{Distance bins defined for the stacking of ISM spectra.}
\label{tab:dist-bin}
\begin{tabular}{rrr}
\hline\hline 
Distance bin (pc) & Pixel number\tablefootmark{a} & $\rm S/N\,{\geqslant}\,200$ (\%)  \\ [0.05ex]
\hline        
    0--130 & 11\,179 & 48.7\% \\
  130--200 & 11\,202 & 67.3\% \\
  200--270 & 11\,204 & 68.0\% \\
  270--350 & 11\,207 & 66.6\% \\
  350--430 & 11\,203 & 52.9\% \\
  430--530 & 11\,204 & 50.2\% \\
  530--650 & 11\,200 & 48.4\% \\
  650--780 & 11\,189 & 51.3\% \\
  780--950 & 11\,167 & 64.8\% \\
 950--1150 & 11\,146 & 66.1\% \\
1150--1400 & 11\,135 & 63.4\% \\
1400--1670 & 11\,093 & 53.0\% \\
1670--2000 & 11\,008 & 46.3\% \\
2000--2250 & 10\,656 & 30.9\% \\
2250--2550 & 10\,416 & 27.3\% \\
2550--2950 & 10\,050 & 27.3\% \\
2950--3500 &  9\,464 & 26.1\% \\
3500--4500 &  9\,060 & 25.0\% \\
4500--5000 &  6\,659 &  7.9\% \\
5000--5500 &  5\,943 &  5.2\% \\
5500--6000 &  5\,164 &  3.0\% \\
6000--6500 &  4\,528 &  1.2\% \\
6500--7000 &  3\,876 &  0.5\% \\
7000--7500 &  3\,392 &  0.3\% \\
7500--8000 &  2\,959 &  0.1\% \\
8000--8500 &  2\,541 &  0.0\% \\
8500--9000 &  2\,147 &  0.1\% \\
9000--9500 &  1\,838 &  0.0\% \\
9500--10000 & 1\,598 &  0.1\% \\ [0.05ex]
\hline
\end{tabular}
\tablefoot{ \\
\tablefoottext{a}{Number of HEALPixels with $N_{\rm tar}\,{\geqslant}\,1$ in each distance bin.} 
}
\end{table}
%--------------------------------------------------------------------------------------------------------------------------------

The ISM spectra are stacked in each voxel by taking, in each wavelength bin, the median value of the fluxes in order to reduce the 
influence of the outliers. The flux uncertainty at each wavelength bin is taken as the mean of the individual flux uncertainties 
divided by $\sqrt{N_{\rm tar}}$. The S/N of the stacked ISM spectra is calculated between 860.2 and 861.2 nm as $\rm mean(flux)/std(flux)$. 
%--------------------------------------------------------------------------------------------------------------------------------

For the profiles of the two DIBs, $\lambda$862.1 was usually assumed to have a Gaussian profile in previous studies 
\citep[e.g.][]{Kos2013,hz2021a}. Although some departures from a Gaussian profile were reported \citep[e.g.][]{Krelowski2019b}, 
the origin is more like the superposition of multiple DIB clouds, as no evidence supports an intrinsic asymmetry of the $\lambda$862.1
profile. Thus, we still assume a Gaussian profile for $\lambda$862.1 in this work and treat the possible departures as a source of
uncertainty. Rare studies of $\lambda$864.8 make it harder to determine the shape of its profile. \citet{Zhao2022} chose a Lorentzian
profile as it showed smaller residuals on the ISM spectra compared to a Gaussian profile. Additionally, the Lorentzian profile has been proved 
to be appropriate for the very broad DIB\,$\lambda$442.8 \citep{Snow2002} while $\lambda$864.8 has a broad profile as well. Therefore,
{\dibspec} models the profiles of the two DIBs in stacked ISM spectra by a Gaussian function (Eq. \ref{eq:gauss}) for $\lambda$862.1, 
a Lorentzian function (Eq. \ref{eq:lorentz}) for $\lambda$864.8, and a linear function for continuum (Eq. \ref{eq:cont}):

\begin{equation} \label{eq:gauss}
    G(\lambda;\dibdepth,\diblambda,\dibwidth) = -\dibdepth \times {\rm exp}\left(-\frac{(\lambda-\diblambda)^2}{2\dibwidth^2}\right),
\end{equation}

\begin{equation} \label{eq:lorentz}
    L(\lambda;\dibdepth,\diblambda,\dibwidth)= \frac{-(\dibdepth \dibwidth^2)}{(\lambda-\diblambda)^2+\dibwidth^2},
\end{equation}

\begin{equation} \label{eq:cont}
    C(\lambda;a_0,a_1)= a_0 \times \lambda + a_1,\end{equation}
%--------------------------------------------------------------------------------------------------------------------------------

where $\dibdepth$ and $\dibwidth$ are the depth and width of the DIB profile, $\diblambda$ is the measured central 
wavelength, $a_0$ and $a_1$ describe the linear continuum, and $\lambda$ is the wavelength. Subscripts `862.1' and `864.8' are 
used to distinguish the profile parameters of the two DIBs. The full DIB model, $M_{\Theta}$, is the sum of Eqs. \ref{eq:gauss}--\ref{eq:cont},
where $\Theta=\{\mathcal{D}_{862.1},\lambda_{862.1},\sigma_{862.1},\mathcal{D}_{864.8},\lambda_{864.8},\sigma_{864.8},a_0,a_1\}$ 
are the adjusted model parameters. Given the stacked ISM spectrum $\{\lambda,y,\sigma_y\}$, where $y$ is the normalized flux and 
$\sigma_y$ is the flux uncertainty, and the unnormalized posterior probability density function (PDF) is $P(\Theta|y) \propto P(y|\Theta)P(\Theta)$. 
$P(y|\Theta)$ is the likelihood:

\begin{equation} \label{eq:llh}
    P(y|\Theta) = \frac{1}{\sqrt{2\pi}\sigma_y} {\rm exp}\left[-\frac{1}{2\sigma_y^2}\left(y-M_{\Theta}\right)^2\right],
\end{equation}

\noindent and $P(\Theta)$ represents the prior distributions of the parameters. Flat and independent priors were applied for the 
DIB parameters, that is 0--0.2 for $\mathcal{D}_{862.1}$ and $\mathcal{D}_{864.8}$, 861.2--863.0\,nm for $\lambda_{862.1}$,
863.0--866.0\,nm for $\lambda_{864.8}$, 0.01--0.5\,nm for $\sigma_{862.1}$, and 0.1--1.5\,nm for $\sigma_{864.8}$. No priors were
used for $a_0$ and $a_1$.
%--------------------------------------------------------------------------------------------------------------------------------

The optimization of the eight parameters -- three for each DIB plus two for the continuum -- was done by sampling their full posterior
distributions. We note that during the optimization, a masked region was applied between 866.0 and 866.8\,nm for the residuals 
\ion{Ca}{ii} that were caused by downweighting the central regions of \ion{Ca}{ii} lines in the BNM (see Sect. \ref{subsect:ism-spec}).
A Markov chain Monte Carlo (MCMC) procedure \citep{Foreman-Mackey13} was performed to implement the parameter estimates. The initial 
guess of $\dibdepth$ and $\diblambda$ were determined by averaging flux near the lowest point (5 pixels) within 861.2--863.0\,nm 
and 863.5--866.0\,nm for $\lambda$862.1 and $\lambda$864.8, respectively. And initial $\dibwidth$ was fixed as 0.12\,nm for 
$\lambda$862.1 and 0.4\,nm for $\lambda$864.8. $a_0$ and $a_1$ were initially set as 0 and 1 as well. One hundred walkers were 
progressed for 250 steps to complete the burn-in stage. The best fits derived by the last 200 steps were then used as the initial 
conditions to sample the posterior with 100 walkers and 2000 steps. The best estimates and their statistical uncertainties were 
taken in terms of the 50th, 16th, and 84th percentiles of the posterior PDF drawn from the last 1500 steps. 
According to Eqs. \ref{eq:gauss} and \ref{eq:lorentz}, the equivalent width (EW) for $\lambda$862.1 is calculated as 
%--------------------------------------------------------------------------------------------------------------------------------

\begin{equation} \label{eq:ew8621}
    \rm EW_{862.1} =\sqrt{2\pi} \times \mathcal{D}_{862.1} \times \sigma_{862.1},
\end{equation}

\noindent and for $\lambda$864.8 as 

\begin{equation} \label{eq:ew8648}
    \rm EW_{864.8}=\pi \times \mathcal{D}_{864.8} \times \sigma_{864.8}.
\end{equation}

The lower (16\%) and upper (84\%) confidence levels of EW were estimated by $\dibdepth$ and $\dibwidth$ 
drawn from the MCMC posterior samplings. Specifically, each $\{\dibdepth,\dibwidth\}$ pair sampled during the MCMC fitting
was used to calculate one value of EW, and finally we had an EW distribution from which the 16th, 50th, and 84th percentile values
were calculated.
%--------------------------------------------------------------------------------------------------------------------------------

\begin{figure}[!h]
  \centering
  \includegraphics[width=8.4cm]{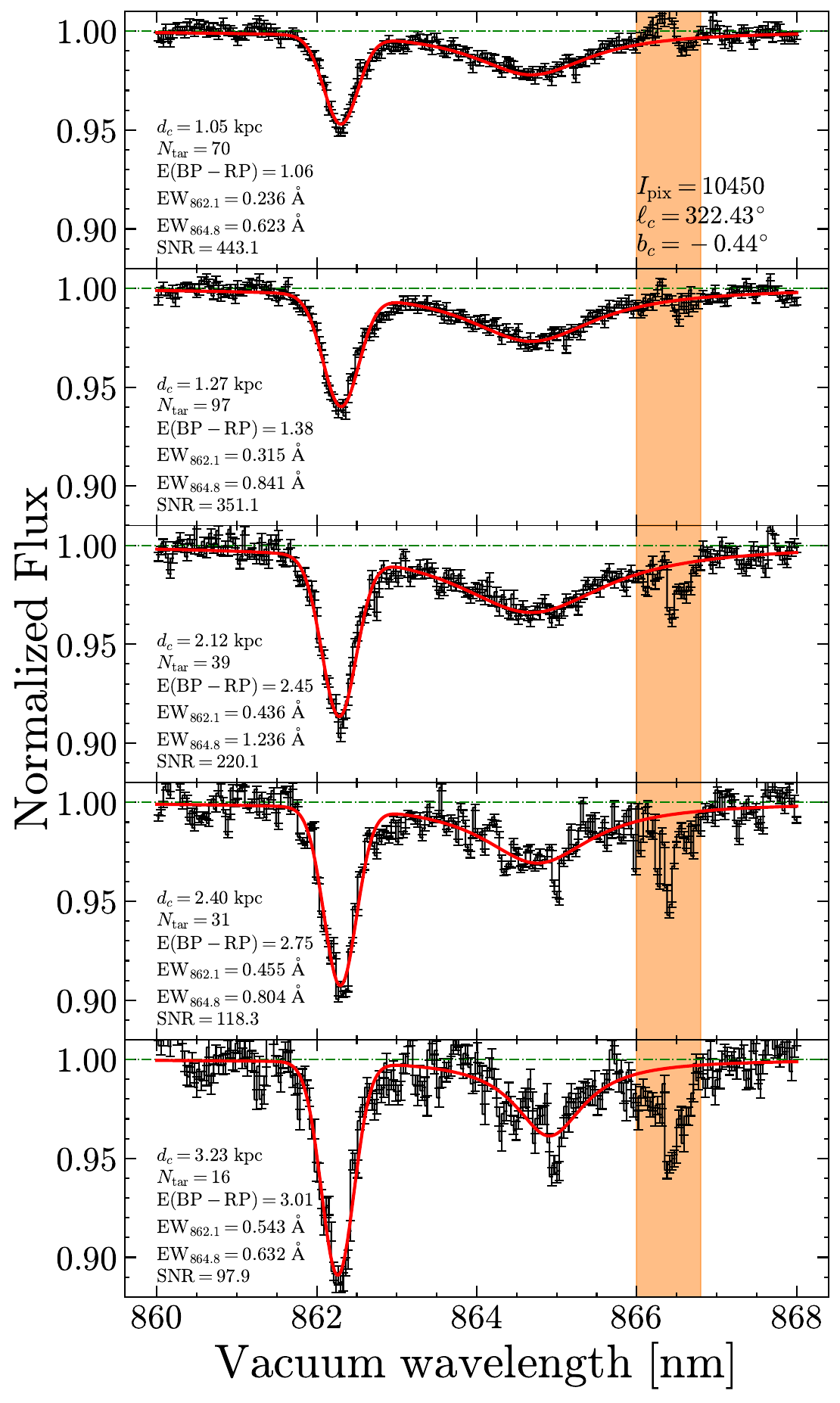}
  \caption{Examples of the fits to DIBs $\lambda$862.1 and $\lambda$864.8 in stacked ISM spectra in five voxels in the same
  direction, whose HEALPix number ($I_{\rm pix}\,{=}\,10\,450$) and GC $(\ell_c,b_c)\,{=}\,(322.43^{\circ},{-}0.44^{\circ})$
  are marked in the top panel. The black and red lines are the ISM spectra and fitted DIB profiles, respectively, normalized 
  by the fitted linear continuum. The error bars indicate the flux uncertainties at each pixel. Orange indicates the region 
  that was masked during the fittings. The central heliocentric distance ($d_c$), the number of target spectra ($N_{\rm tar}$), 
  mean $\EBPRP$, EWs of two DIBs, and the S/N of the stacked ISM spectrum in each voxel are indicated as well.}
  \label{fig:stack-fit}
\end{figure}

Figure \ref{fig:stack-fit} shows examples of the fits for five stacked ISM spectra in voxels towards the same direction, that is
HEALPixel number of 10\,450, corresponding to Galactic coordinates of the voxel of $(\ell_c,b_c)\,{=}\,(322.43^{\circ},{-}0.44^{\circ})$. 
With increasing voxel central heliocentric distance ($d_c$) from top to bottom, $\rm EW_{862.1}$ and mean $\EBPRP$ measured
by {\gspphot} \citep{Andrae2022} both increase and show a good correlation with each other (these values are indicated
in Fig. \ref{fig:stack-fit}). The profiles of $\lambda$862.1 are strong and prominent in all these voxels. On the other hand, 
the S/N of the stacked ISM spectra and the number of target spectra ($N_{\rm tar}$) decrease with distance. Consequently, the fit 
to the profile of $\lambda$864.8 becomes worse in voxels with $d_c\,{=}\,2.40$ and 3.23\,kpc (bottom panels in  Fig.~\ref{fig:stack-fit}) 
as the very broad and shallow profile of $\lambda$864.8 is more affected by the stellar residuals and uncertainties introduced by 
BNM and stacking than $\lambda$862.1. 
%--------------------------------------------------------------------------------------------------------------------------------

The corner plots of these examples, presenting the one- and two-dimensional projections of the posterior distributions of the 
fitted parameters, are shown in Figs. \ref{fig:corner0}--\ref{fig:corner4}, respectively. The corner plots clearly show that the 
MCMC fitting is converged and the posterior PDF of all the parameters is Gaussian. $\dibdepth$ and $\dibwidth$ are correlated with 
each other during the fitting for both $\lambda$862.1 and $\lambda$864.8. The depth and central position of $\lambda$862.1 
($\mathcal{D}_{862.1}$ and $\lambda_{862.1}$) are not sensitive to the continuum placement ($a_0$ and $a_1$), while $\sigma_{862.1}$ 
presents a weak correlation with $a_0$ and $a_1$. The continuum placement has a heavier effect on the broad and shallow profile 
of $\lambda$864.8.
%--------------------------------------------------------------------------------------------------------------------------------

%--------------------------------------------------------------------------------------------------------------------------------

\begin{table*}
\centering
\caption{Column names and their descriptions of the parameter table of {\dibspec} (interstellar\_medium\_params).}
\label{tab:catalog-params}
\begin{tabular}{lccl}
\hline\hline 
Column & Name & Unit & Description  \\ [0.05ex]
\hline        
1  & solution\_id  & -     & Solution Identifier \\
2  & healpix       & -     & HEALPix identification \\
3  & lc            & deg   & Central Galactic longitude of each voxel \\
4  & bc            & deg   & Central Galactic latitude of each voxel \\
5  & dc            & kpc   & Central heliocentric distance of each voxel \\
6  & n\_targets    & -     & Number of target stars in each voxel \\
7  & snr           & -     & S/N of the stacked ISM spectrum \\
8  & ew8620        & nm    & Equivalent width of DIB at 862.1\,nm ($\rm EW_{862.1}$) \\
9  & ew8620\_lower & nm    & Lower confidence level (16\%) of equivalent width of DIB\,$\lambda$862.1 \\
10 & ew8620\_upper & nm    & Upper confidence level (84\%) of equivalent width of DIB\,$\lambda$862.1 \\ 
11 & flags8620     & -     & Quality flag of the parameters of DIB at 862.1\,nm \\
12 & p08620        & -     & Depth of DIB at 862.1\,nm ($\mathcal{D}_{862.1}$) \\
13 & p08620\_lower & -     & Lower confidence level (16\%) of depth of DIB\,$\lambda$862.1 \\
14 & p08620\_upper & -     & Upper confidence level (84\%) of depth of DIB\,$\lambda$862.1 \\
15 & p18620        & nm    & Central wavelength of DIB at 862.1\,nm ($\lambda_{862.1}$) \\
16 & p18620\_lower & nm    & Lower confidence level (16\%) of central wavelength of DIB\,$\lambda$862.1 \\
17 & p18620\_upper & nm    & Upper confidence level (84\%) of central wavelength of DIB\,$\lambda$862.1 \\
18 & p28620        & nm    & Gaussian width of DIB at 862.1\,nm ($\sigma_{862.1}$) \\
19 & p28620\_lower & nm    & Lower confidence level (16\%) of Gaussian width of DIB\,$\lambda$862.1 \\
20 & p28620\_upper & nm    & Upper confidence level (84\%) of Gaussian width of DIB\,$\lambda$862.1 \\
21 & ew8648        & nm    & Equivalent width of DIB at 864.8\,nm ($\rm EW_{864.8}$) \\
22 & ew8648\_lower & nm    & Lower confidence level (16\%) of equivalent width of DIB\,$\lambda$864.8 \\
23 & ew8648\_upper & nm    & Upper confidence level (84\%) of equivalent width of DIB\,$\lambda$864.8 \\ 
24 & flags8648     & -     & Quality flag of the parameters of DIB at 864.8\,nm \\
25 & p08648        & -     & Depth of DIB at 864.8\,nm ($\mathcal{D}_{864.8}$) \\
26 & p08648\_lower & -     & Lower confidence level (16\%) of depth of DIB\,$\lambda$864.8 \\
27 & p08648\_upper & -     & Upper confidence level (84\%) of depth of DIB\,$\lambda$864.8 \\
28 & p18648        & nm    & Central wavelength of DIB at 864.8\,nm ($\lambda_{864.8}$) \\
29 & p18648\_lower & nm    & Lower confidence level (16\%) of central wavelength of DIB\,$\lambda$864.8 \\
30 & p18648\_upper & nm    & Upper confidence level (84\%) of central wavelength of DIB\,$\lambda$864.8 \\
31 & p28648        & nm    & Lorentzian width of DIB at 864.8\,nm ($\sigma_{864.8}$) \\
32 & p28648\_lower & nm    & Lower confidence level (16\%) of Lorentzian width of DIB\,$\lambda$864.8 \\
33 & p28648\_upper & nm    & Upper confidence level (84\%) of Lorentzian width of DIB\,$\lambda$864.8 \\
34 & dibcont\_a0   & -     & Slope of the linear continuum fitted to the stacked ISM spectrum ($a_0$) \\
35 & dibcont\_a0\_lower & -  & Lower confidence level (16\%) of the slope of continuum \\
36 & dibcont\_a0\_upper & -  & Upper confidence level (84\%) of the slope of continuum \\ 
37 & dibcont\_a1   & -     & Intercept of the linear continuum fitted to the stacked ISM spectrum ($a_1$) \\ [0.05ex]
\hline
\end{tabular}
\end{table*}
%--------------------------------------------------------------------------------------------------------------------------------

\begin{table*}
\centering
\caption{Column names and their descriptions of the spectra table of {\dibspec} (interstellar\_medium\_spectra).}
\label{tab:catalog-spectra}
\begin{tabular}{lccl}
\hline\hline 
Column & Name & Unit & Description  \\ [0.05ex]
\hline        
1  & solution\_id  & -   & Solution Identifier \\
2  & healpix     & -     & HEALPix identification \\
3  & lc          & deg   & Central Galactic longitude of each voxel \\
4  & bc          & deg   & Central Galactic latitude of each voxel \\
5  & dc          & kpc   & Central heliocentric distance of each voxel \\
6  & lambda      & {\AA} & Wavelength in vacuum \\
7  & flux        & -     & Normalized flux\\
8  & flux\_uncertainty   & -    & Uncertainty of the normalized flux \\
\hline
\end{tabular}
\end{table*}
%--------------------------------------------------------------------------------------------------------------------------------

\section{{\dibspec} outputs} \label{sect:catalog}

% The outputs of {\dibspec} are published as an archive table containing DIB measurements in 235\,428 voxels and the corresponding
% stacked ISM spectra in each voxel. These tables can be queried from the \href{https://gea.esac.esa.int/archive/}{\gaia archive} 
% with the common search and query tools (e.g. ADQL query). The goal of {\dibspec} is to provide a homogeneous DIB catalog containing 
% measurements of $\lambda$862.1 and $\lambda$864.8 in stacked ISM spectra in each voxel. Compared to DR3, we are able to reach a higher 
% SNR by stacking spectra, and thus we can explore a larger Galactic volume albeit at a lower spatial resolution. Moreover, we are 
% free of assumptions on the underlying stellar spectra (i.e. we avoid synthetic stellar spectra).
% This is 
% a different approach from the one used in \citet{GSPspecDR3} with the advantages of increasing the detection distance and containing the
% broad DIB\,$\lambda$864.8, and the disadvantage of decreasing the spatial resolution ($\rm 1.8^{\circ}$ for {\dibspec}).
% %--------------------------------------------------------------------------------------------------------------------------------

\dibspec successfully fitted the two DIBs in 235\,428 voxels. This is less than the total number (356\,352) of a level 5 HEALPix 
binning $\times$ 29 distance bins, which is due to discarding voxels with no target stars (targets with failed generated 
stellar templates were not included as well). Each voxel on average contains 20 target stars, with a maximum number of 386. There 
are 23\,875 voxels (10.1\%) that only contain one target spectrum. The outputs of {\dibspec} are arranged into two tables in the 
\gaia Archive\footnote{\url{https://gea.esac.esa.int/archive/}}: (1) `interstellar\_medium\_params': the parameter table listing 
the fitted DIB parameters in each voxel; (2) `interstellar\_medium\_spectra': the spectra table containing the stacked ISM spectra 
in each voxel. The column names, units, and descriptions of the two tables are given in Tables \ref{tab:catalog-params} and 
\ref{tab:catalog-spectra} respectively. In Table \ref{tab:catalog-params}, the symbols of the full DIB parameters (Eqs. 
\ref{eq:gauss}--\ref{eq:cont}) and EWs are indicated in the description of the corresponding columns. 
%--------------------------------------------------------------------------------------------------------------------------------

There are some notes for the {\dibspec} outputs:
\begin{enumerate}
    \item As the column names were pre-defined, the DIB at 862.1\,nm was cited as `8620' in related names, but we prefer to use 
          `862.1' in the descriptions and in the context of this Gaia product.
    \item Second, the lower and upper confidence levels of the intercept of the linear continuum (`dibcont\_a1') are not included 
          in the parameter table because they were not defined at the time of the processing and archive table definition.
    \item As in {\dibspec} the HEALPix binning was done in the equatorial system, following the \gaia convention\footnote{See 
          Bastian and Portell (2020): Source Identifiers -- Assignment and Usage throughout DPAC (GAIA-C3-TN-ARI-BAS-020), 
          which can be accessed through https://www.cosmos.esa.int/web/gaia/public-dpac-documents.}, the Galactic coordinates 
          of the voxel centre (`lc' and `bc' in the table) were converted from the equatorial coordinates of the centre of each HEALPixel.
    \item The fitted DIB parameters result from the integration of their carriers from the voxels to the observer, like dust extinction, 
          rather than from one voxel to the next. 
    \item About 5.4\% (12\,692) of stacked spectra have null flux uncertainties. This is due to the fact that the first pixels of the 
          individual RVS spectra for stacking do have zero values. Their flux uncertainties are fixed as 0.01 for the MCMC fittings.
    \item The spectra table contains 62\,859\,276 rows which equal 235\,428 voxels $\times$ 267 wavelength bins of the stacked spectra, 
          that is each row in the spectra table contains information of one wavelength bin. A simple Python script shown in Sect. 
          \ref{app:python-code} can be used to convert the spectra table to a fits file, in which each row stands for one stacked ISM spectra.
\end{enumerate}

Below, we describe and discuss the fitted DIB parameters and their uncertainties.

\subsection{S/N and DIB quality flag (QF)} \label{subsect:qf}

\begin{figure}
\centering
\includegraphics[width=8.4cm]{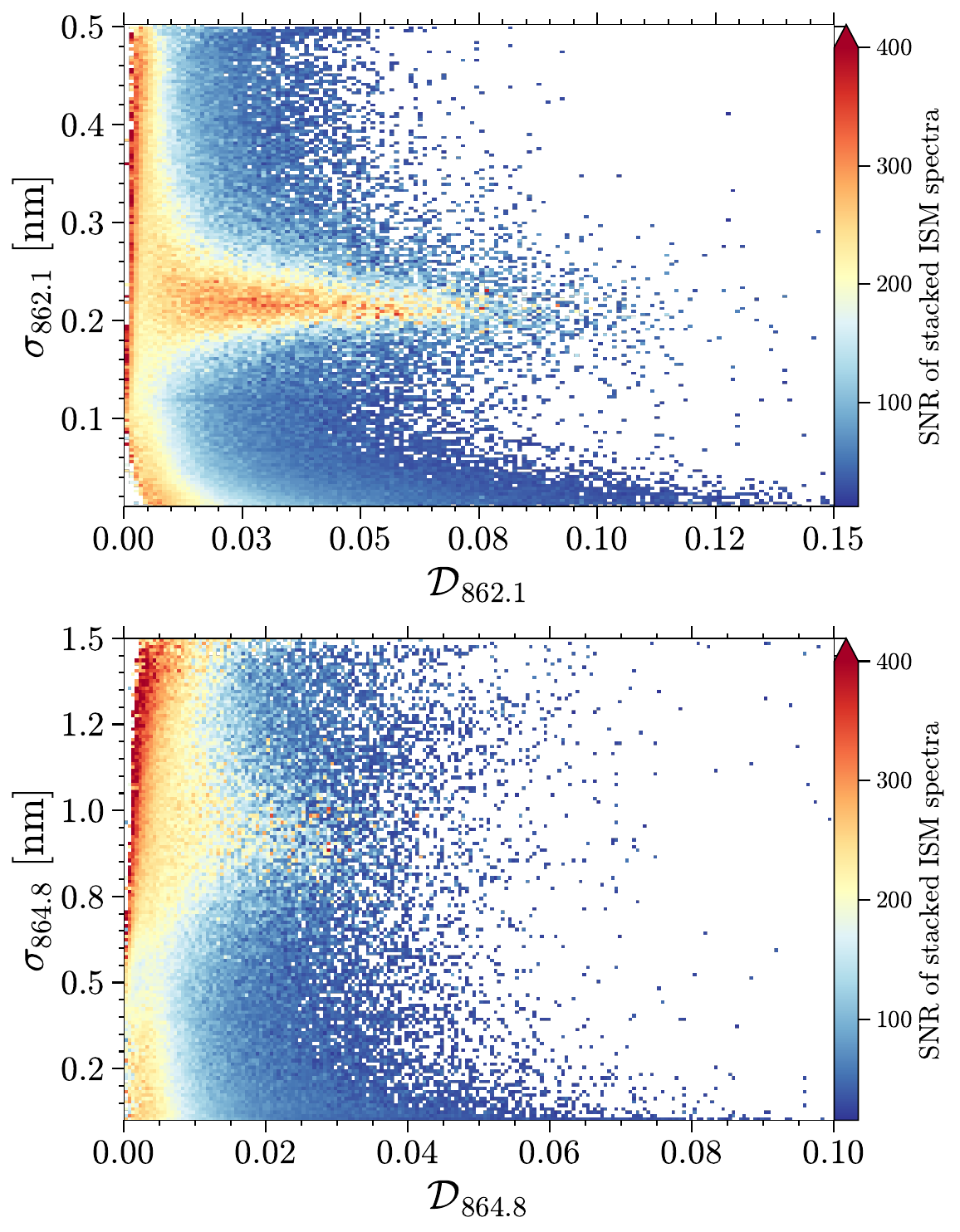}
\caption{Distribution of the S/N of the stacked ISM spectra in the DIB depth--width plane ($\dibdepth$ vs. $\dibwidth$) for DIBs
$\lambda$862.1 and $\lambda$864.8, respectively. The colour represents the mean S/N in each $0.001 \times 0.003$\,nm bin for 
$\lambda$862.1 and $0.001 \times 0.01$\,nm bin for $\lambda$864.8.}
\label{fig:snr} 
\end{figure}

The S/N of the stacked ISM spectra strongly affects the quality of the DIB fit. S/N is determined by the quality of individual RVS 
spectra, the number of target stars in each voxel, and the performance of the BNM on target spectra. For {\dibspec} results, 42\% 
of the voxels have a stacked spectrum $\rm S/N\,{>}\,200$, but 59\% of these voxels are within 1\,kpc. The DIB signal will be more 
easily detected in spectra with higher S/N, while the DIB depth and strength are generally smaller in the solar neighbourhood than 
in distant zones. Thus, the goodness-of-fit for DIBs cannot be determined simply by S/N. On one hand, the distribution of S/N 
shown in Fig. \ref{fig:snr} presents a dependence on both $\dibdepth$ and $\dibwidth$. A main part of high S/N ($\gtrsim$200) is 
found with very small $\dibdepth$, corresponding to nearby voxels containing weak or no DIB signals. On the other hand, large 
$\dibdepth$ found in noisy spectra (low S/N) would indicate a needle-like spurious DIB signal caused by the random noise or the 
stellar residuals, especially in the regions with relatively small $\dibwidth$. Nevertheless, one can also find some regions with 
both high S/N and large $\dibdepth$, indicating a pronounced DIB signal. Such regions are evident near $\dibwidth\,{\sim}\,0.22$\,nm 
for DIB\,$\lambda$862.1, and between 0.8 and 1.0\,nm for DIB\,$\lambda$864.8, but not particularly obvious due to the fact that 
$\lambda$864.8 is much broader and shallower than $\lambda$862.1. 
%--------------------------------------------------------------------------------------------------------------------------------

%--------------------------------------------------------------------------------------------------------------------------------

\begin{figure*}
\centering
\includegraphics[width=13cm]{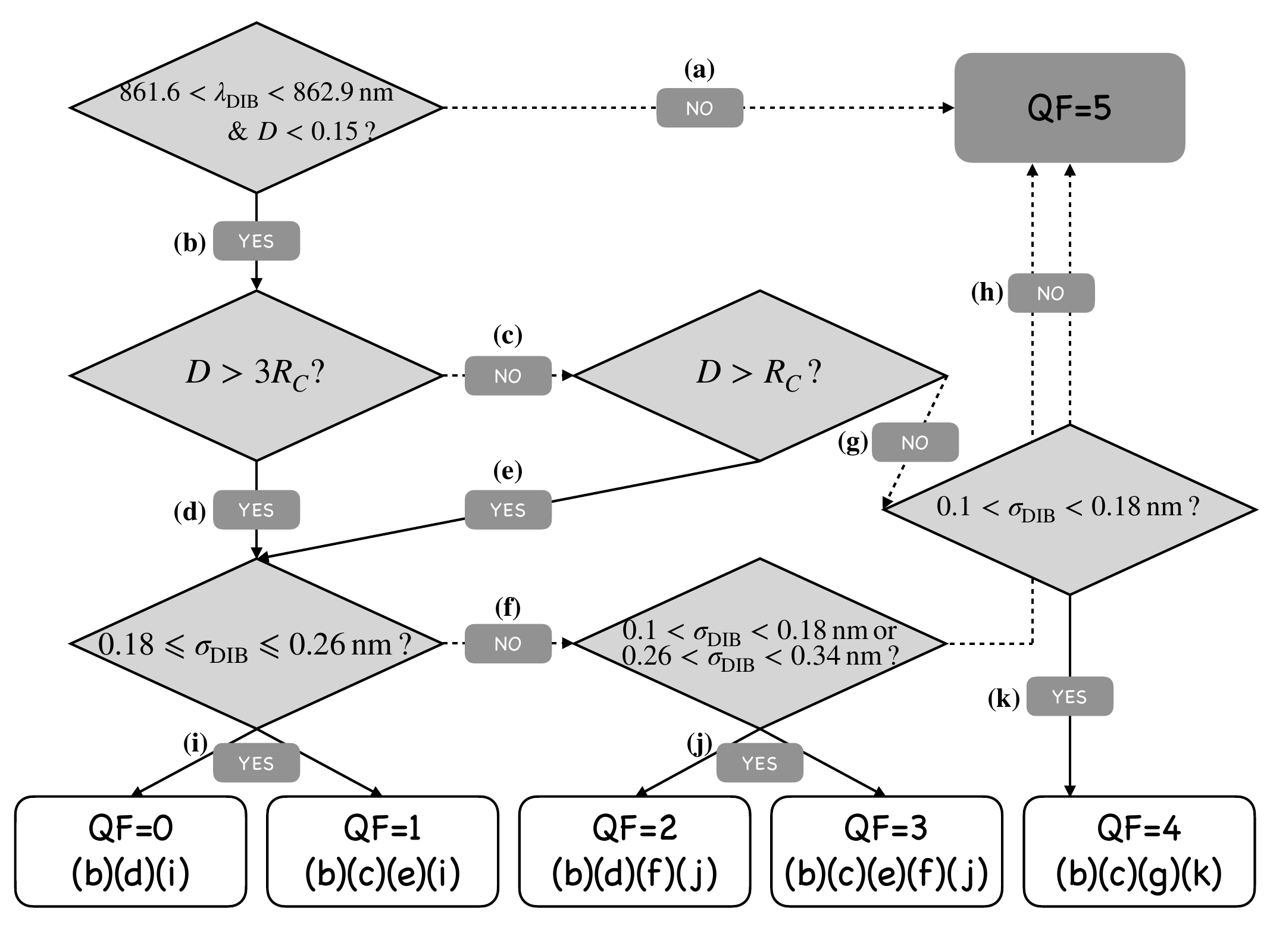}
\includegraphics[width=13cm]{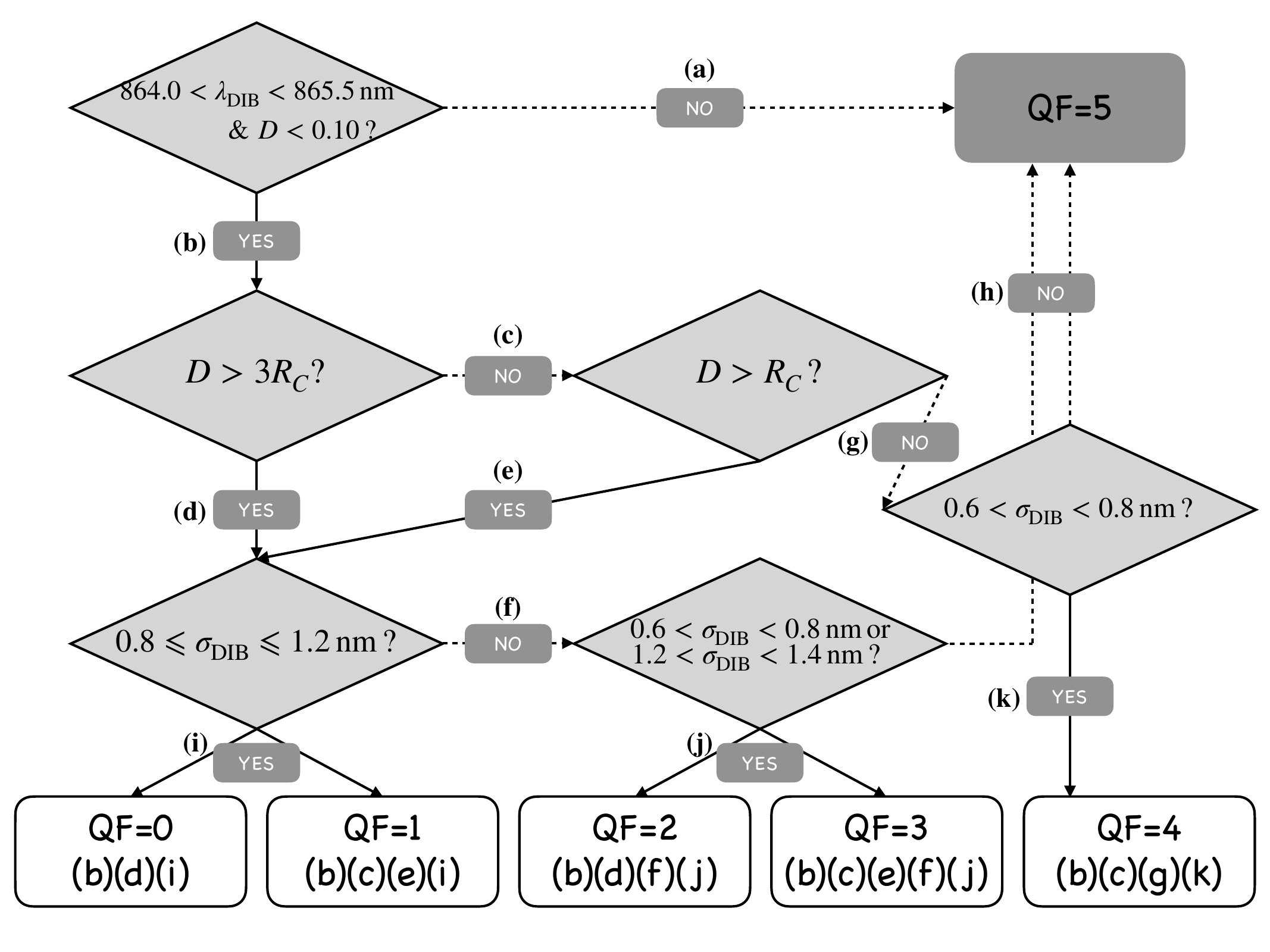}
\caption{Flowchart of the criteria to generate quality flags (QFs) for DIBs\,$\lambda$862.1 (upper panel) and $\lambda$864.8
(lower panel), respectively. The flag numbers and the corresponding arrow paths for classification are listed in the bottom box 
in each panel.}
\label{fig:qf} 
\end{figure*}
%--------------------------------------------------------------------------------------------------------------------------------

Considering both S/N and the shape of the fitted DIB profile ($\{\dibdepth,\dibwidth\}$), we generated quality flags (QF) to describe 
the quality of the fits. This idea comes from \citet{Elyajouri2016} and was applied to DIB\,$\lambda$862.1 in 
\gaia DR3 \citep{GSPspecDR3,PVP} as well. In the present work, we follow the same scheme for $\lambda$862.1 as in DR3, but some borders 
of the DIB parameters are redefined according to the {\dibspec} results. The scheme for $\lambda$864.8 contains newly defined 
borders, as it has been little investigated to date.
%--------------------------------------------------------------------------------------------------------------------------------

\begin{figure}
\centering
\includegraphics[width=8cm]{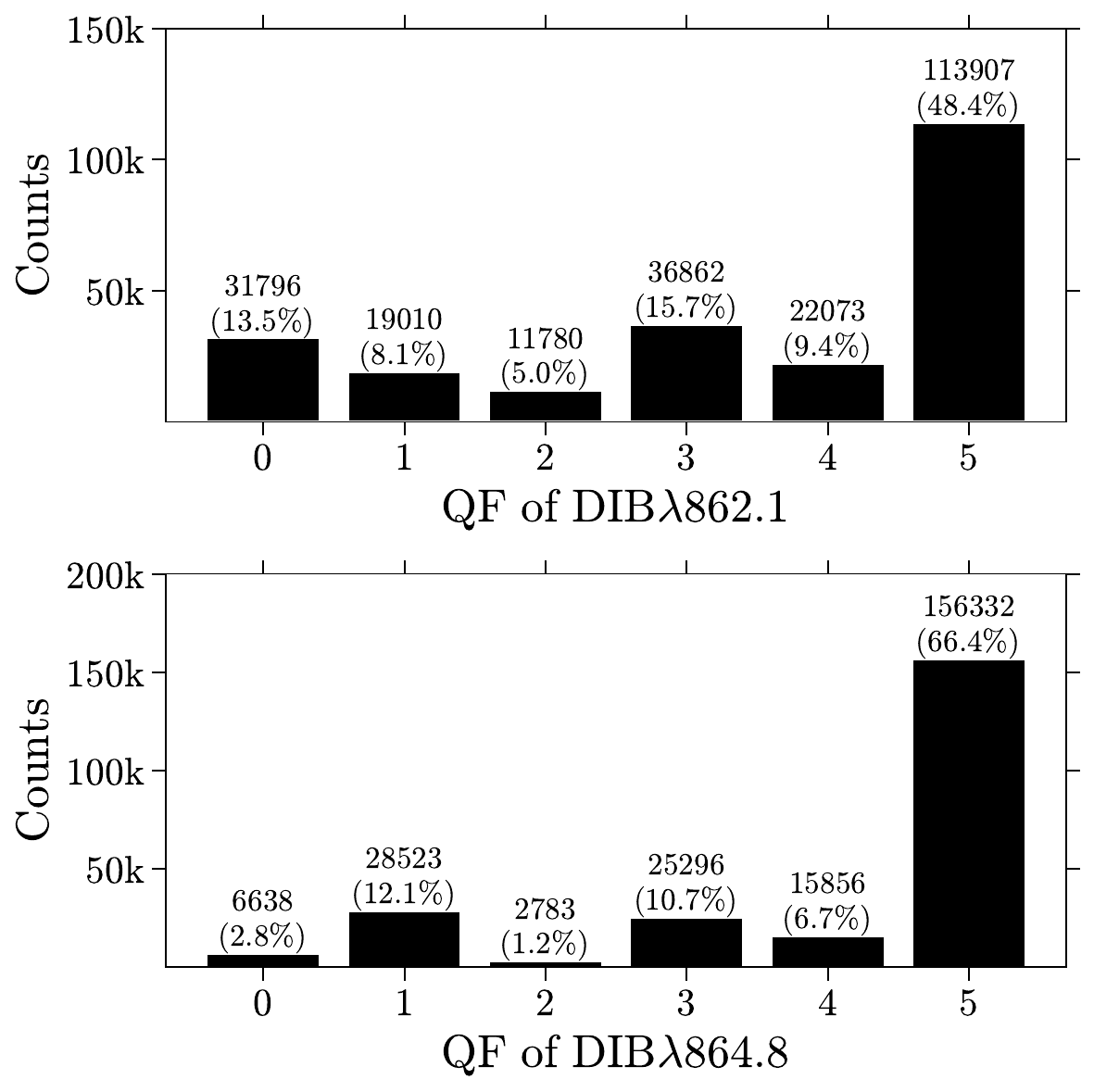}
\caption{Distribution of DIB quality flags (QFs) at each level (0--5, 0 is the best and 5 is the worst) for DIBs\,$\lambda$862.1 
(upper panel) and $\lambda$864.8 (lower panel). The number of DIB detections and the fraction are indicated at the top of each bar.}
\label{fig:qf-distrib} 
\end{figure}
%--------------------------------------------------------------------------------------------------------------------------------

\begin{figure*}[!h]
   \centering
   \includegraphics[width=14cm]{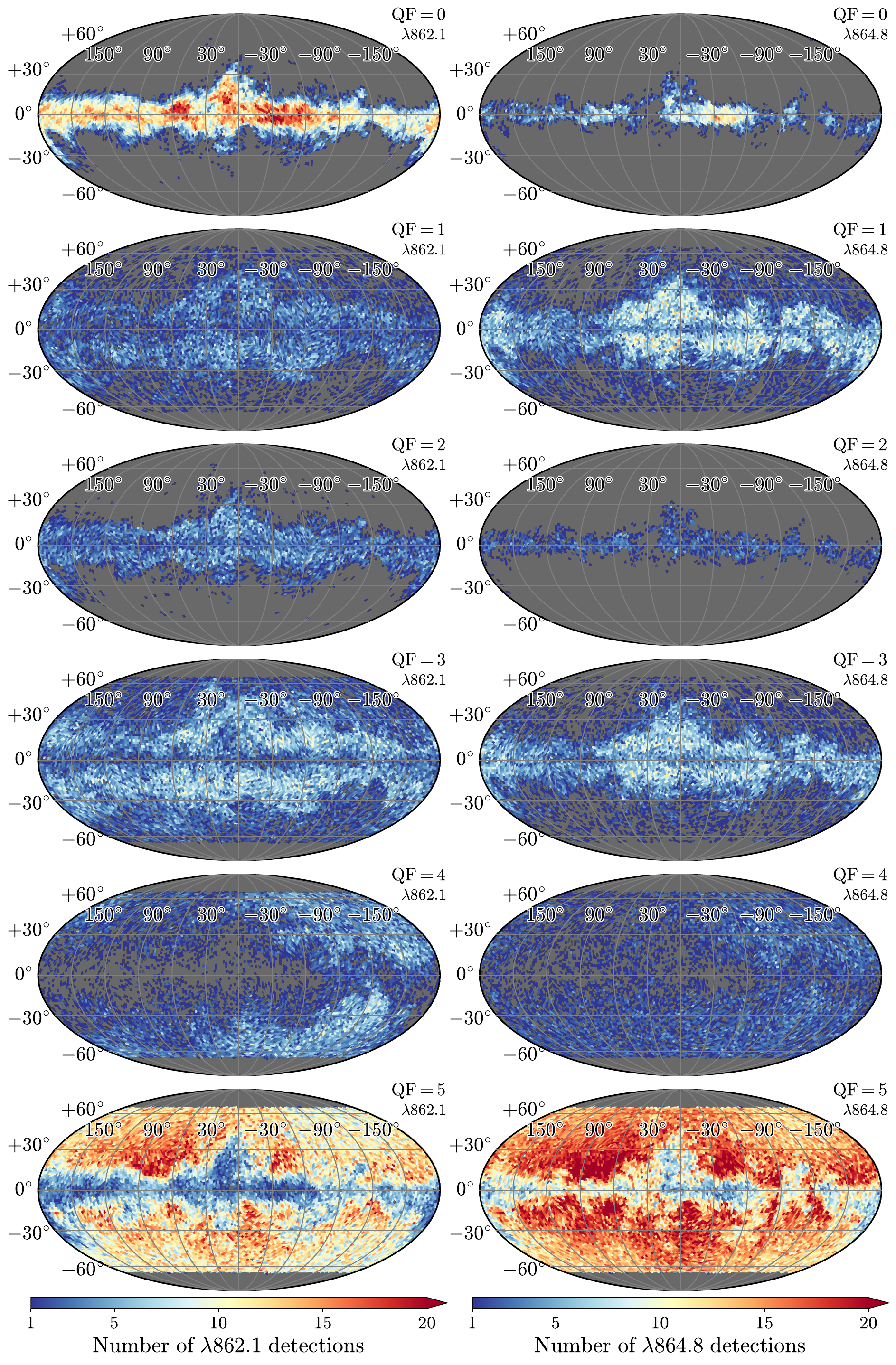}
   \caption{Galactic spatial distribution of each level of DIB QF (0--5) for $\lambda$862.1 ({\it left panels}) and $\lambda$864.8 
   ({\it right panels}). The colour represents the counts of DIB detections in each HEALPixel at various selected distances. This HEALPix 
   map has a level of 5, according to a spatial resolution of $1.8^{\circ}$. }
   \label{fig:qf-Gmap}
\end{figure*}
%--------------------------------------------------------------------------------------------------------------------------------

Figure \ref{fig:qf} shows the flowchart for the generation of the quality flag for the two DIBs, respectively, with $\rm QF\,{=}\,0$ 
indicating the best fit and $\rm QF\,{=}\,5$ the worst fit. First, we require the measured $\diblambda$ to be consistent with a 
Doppler wavelength shift  within about $\pm200\,{\rm km\,s^{-1}}$ for the radial velocity of the two DIBs, which should be a 
reasonable velocity range for the ISM within 3\,kpc from the Sun where most of the DIBs were detected. The DIB radial velocity is 
calculated from the estimated $\diblambda$ and the rest-frame wavelength $\lambda_0$ of the two DIBs reported in \citetalias{PVP} 
and \citetalias{Zhao2022}. We note that the upper limit of $\diblambda$ for $\lambda$864.8 is a bit larger than a wavelength 
corresponding to $200\,{\rm km\,s^{-1}}$ because in the literature this DIB was suggested to be located around 8648\,{\AA} in air. 
Secondly, based on  \citetalias{PVP} and \citetalias{Zhao2022}, $\dibdepth$ is required to be smaller than 0.15 for $\lambda$862.1 
and 0.10 for $\lambda$864.8, respectively. $\dibdepth$ is then compared to $R_C$, which is defined as the standard deviation of the 
difference between observed and modelled flux of the ISM spectra, $\rm std(data-model)$, in a range from $\diblambda\,{-}\,3\dibwidth$ 
to $\diblambda\,{+}\,3\dibwidth$ to represent the noise level around the position of the DIB feature. We note that $R_C$ is not 
published in the {\dibspec} tables, but one can recalculate $R_C$ from the DIB parameters and the stacked ISM spectra. Last, we need 
to define some acceptable values of $\dibwidth$ in order to evaluate the QFs, that is a ``best range'' of $\dibwidth$ for $\rm 
QF\,{=}\,0,1$ and a ``secondary range'' for $\rm QF\,{=}\,2,3,4$. DIBs with $\dibwidth$ out of these two ranges will be marked as 
$\rm QF\,{=}\,5$. The best range of $\sigma_{862.1}$ is set as 0.18--0.26\,nm, a width of 0.04\,nm to the median $\sigma_{862.1}$ 
(0.22\,nm) that is about 2.5 times the average error of $\sigma_{862.1}$ for detections with $\rm QF\,{=}\,0$. The secondary range 
of $\sigma_{862.1}$ is 0.1--0.34\,nm, threefold the span of the best range. Determining the ranges for $\sigma_{864.8}$ is much harder, 
as very few previous studies can serve as a reference. We determined a median $\sigma_{864.8}\,{=}\,0.81$\,nm in \citetalias{Zhao2022}, 
but the median $\sigma_{864.8}$ in {\dibspec} is close to 1.0\,nm when we impose $\rm S/N\,{>}\,200$. Thus, we select a loose range, 
0.8--1.2\,nm, for the best range of $\sigma_{864.8}$, and 0.6--1.4\,nm as the secondary range. The distribution of $\dibwidth$ is 
discussed in detail in Sect. \ref{subsect:dib-width}. This suggests  that careful analysis of the {\dibspec} results is needed to 
refresh our knowledge about $\lambda$864.8. It should be noted that when $\dibdepth\,{<}\,R_C$ (case (g), Fig.~\ref{fig:qf}), 
$\dibwidth$ is only compared with the range of 0.1--0.18\,nm for $\lambda$862.1 and 0.6--0.8\,nm for $\lambda$864.8 (case (k), Fig.
\ref{fig:qf}). This choice follows the idea that when $\dibdepth$ is smaller than the noise level, a large $\dibwidth$ is more 
likely a spurious feature caused by the fit to the successive correlated noise.  
%--------------------------------------------------------------------------------------------------------------------------------

Figure \ref{fig:qf-distrib} shows the distribution of the QF for DIBs\,$\lambda$862.1 and $\lambda$864.8. Nearly half of $\lambda$862.1 
have $\rm QF\,{=}\,5$, and this fraction reaches over 66\% for DIB\,$\lambda$864.8, which dramatically reduces the DIB sample size 
for further analysis, especially for $\lambda$864.8. The much lower proportion of $\rm QF\,{=}\,0,2$ for $\lambda$864.8 than $\rm 
QF\,{=}\,1,3$ is due to its very small $\dibdepth$, since only 4.3\%  of $\mathcal{D}_{864.8}$ is larger than $3R_C$. Thus, only 16.1\% of  
DIB\,$\lambda$864.8 have $\rm QF\,{\leqslant}\,2$, the recommended high quality in \gaia DR3 \citepalias{PVP}. DIB\,$\lambda$862.1 
has a larger proportion (26.6\%) because $\lambda$864.8 is much broader than $\lambda$862.1 and therefore under a heavier effect of 
the noise. The QF provides us with an evaluation of the fitted profile of the DIB feature and consequently the goodness-of-fit for 
the ISM spectra. But the present QF still contains some shortcomings, for example, the QF has discrete values and has hard and 
artificial borders for $\dibwidth$. Therefore, the QF should be used with caution for DIB $\lambda$864.8, which is not well-studied. 
%--------------------------------------------------------------------------------------------------------------------------------

Figure \ref{fig:qf-Gmap} presents the distribution on the sky of QFs for the two DIBs in  Galactic coordinates. Each level of QF
is shown in one HEALPix map with level 5. The colour scale represents the count of DIB detections in each HEALPixel at different
distances. Because QFs 0--5 are classified by $\{\dibdepth,\dibwidth,R_C\}$, the QF sky distribution would be related to DIB
properties. At high latitudes, the stacked ISM spectra generally have small S/N due to the decreasing density of RVS objects (S/N 
is also affected by distance), which results in a large $R_C$. Moreover, $\dibdepth$ is also expected to decrease with $|b|$. Thus, we can
find numerous detections with $\rm QF\,{=}\,1$ and 3 at $|b|\,{\gtrsim}\,20^{\circ}$ but rare with $\rm QF\,{=}\,0$ and 2. The
detections with $\rm QF\,{=}\,0$ and 2 mainly occupy the Galactic middle plane where one expects more abundant ISM and stronger 
DIB signals (more RVS objects as well) and only extend to higher latitudes ($|b|\,{\sim}\,30^{\circ}$) towards the inner disc 
($|\ell|\,{<}\,30^{\circ}$) and the Galactic anti-centre ($\ell\,{\sim}\,180^{\circ}$) where numerous molecular clouds exist.
The two DIBs show similar QF distribution, but $\lambda$862.1 has a wider latitude distribution than $\lambda$864.8 for $\rm 
QF\,{=}\,0$ and 2 because $\mathcal{D}_{864.8}$ is only $\sim$35\% of $\mathcal{D}_{862.1}$ (see Sect. \ref{subsect:dib-dust}).
%--------------------------------------------------------------------------------------------------------------------------------

On the other hand, the QF sky distribution is similar between $\rm QF\,{=}\,0$ and $\rm QF\,{=}\,2$, although they contain
different ranges of $\dibwidth$ (the same for $\rm QF\,{=}\,1$ and $\rm QF\,{=}\,3$). Moreover, we can find more detections of
$\lambda$862.1 with $\rm QF\,{=}\,0$ than $\rm QF\,{=}\,2$ and less with $\rm QF\,{=}\,1$ than $\rm QF\,{=}\,3$, which indicates
that the fit of $\dibwidth$ is heavily affected by $\dibdepth/R_C$ which can be treated as a measure of the S/N of DIB signal.
%--------------------------------------------------------------------------------------------------------------------------------

The detected DIB signals with $\rm QF\,{=}\,4$ are weak and noisy with a main distribution out of the middle plane. DIB detections
with $\rm QF\,{=}\,5$ are complicated, containing the cases with $\diblambda$ out of the reasonable range, too small or too big
$\dibwidth$, and too low $\dibdepth$. $\rm QF\,{=}\,5$ is distributed in almost the full sky with $|b|\,{<}\,65^{\circ}$ as it 
takes half of the DIB sample. It is interesting that the empty or low-density regions in $\rm QF\,{=}\,4$ and 5 are complementary
to those in $\rm QF\,{=}\,0$. 
%--------------------------------------------------------------------------------------------------------------------------------

QFs of 0 and 2 are recommended as the best level, and 1 and 3 are the secondary level. While 4 and 5 are the worst and are 
suggested to be better used in a statistical way.
%--------------------------------------------------------------------------------------------------------------------------------

\begin{figure}[!h]
\centering
\includegraphics[width=8.4cm]{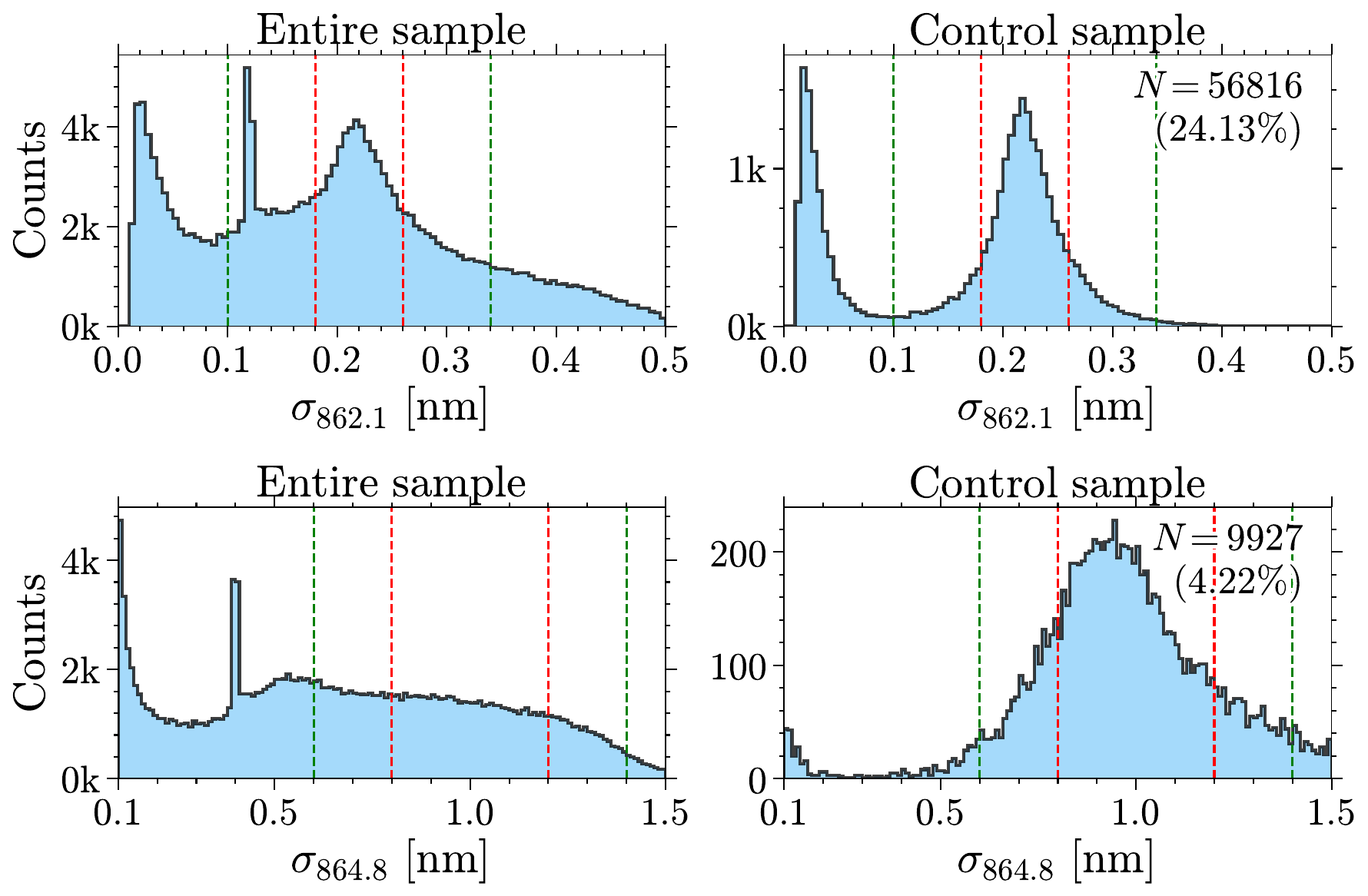}
\caption{Distribution of $\dibwidth$ for DIBs $\lambda$862.1 and $\lambda$864.8 (top and bottom, respectively). The dashed red and 
green lines correspond to the ``best range'' and the ``secondary range'' of $\dibwidth$ defined in Sect. \ref{subsect:qf} for the 
DIB QF, respectively. The distribution of the full {\dibspec} results (235\,428 voxels) is shown in the left panels, while the right
panels show a quality-controlled sample with $\rm S/N\,{>}\,100$ and $\dibdepth\,{>}\,3R_C$ (see Sect. \ref{subsect:dib-width}). 
The latter criterion implies QF values of 0 or 2. The number of detected DIBs and the percentage after the quality control 
is indicated as well.}
\label{fig:dib-width} 
\end{figure}
%--------------------------------------------------------------------------------------------------------------------------------

\subsection{DIB width} \label{subsect:dib-width}

The width of the DIB profile contains vast information about the properties of the intervening ISM clouds and the DIB carriers,
such as the profile broadening that is related to the gas kinetic and rotational temperature in different physical environments
\citep[e.g.][]{Lai2020,Krelowski2021}. The fitted $\dibwidth$ in {\dibspec} is a measure of the average width under different ISM 
environments and may be affected by Doppler splitting or broadening, especially for distant voxels. An investigation of the Doppler 
effect and the decomposition of multiple velocity components with the {\dibspec} results will be done in a forthcoming work.
%--------------------------------------------------------------------------------------------------------------------------------

The distribution of $\dibwidth$ of the two DIBs with the full {\dibspec} results (235\,428 voxels) is shown in the left panels in
Fig. \ref{fig:dib-width} (entire sample). The histogram of the width of $\lambda$862.1 contains three peaks (upper left panel). 
The first peak at $\sigma_{862.1}\,{=}\,0.02$\,nm indicates ISM spectra with low S/N or extremely weak DIB signals. The second peak 
at $\sigma_{862.1}\,{=}\,0.12$\,nm, the initial guess of $\sigma_{862.1}$, is also caused by the fit to low-S/N spectra, where the 
fitted parameters are just around their initial values. The third peak at $\sigma_{862.1}\,{=}\,0.22$\,nm, as we have seen in Fig. 
\ref{fig:snr}, marks the best fits of DIB\,$\lambda$862.1. For $\lambda$864.8 (lower left panel), the first ($\sigma_{864.8}\,{=}\,0.1$\,nm) 
and second ($\sigma_{864.8}\,{=}\,0.4$\,nm) peaks can also be found. But the width distribution is then very flat for larger 
$\sigma_{864.8}$, and the peak of $\sigma_{864.8}$ that indicates the best $\lambda$864.8 profiles is smoothed by the noisy detections,
which could only be seen in the control sample (see below).
%--------------------------------------------------------------------------------------------------------------------------------

\begin{figure}[!h]
\centering
\includegraphics[width=8cm]{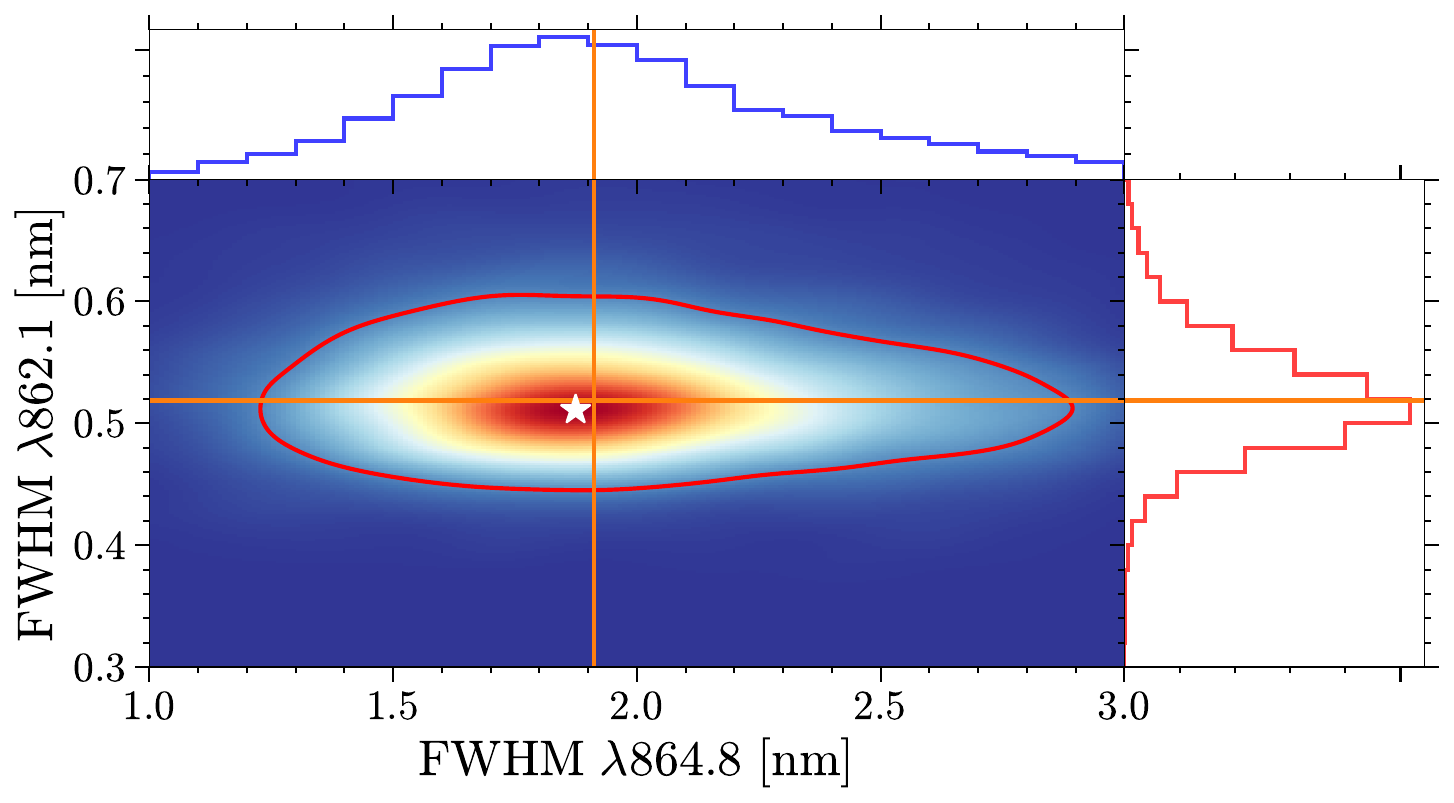}
\caption{Joint distribution of the FWHM of DIBs $\lambda$862.1 and $\lambda$864.8, measured in 9778 stacked ISM spectra,
with $\rm S/N\,{>}\,100$ and $\dibdepth\,{>}\,3R_C$ (see Sect. \ref{subsect:dib-width}), generated by a Gaussian kernel 
density estimation (KDE). The white star indicates the peak density, and the red line in the central panel indicates the contour of the 
2$\sigma$ level. The orange lines indicate the median FWHM of the two DIBs.}
\label{fig:width-kde} 
\end{figure}
%--------------------------------------------------------------------------------------------------------------------------------

\begin{figure}
\centering
\includegraphics[width=8cm]{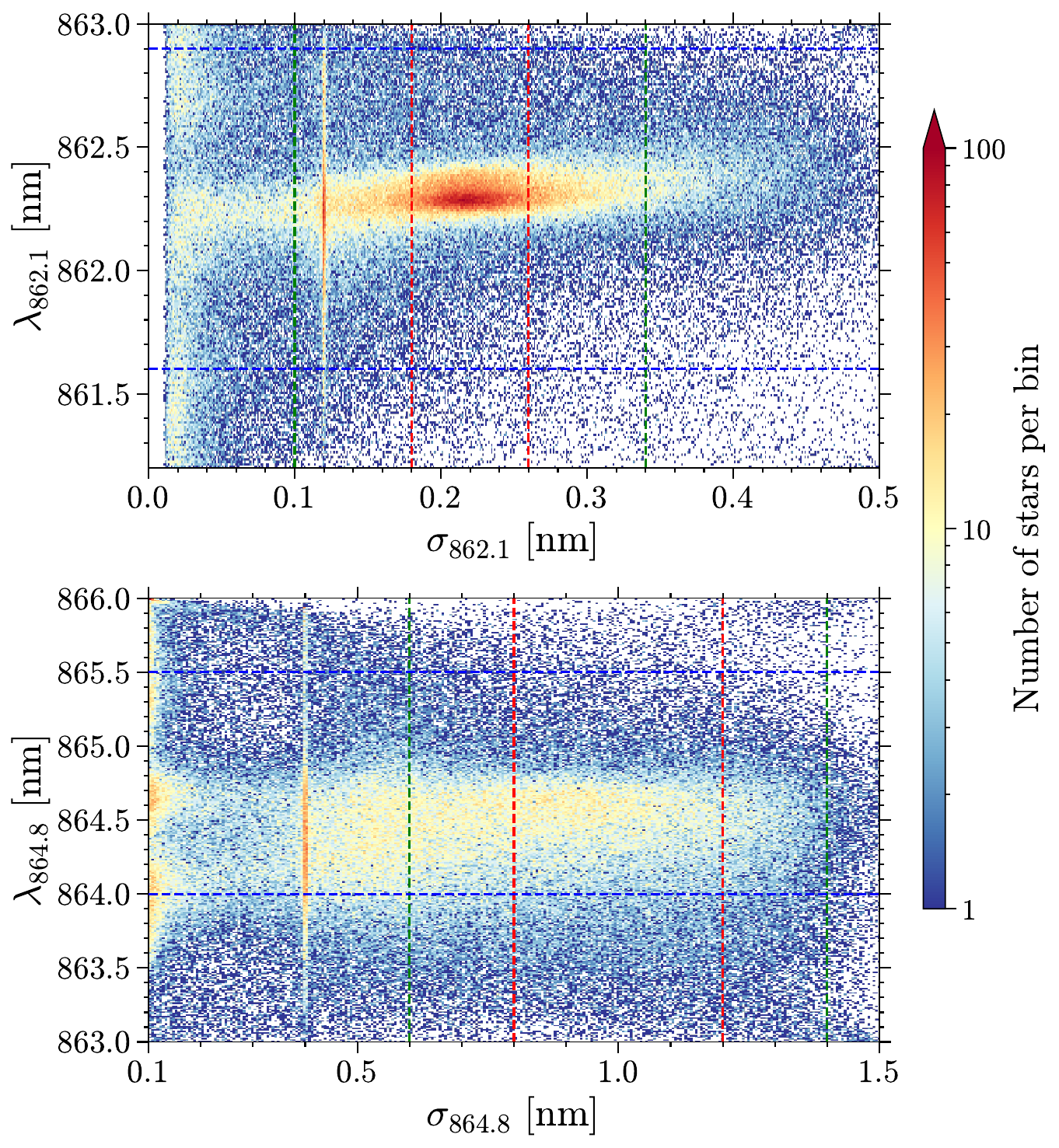}
\caption{Number density of DIB detections in {\dibspec} as a function of $\diblambda$ and $\dibwidth$, without any cuts,
for $\lambda$862.1 ({\it upper panel}) and $\lambda$864.8 ({\it lower panel}). The dashed blue lines indicate the range 
of permitted $\diblambda$ in the QF evaluation (see Sect. \ref{subsect:qf}), and the dashed red and green lines correspond 
to the ``best range'' and the ``secondary range'' of $\dibwidth$, respectively. }
\label{fig:l-s} 
\end{figure}
%--------------------------------------------------------------------------------------------------------------------------------

% \Tomaz{Would it be worth to add panels with applied quality cuts? It would illustrate that stellar contamination is not problematic if a high-quality sample is used.} 
% \hzhao{There are already many figures, maybe we can clarify this in the context?}

To determine the ``best'' and ``secondary'' ranges of $\dibwidth$ for QF evaluation (Sect. \ref{subsect:qf}), we apply a quality 
control of $\rm S/N\,{>}\,100$ and $\dibdepth\,{>}\,3R_C$, and get 56\,816 fit results for $\lambda$862.1 and 9927 for $\lambda$864.8, 
whose width distributions are shown in the right panels in Fig. \ref{fig:dib-width}. The peaks at the initial guess of $\dibwidth$
disappear, and a quasi-Gaussian distribution of $\dibwidth$ can now be found for both DIBs with a lower cut that excludes 
very narrow spurious features. The selected ranges of $\dibwidth$ for QF determination are marked in Fig. 
\ref{fig:dib-width} and discussed in Sect. \ref{subsect:qf}. We note that the upper limit of the acceptable range for $\sigma_{864.8}$ 
in QF evaluation is enlarged because of the long tail of the distribution of $\sigma_{864.8}$ showing more detections with 
$\sigma_{864.8}\,{\gtrsim}\,1.1$\,nm than with $\sigma_{864.8}\,{\lesssim}\,0.8$\,nm. The cause is unknown and DIB\,$\lambda$864.8 
with a very broad profile needs further exploration. 
%--------------------------------------------------------------------------------------------------------------------------------
% The significantly shallower profile of $\lambda$864.8, compared to $\lambda$862.1, means that the 
% former would need a much higher SNR to fulfill $\dibdepth\,{>}\,3R_C$. Therefore, DIB\,$\lambda$864.8 has far fewer detections with 
% very small $\dibwidth$ than does $\lambda$862.1.

\begin{figure*}
\centering
\includegraphics[width=14cm]{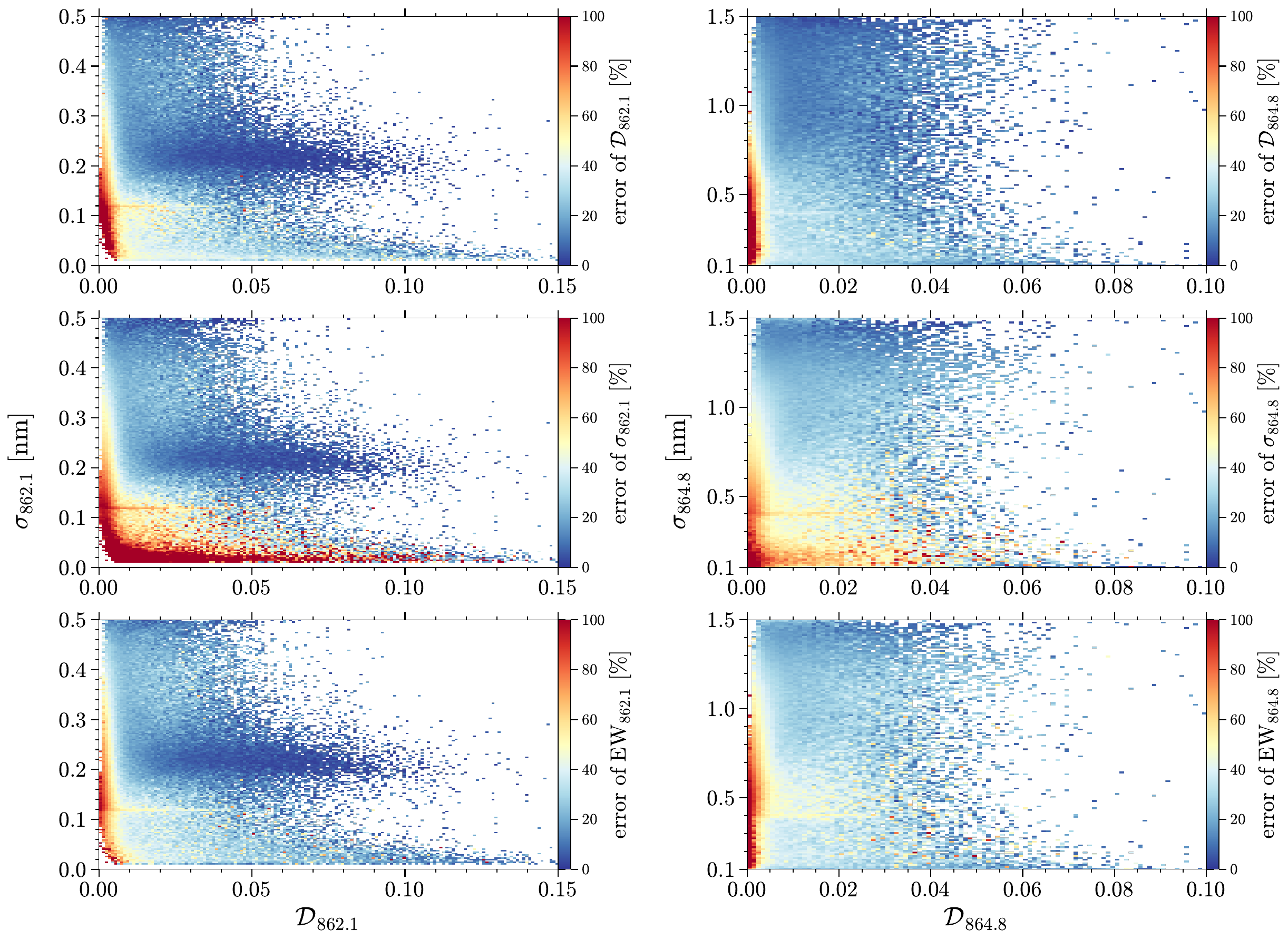}
\caption{Distributions of the fractional uncertainties of $\dibdepth$, $\dibwidth$, and DIB EW as a function of $\dibdepth$ and 
$\dibwidth$ for $\lambda$862.1 ({\it left panels}) and $\lambda$864.8 ({\it right panels}). Colour represents the mean fractional 
errors in each $0.001 \times 0.003$\,nm bin for $\lambda$862.1 and $0.001 \times 0.01$\,nm bin for $\lambda$864.8.}
\label{fig:dib-para-err}
\end{figure*}
%--------------------------------------------------------------------------------------------------------------------------------

\begin{figure}
\centering
\includegraphics[width=7cm]{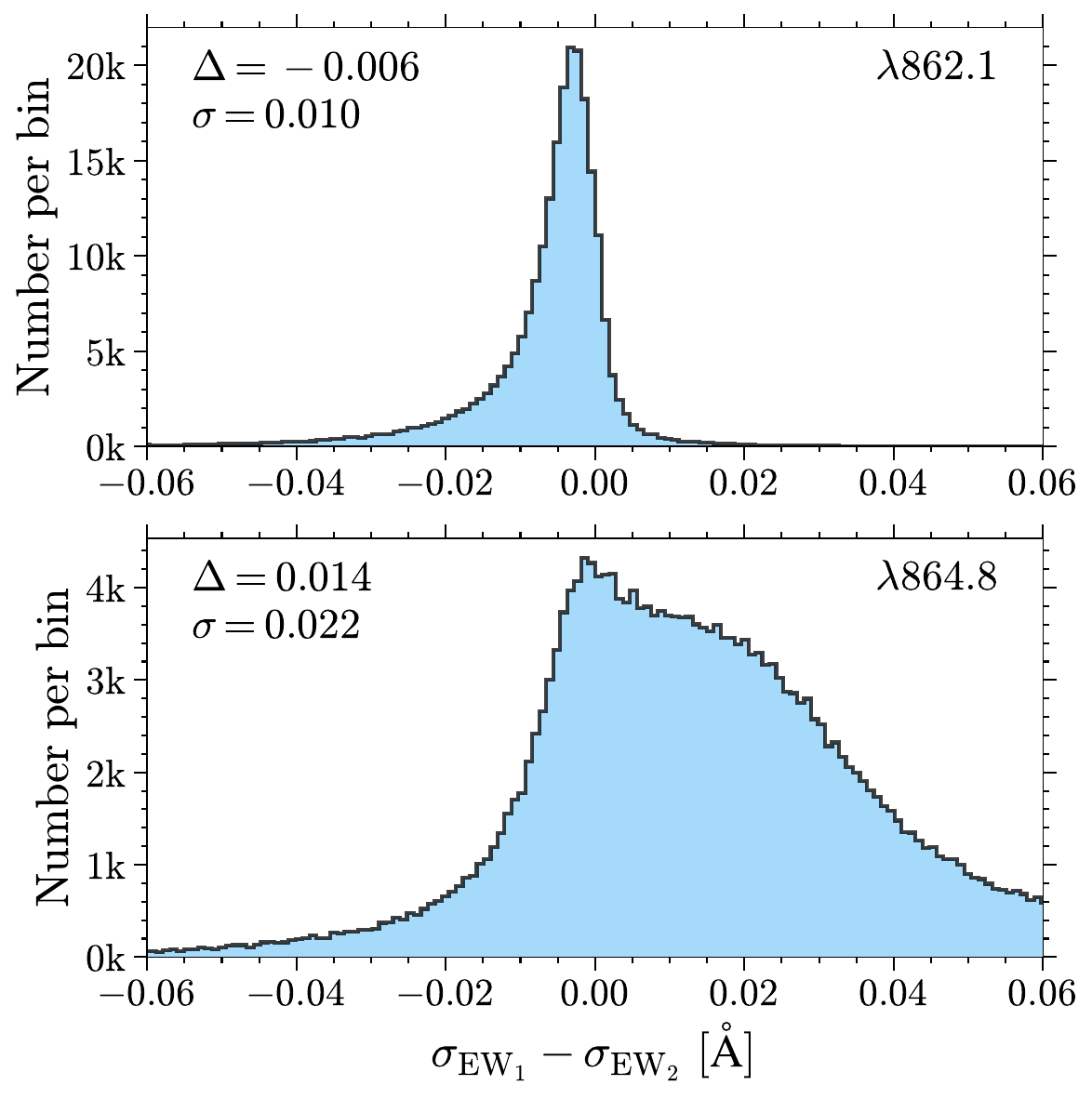}
\caption{Histogram of the difference between the EW uncertainty estimated in this work ($\sigma_{\rm EW_1}$) and the one
calculated by the formula $\sigma_{\rm EW_2}\,{=}\,\sqrt{6\,{\rm FWHM}\,\delta \lambda} \times {\it R_C}$ (see Sect. 
\ref{subsect:params}) for $\lambda$862.1 ({\it upper panel}) and $\lambda$864.8 ({\it lower panel}), respectively. The mean
($\Delta$) and standard deviation ($\sigma$) of the differences are also indicated.  }
\label{fig:ew-error}
\end{figure}
%--------------------------------------------------------------------------------------------------------------------------------

% \Tomaz{I find powers $^1$ and $^2$ somewhat confusing. What about $\sigma_{EW1}$ and $\sigma_{EW2}$?}
% \hzhao{Changed.}

The full width at half maximum (FWHM) of the two DIBs is calculated as $\rm FWHM_{862.1}=\sqrt{8{\rm ln}(2)} \times \sigma_{862.1}$ 
for the Gaussian profile (Eq. \ref{eq:gauss}) of $\lambda$862.1 and $\rm FWHM_{864.8}=2 \times \sigma_{864.8}$ for the Lorentzian
profile (Eq. \ref{eq:lorentz}) of $\lambda$864.8.  The joint distribution of $\rm FWHM_{862.1}$ and $\rm FWHM_{864.8}$ for 9778 
detections with $\rm S/N\,{>}\,100$ and $\dibdepth\,{>}\,3R_C$ (for the two DIBs) is shown in Fig. \ref{fig:width-kde}. The 
distribution is Gaussian with a long tail for $\rm FWHM_{864.8}$. The median FWHM of DIB\,$\lambda$862.1 is $0.52\,{\pm}\,0.05$\,nm, 
which is consistent with \citet[][$0.55\,{\pm}\,0.06$\,nm]{Zhao2022} but larger than previous results based on early-type stars, 
that is 4.3\,{\AA} of \citet{HL1991}, 4.38\,{\AA} of \citet{JD1994}, and 4.69\,{\AA} of \citet{Maiz-Apellaniz2015a}. As discussed
in \citetalias{Zhao2022}, the increase in $\rm FWHM_{862.1}$ of our result could be explained by a Doppler broadening caused by 
our stacking strategy. Other effects, such as the observational instrument and the stellar residuals, may be proposed as well.
With a machine-learning approach on published RVS spectra, \citet{Saydjari2022} got a $\sigma_{862.1}=1.9$\,{\AA}, corresponding 
to a FWHM of 4.47\,{\AA}. This consistency indicates that {\it Gaia} has no significant instrumental effects on the DIB measurement.
Furthermore, a data-driven method could significantly reduce the influence of the residuals of specific stellar lines (e.g., 
\ion{Fe}{i} for $\lambda$862.1) on the DIB width. On the other hand, \citet{Hobbs2009} and \citet{Fan2019} reported a FWHM of 3.56\,{\AA}
and 3.98\,{\AA} for $\lambda$862.1, respectively, which are smaller than mentioned results above. \citet{Puspitarini2015} mentioned
that many {\it Gaia}--ESO spectra of the $\lambda$862.1 region are contaminated by sky emission residuals which fall within the 
red wing of $\lambda$862.1. In principle, any emission residuals could make the DIB appear narrower than it actually is, even in 
the case of hot stars. Therefore, we propose that $\rm FWHM\,{\sim}\,4.7$\,{\AA} ($\sigma_{862.1}$ around 2\,{\AA}) could be 
proper for DIB\,$\lambda$862.1. But we should keep in mind that the DIB width could vary under different ISM environments. We note 
that DIB\,$\lambda$862.1 could contain multiple components \citep{JD1994} and present an asymmetric profile \citep[e.g.][]{Krelowski2019b}.
But considering the resolution of RVS spectra, our study cannot reveal an accurate shape of $\lambda$862.1. Nevertheless, the high
S/N of the stacked spectra would allow us to resolve different DIB velocity components in further analysis.
%--------------------------------------------------------------------------------------------------------------------------------

The median FWHM of DIB\,$\lambda$864.8 is $1.91\,{\pm}\,0.44$\,nm, larger than $1.62\,{\pm}\,0.33$\,nm in \citetalias{Zhao2022}, 
1.4\,nm in \citet{HL1991} and 0.42\,nm in \citet{JD1994}. Doppler broadening should have much less effect on the FWHM of
$\lambda$864.8 due to its very large breadth. \citetalias{Zhao2022} also measured the median $\sigma_{864.8}$ by stacking RVS 
spectra but in a much smaller sample (1103 detections) with weaker quality control. Figure \ref{fig:dib-width} shows that involving 
noisy detections would reduce and smooth the peak of $\sigma_{864.8}$. The measurements in early studies could be questionable 
due to the broad span of $\lambda$864.8 and its superposition with some blended stellar lines. The median $\dibwidth$ reported 
in this work is from a quasi-Gaussian distribution. 
%--------------------------------------------------------------------------------------------------------------------------------
% The variation of $\dibwidth$ for $\lambda$862.1 or $\lambda$864.8 has not been reported before.
% which could be caused by a light Doppler broadening for stacked ISM spectra.

% \Tomaz{It may be worth to mention that  difference in the numerical factor (2.35 vs. 2) comes from different adopted line profiles, as stated in eqs. 1 and 2.}
% \hzhao{Added.}

\subsection{Scaling factor for the lower and upper limits of DIB EW} \label{subsect:dib-ew}

The EW of the DIBs $\lambda$862.1 and $\lambda$864.8 (`ew8620' and `ew8648' in Table \ref{tab:catalog-params}) was calculated
by Eqs. \ref{eq:ew8621} and \ref{eq:ew8648}, respectively. Nevertheless, when estimating the lower and upper confidence levels of
EW by the $\{\dibdepth,\dibwidth\}$ pairs from MCMC samplings (see Sect. \ref{subsect:stack-fit}), the EW values of each pair were
calculated by numerical integration of Eqs. \ref{eq:gauss} and \ref{eq:lorentz} from $\dibdepth\,{-}\,3\dibwidth$ to 
$\dibdepth\,{+}\,3\dibwidth$ for the two DIBs, respectively, instead of using Eqs. \ref{eq:ew8621} and \ref{eq:ew8648}. 
This approximation is suitable for a Gaussian function, as the integration equals to ${\rm erf}(3/\sqrt{2}) \times \sqrt{2\pi} 
\times \mathcal{D}_{862.1} \times \sigma_{862.1} \approx 0.997\,{\rm EW_{862.1}}$. But it is problematic for the Lorentzian 
profile of $\lambda$864.8 for the integration equals to $2{\rm arctan}(3) \times \mathcal{D}_{864.8} \times \sigma_{864.8} 
\approx 0.795\,{\rm EW_{864.8}}$. Therefore, we propose a scaling factor of 1.258 for `ew8648\_lower' and `ew8648\_upper'
in Table \ref{tab:catalog-params}, that is the table values times 1.258 are the correct lower and upper levels for $\rm EW_{864.8}$.
A scaling factor of 1.003 can also be used for `ew8620\_lower' and `ew8620\_upper'.

\subsection{Effect of stellar residuals} \label{subsect:stellar-effect}

Because most of the target stars processed by {\dibspec} are late-type  stars whose spectra contain abundant stellar components, 
the stellar residuals from the unresolved stellar features or the not well-modelled stellar lines would affect the DIB detection and 
measurement, such as the shift of $\diblambda$ or the broadening of $\dibwidth$. In extreme cases, the detected signal is an artefact that 
comes from the stellar residuals. As pointed out by \citet{Saydjari2022}, for example, the DIB detections in the \gaia DR3 catalogue 
would be contaminated by not perfectly modelled lines, such as \ion{Fe}{i}, when $\dibwidth$ is very small. 
% But we have to note that some of the semi-forbidden lines, such as XX, have not been detected in astronomical spectra so they 
% cannot be used to account for the clumpy small DIB features. 
%--------------------------------------------------------------------------------------------------------------------------------

Figure \ref{fig:l-s} shows the distribution of DIB detections as a function of $\diblambda$ (in stellar frame) and $\dibwidth$, 
which can straightforwardly identify the effect of stellar residuals. A vertical stripe at the initial guess of $\dibwidth$ can
be seen for both the two DIBs that corresponds to the noisy cases, mostly with $\dibdepth\,{<}\,R_C$. $\lambda_{862.1}$ is
uniform and widespread when $\sigma_{862.1}\,{<}\,0.04$\,nm, where DIB detection is noise dominated. For larger $\sigma_{862.1}$,
especially 0.16--0.28\,nm, $\lambda_{862.1}$ is concentrated between 862.2 and 862.5\,nm, representing the reliable measurements.
Our detection of $\lambda$862.1 is not or at most very weakly affected by the stellar residuals, as no clustering near the 
stellar lines is found in the $\sigma_{862.1}-\lambda_{862.1}$ plane. This is because the stacking increases the S/N of the ISM 
spectra and averages out the large residuals in individual spectra.
%--------------------------------------------------------------------------------------------------------------------------------

The very broad profile of $\lambda$864.8 makes it difficult for accurate measurements of $\lambda_{864.8}$ and $\sigma_{864.8}$,
resulting in a very scattered $\lambda_{864.8}$ versus $\sigma_{864.8}$ distribution. However, the large $\sigma_{864.8}$ 
can hardly be affected by the stellar residuals. 
% Two clumps with small $\sigma_{864.8}$ ($<$0.2\,nm) are located around 864.0 and 864.6\,nm with a span of $\sim$0.4\,nm. 
%--------------------------------------------------------------------------------------------------------------------------------

\subsection{Uncertainties of DIB parameters} \label{subsect:params}

Since $\dibdepth$ and $\dibwidth$ are correlated with each other during the MCMC fitting, the distribution of the uncertainty of
the DIB parameters presents a dependency on both $\dibdepth$ and $\dibwidth$. Figure \ref{fig:dib-para-err} shows the distribution 
of the fractional uncertainties of $\dibdepth$, $\dibwidth$, and EW in the $\dibdepth-\dibwidth$ plane for $\lambda$862.1 (left 
panels) and $\lambda$864.8 (right panels). The uncertainty of $\dibdepth$, $\dibwidth$, and EW are taken as the mean difference 
between their lower, median, and upper values. 
%--------------------------------------------------------------------------------------------------------------------------------

Large uncertainties can generally be found in regions with small $\dibdepth$ or $\dibwidth$. DIB detections with very small
$\dibdepth$ will get  large QF (low reliability) by comparing with $R_C$, and those with small $\dibwidth$ are ruled out by the
width border defined in the QF evaluation. It is noted that detections with small uncertainties can be found very close to the
upper limit of $\dibwidth$ in priors, 0.5\,nm for $\lambda$862.1 and 1.5\,nm for $\lambda$864.8. These small uncertainties are
because of the narrow sampling range in the MCMC fitting and do not represent good measurements. Figure \ref{fig:dib-width} also
shows that the number of detections with extremely large $\dibwidth$ will significantly decrease when applying strict constraints.
For $\lambda$862.1, detections with low uncertainty ($<$10\%) are located in similar fields for $\mathcal{D}_{862.1}$, $\sigma_{862.1}$,
and $\rm EW_{862.1}$ and correspond to the ``best range'' of $\sigma_{862.1}$ defined in Sect. \ref{subsect:qf}. On the other hand,
low-uncertainty regions for $\lambda$864.8 do not present a well-constrained range for $\sigma_{864.8}$.
%--------------------------------------------------------------------------------------------------------------------------------

The uncertainty of EW in some regions, mainly the bottom in each panel with small $\dibwidth$ and large $\dibdepth$, is much smaller
than that of $\dibwidth$, although EW was calculated by $\dibdepth$ and $\dibwidth$. The reason is that in each MCMC chain, in 
spite of the width with large uncertainty, the depth could be stable with small uncertainty. Consequently, the calculated EW also
concentrates and has smaller relative uncertainty. Besides the method used in {\dibspec}, the EW uncertainty is also proposed to
be calculated by the span of the profile (3$\times$FWHM), the pixel resolution ($\delta \lambda\,{=}\,0.03$\,nm$/$pixel), and the 
noise level of the line centre ($R_C\,{=}\,{\rm std(data-model)}$), as $\sigma_{\rm EW}\,{=}\,\sqrt{6\,{\rm FWHM}\,\delta \lambda} 
\times {\it R_C}$ \citepalias{Zhao2022}. Similar formulas were also given by \citet{Vos2011} and \citet{VE2006}. Figure \ref{fig:ew-error}
shows the difference between these two kinds of EW uncertainty. The difference, in general, is small (over 90\% within 0.02\,{\AA}
for $\lambda$862.1 and within 0.07\,{\AA} for $\lambda$864.8), but our method systematically gives out smaller $\sigma_{\rm EW}$ 
for $\lambda$862.1 and larger $\sigma_{\rm EW}$ for $\lambda$864.8.
%--------------------------------------------------------------------------------------------------------------------------------

\begin{figure}
\centering
\includegraphics[width=7cm]{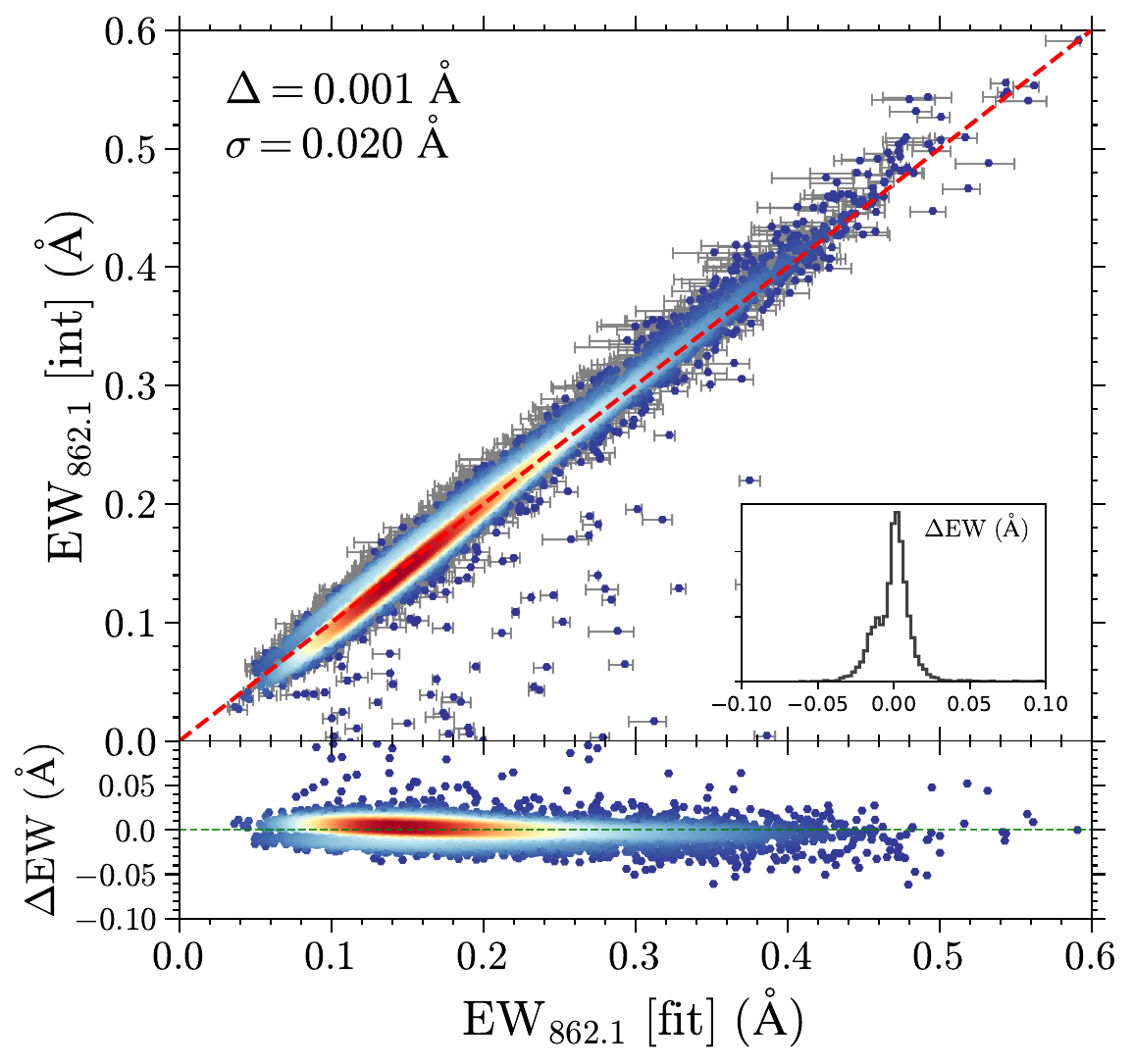}
\caption{   Upper panel: Comparison between the fitted and integrated $\rm EW_{862.1}$. The colour represents the number density 
(estimated by a Gaussian KDE) of the data points in a linear scale. The grey colour bars show the uncertainty of fitted $\rm EW_{862.1}$.
The dashed red line traces the one-to-one correspondence. A zoom-in panel shows the distribution of the EW difference ($\rm 
\Delta EW=EW_{fit}-EW_{int}$). The mean ($\Delta$) and standard deviation ($\sigma$) of the EW difference are indicated. 
{\it Lower panel:} The distribution EW difference as a function of fitted $\rm EW_{862.1}$.}
\label{fig:fit-int} 
\end{figure}
%--------------------------------------------------------------------------------------------------------------------------------

\begin{figure}
\centering
\includegraphics[width=7cm]{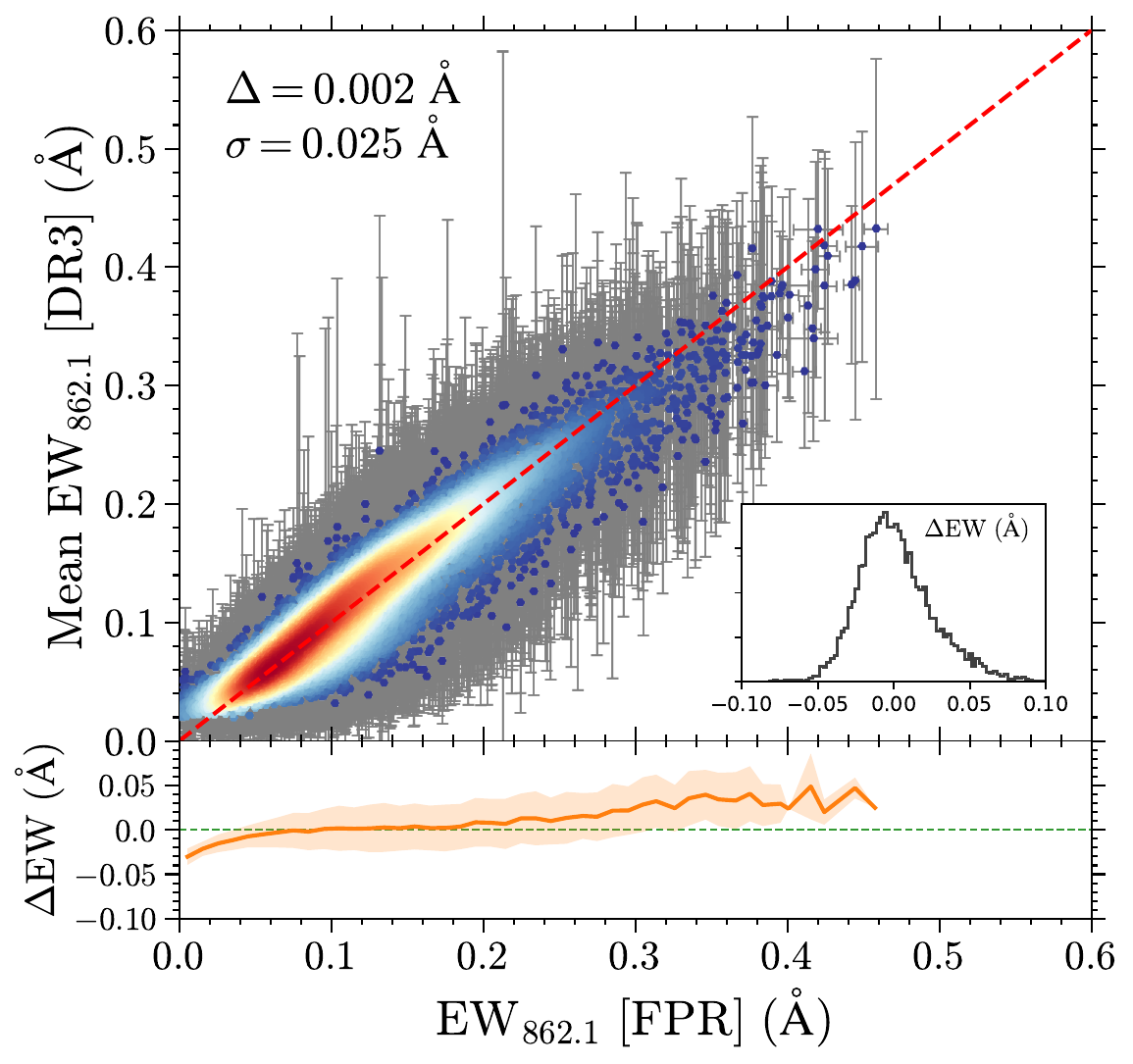}
\caption{{\it Upper panel:} Comparison between the $\rm EW_{862.1}$ measured by {\dibspec}  and the mean $\rm EW_{862.1}$ 
in each voxel taken from the DIB catalogue in DR3 for 8963 voxels. The colour represents the number density (estimated by a Gaussian 
KDE) of the data points in a linear scale. The grey colour bars show the EW uncertainty in {\dibspec} and the standard deviation of EW
in DR3. The dashed red line traces the one-to-one correspondence. A zoom-in panel shows the distribution of the EW difference ($\rm 
\Delta EW=EW_{FPR}-EW_{DR3}$). The mean ($\Delta$) and standard deviation ($\sigma$) of the EW difference are indicated. 
{\it Lower panel:} The variation of the mean EW difference in each EW bin with a step of 0.01\,{\AA} with $\rm EW_{862.1}$ in FPR.
The orange shades the range of 1$\sigma$.}
\label{fig:ew} 
\end{figure}
%--------------------------------------------------------------------------------------------------------------------------------

\section{Validation tests} \label{sect:validation}
% Put here some parts of the validation report here

We present in this Section a number of validation tests for the DIB results from {\dibspec}, including the comparison with \gaia DR3
DIB catalogue for the measurements of $\rm EW_{862.1}$, the EW maps of the two DIBs in a Galactic view at different distances, and
the correlation between two DIBs and dust reddening.
%--------------------------------------------------------------------------------------------------------------------------------

\begin{figure*}
  \centering
  \includegraphics[width=8cm]{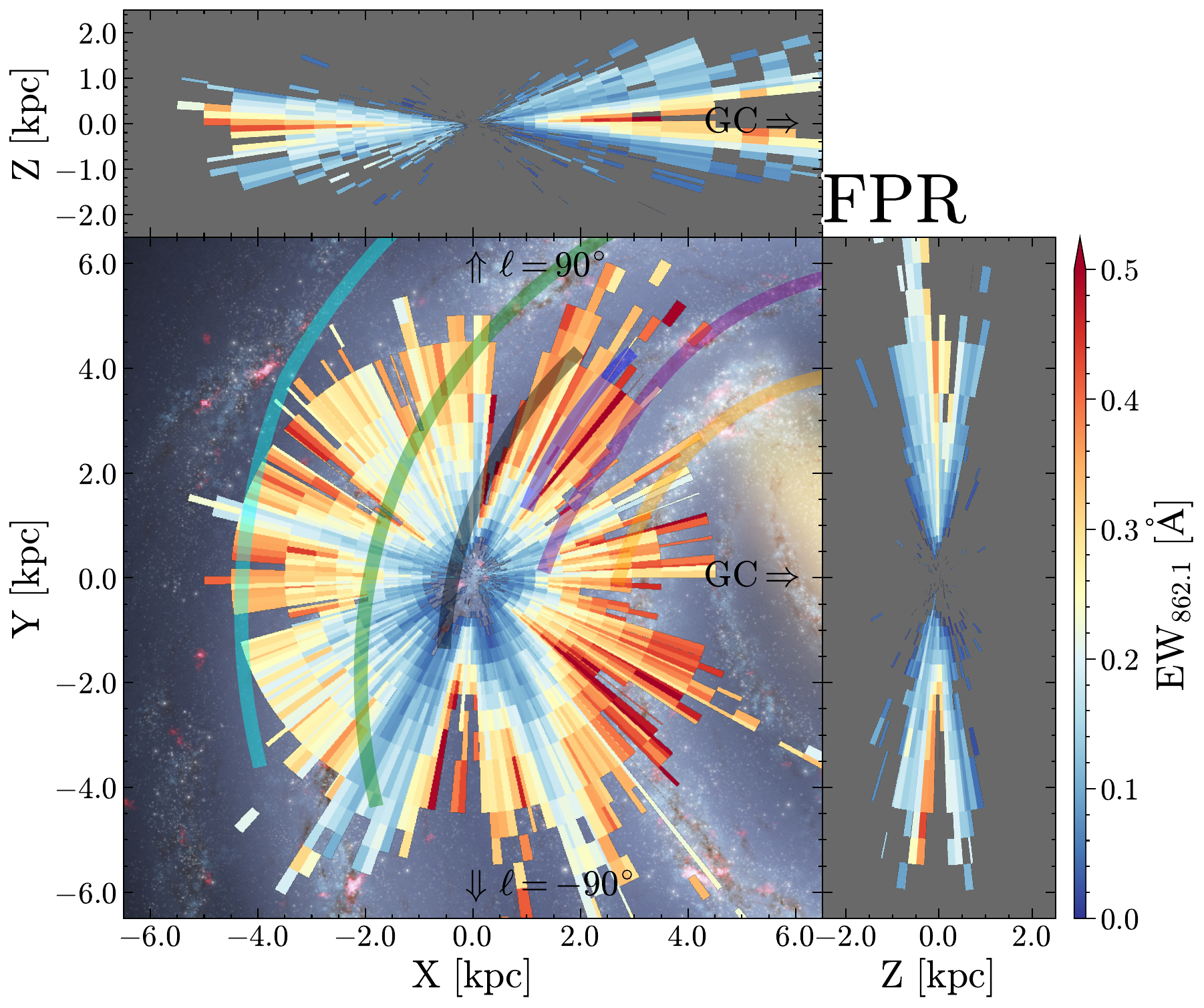}
  \includegraphics[width=8cm]{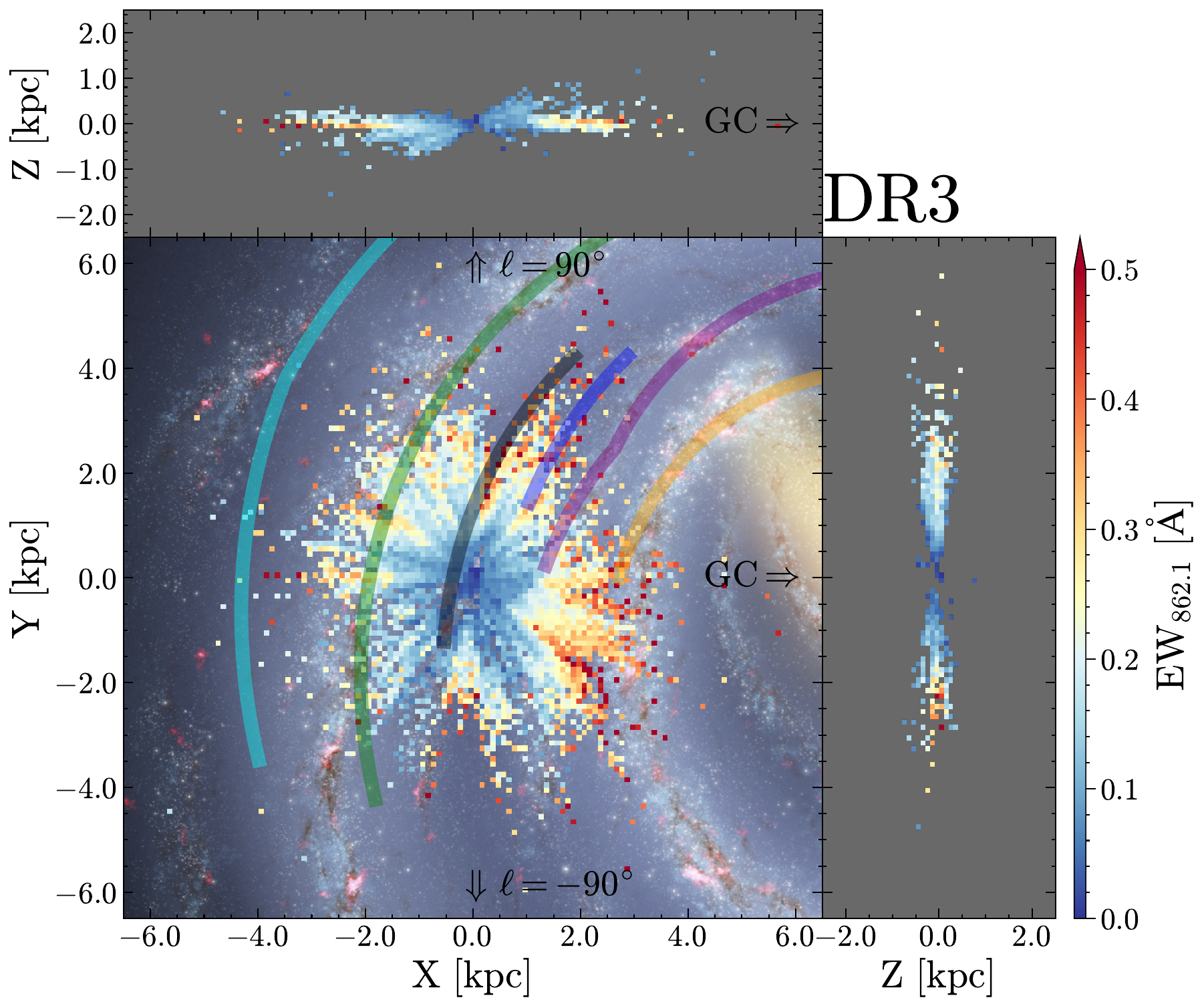}
  \caption{Distribution of $\rm EW_{862.1}$ in the Galactic plane (XY), meridian plane (XZ), and rotational plane (YZ) for the FPR 
  results ({\it left panel}) and the DR3 results ({\it right panel}), respectively, plotted over the Milky Way sketch created by 
  Robert Hurt and Robert Benjamin \citep{Churchwell2009}. Some log-periodic spiral arms described in \citet{Reid2019} are also 
  presented by coloured lines: Scutum--Centaurus Arm in orange; Sagittarius--Carina Arm in purple; Local Arm in black; Perseus Arm
  in green; Outer Arm in cyan; and the spur between the Local and Sagittarius--Carina arms in blue. The Galactic centre is located 
  at $(X,Y,Z)=(8.15,0,0)$.}
  \label{fig:Fmap-8621}
\end{figure*}
%--------------------------------------------------------------------------------------------------------------------------------

\subsection{Comparison between fitted and integrated DIB EW} \label{subsect:fit-int}

Differing from the direct measurement of DIB EW by integrating the ISM spectrum \citep[e.g.][]{Hobbs2008,Fan2019}, the DIB EW in
this work was calculated by the analytic function. The integrated EW is not affected by the asymmetry of the DIB profile, while
the fitted EW is less affected by the noise and the overlapping of different DIB profiles. Specifically, the EW of $\lambda$862.1
cannot be directed integrated because the profiles of $\lambda$862.1 and $\lambda$864.8 could be overlapped. Therefore, we first
subtracted the fitted profile of $\lambda$864.8 from the ISM spectrum and then integrated the rest part of the ISM spectrum within
$\lambda_{862.1} \pm 3\sigma_{862.1}$ to calculate the integrated EW of $\lambda$862.1. Figure \ref{fig:fit-int} shows the 
comparison between fitted and integrated $\rm EW_{862.1}$ for 6240 cases with $\rm S/N\,{\geqslant}\,100$, $\rm QF_{862.1}\,{=}\,0$,
and $\rm QF_{864.8}\,{=}\,0$. We did not do such a comparison for $\lambda$864.8 as the spectra suffer there the residuals of the CaII line. 
The fitted and integrated $\rm EW_{862.1}$ are highly consistent with each other with a mean difference of
only 0.001\,{\AA} and a standard deviation of 0.020\,{\AA}. This proves that the possible asymmetry of the DIB profile has little
effect on $\rm EW_{862.1}$. After the check of the ISM spectra, we find that the outliers with much larger fitted $\rm EW_{862.1}$ 
than integrated $\rm EW_{862.1}$ are caused by the wrongly fitted central position of DIB\,$\lambda$862.1. The $\lambda_{862.1}$
of these cases are close to 863\,nm which is the upper limit of $\lambda_{862.1}$ during the MCMC fitting. A larger upper limit
could improve the results of these fittings. 
%--------------------------------------------------------------------------------------------------------------------------------

\subsection{Comparison with DIB measurements in \gaia DR3} \label{subsect:fpr-dr3}

To make a direct comparison of the measurements of $\rm EW_{862.1}$ between FPR and DR3, the mean $\rm EW_{862.1}$ of the full DIB
catalogue in DR3 is calculated in each voxel defined in {\dibspec}. Figure \ref{fig:ew} presents the comparison between $\rm EW_{862.1}$ 
from FPR and mean $\rm EW_{862.1}$ from DR3 for 8963 voxels (3.81\% of the amount in FPR) that contain at least ten DIB detections
in DR3. The difference of $\rm EW_{862.1}$ (FPR--DR3) presents a quasi-Gaussian distribution with an extremely small mean value 
of 0.002\,{\AA} and a standard deviation of 0.025\,{\AA}. The EW difference is very close to zero between 0.06 and 0.2\,{\AA} of 
$\rm EW_{862.1}$ from FPR. For smaller realms, the DIB results in DR3 present larger values of EW than those in FPR,
and their mean difference increases with the decreasing EW. This is because in DR3, we only successfully detected relatively strong 
DIB signals limited by the S/N of individual RVS spectra, whereas the EW measured in FPR represents an average DIB strength in each 
voxel by stacking a set of ISM spectra. This difference, between the strong signals captured in DR3 and the mean strength measured 
in FPR, also implies a variation in DIB strength in the solar neighbourhoods. On the contrary, FPR gets a systematically larger $\rm 
EW_{862.1}$ than DR3 when $\rm EW_{862.1}\,{\gtrsim}\,0.2$\,{\AA}, because large EW generally comes from distant voxels (or dense 
clouds) where detections in FPR and DR3 may trace different ISM environments. The DIB detections made in individual RVS spectra are 
mainly located in diffuse and intermediate regions, while DIB signals in denser regions can be measured in stacked spectra with higher 
S/N. Moreover, $\lambda$864.8 may also contribute to the difference as in DR3 $\lambda$864.8 was not considered for fittings. Thus, 
if the profile of $\lambda$864.8 is treated as the continuum placement for normalization (the two DIBs are close to each other, 
see Fig. \ref{fig:stack-fit} for example), $\rm EW_{862.1}$ would be underestimated. This effect will be investigated in detail 
in follow-up work.
%--------------------------------------------------------------------------------------------------------------------------------

To compare the spatial distribution of $\rm EW_{862.1}$ between FPR and DR3, Fig. \ref{fig:Fmap-8621} shows the EW maps in the
Galactic (XY), meridian (XZ), and rotational (YZ) planes where the Sun is located at the origin with the GC as the primary direction.
The two EW maps (FPR on the left and DR3 on the right) were constructed in different ways. For DR3, we make use of the high-quality 
sample (see its definition in \citetalias{PVP}) with an additional constraint of $0.18\,{\leqslant}\,{\sigma_{\rm DIB}}\,{\leqslant}0.26$\,nm 
(the best range of $\sigma_{862.1}$ determined in {\dibspec}), resulting in 52\,180 DIB detections. Then the median $\rm EW_{862.1}$
was taken in $0.1 \times 0.1$\,kpc bins for the Galactic, meridian, and rotational planes, respectively, in the Cartesian system 
using detections within $\pm50$\,pc above and below each corresponding plane. For FPR, $\rm EW_{862.1}$ in Galactic, meridian, 
and rotational planes were taken from the voxels that are crossed with each corresponding plane, and the crossed sections were 
painted by the $\rm EW_{862.1}$ in that voxel. To have a clean map with reliable detections, we require $\rm S/N\,{\geqslant}\,100$ 
and $\rm QF_{862.1}\,{\leqslant}\,2$.
%--------------------------------------------------------------------------------------------------------------------------------

Similar large-scale structures can be found in each plane between FPR and DR3 results, while $\lambda$862.1 in FPR can be detected 
in more distant regions, such as between the Perseus Arm and the Outer Arm and beyond the Scutum--Centaurus Arm. Two sightlines, 
$\ell\,{\sim}\,{-}70^{\circ}$ and $\ell\,{\sim}\,{-}117^{\circ}$, have significant low $\rm EW_{862.1}$ reaching over 4\,kpc, 
indicating two void regions with less abundant ISM between the Galactic main arms. On the other hand, much fewer DIB signals 
can be detected towards $\sim 80^{\circ} - \sim 90^{\circ} $
and  $\sim -100^{\circ} - \sim -90^{\circ} $ for both FPR and DR3 because  in these two ranges, we are looking to directions 
that are parallel to the Local Arm with less intervening DIB clouds.
% these directions are near the tangent point of the Local Arm with numerous and massive star-forming regions like the Cygnus complex. 
% \Tomaz{I would say in these two ranges we are looking to directions which are parallel to the Local Arm, so we see less intervening DIB clouds.}
%--------------------------------------------------------------------------------------------------------------------------------

%--------------------------------------------------------------------------------------------------------------------------------

\subsection{Equivalent width (EW) map: In a Galactic view} \label{subsect:ew-map}

\begin{figure*}
\centering
\includegraphics[width=14cm]{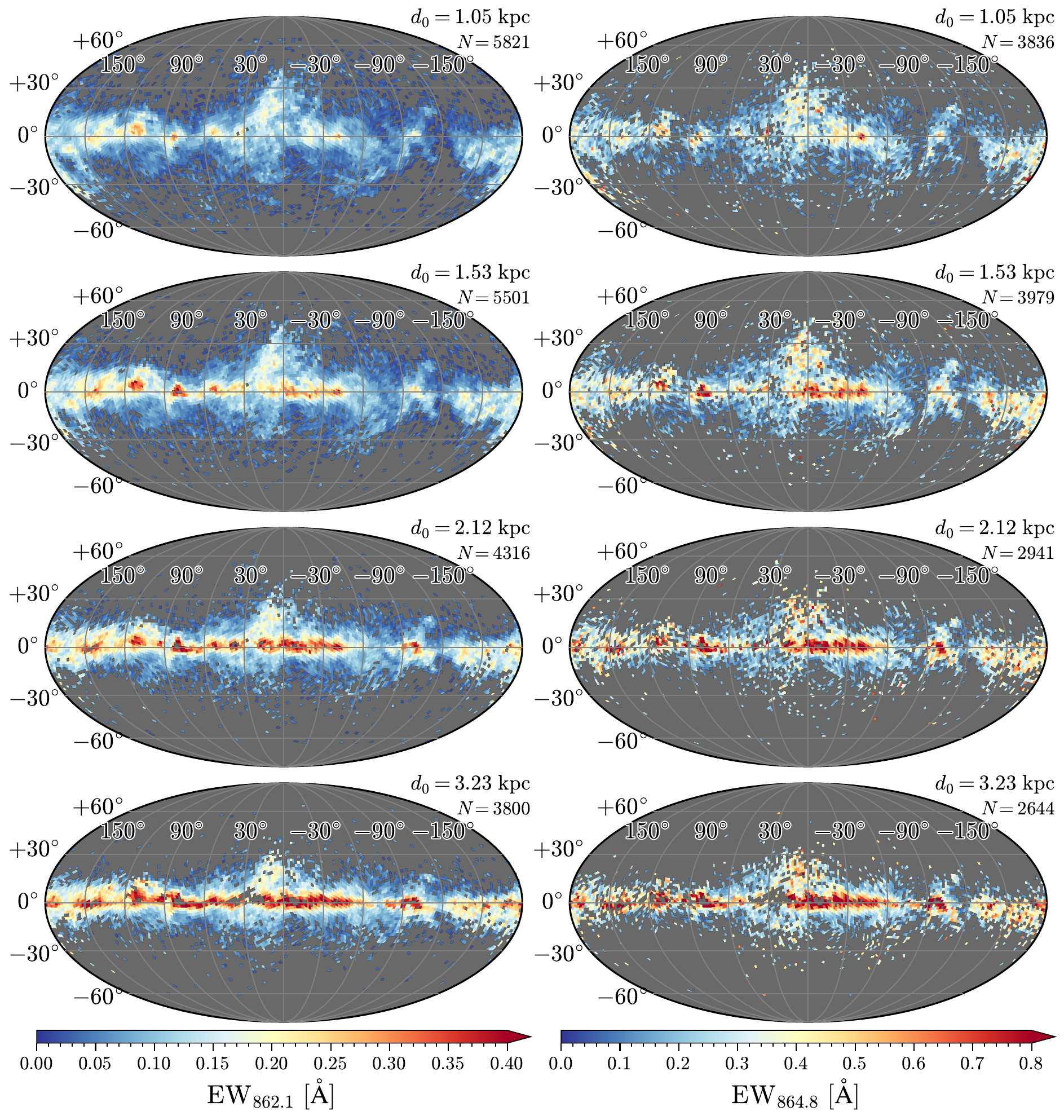}
\caption{Galactic distributions of $\rm EW_{862.1}$ ({\it left panels}) and $\rm EW_{864.8}$ ({\it right panels}) from DIB
results in FPR in Mollweide projection at HEALPix level 5, at four distances. The distance ($d_c$) and the number of voxels ($N$) 
are indicated in each subpanel.}
\label{fig:Gmap}
\end{figure*}

Figure \ref{fig:Gmap} shows the Galactic distribution of $\rm EW_{862.1}$ (left panels) and $\rm EW_{864.8}$ (right panels) at 
four distances, $d_c\,{=}\,1.05$, 1.53, 2.12, and 3.23\,kpc from the DIB results in FPR. As {\dibspec} stacked spectra in 
$(\alpha,\delta,d)$ voxels, the distances indicated in the figure are the centre of the voxels from the Sun. $\rm S/N\,{\geqslant}\,100$ 
and $\rm QF\,{\leqslant}\,3$ (for $\lambda$862.1 and $\lambda$864.8, respectively) are used to control the quality of the DIB detections. 
The number of selected DIB detection decreases with distance for both $\lambda$862.1 and $\lambda$864.8 (from $d_c\,{=}\,1.05$\,kpc 
to $d_c\,{=}\,1.53$\,kpc for $\lambda$864.8 is an exception). And at 2.12\,kpc, there are 4316 detections of $\lambda$862.1 that 
cover 35\% of the full sky. The amount of $\lambda$864.8 is around 70\% of $\lambda$862.1 at each $d_c$.
%--------------------------------------------------------------------------------------------------------------------------------

%--------------------------------------------------------------------------------------------------------------------------------

The sky coverage of $\lambda$862.1 with reliable detections in FPR is similar to that in DR3 (see Fig. 6 in \citetalias{PVP} for a
median EW map at HEALPix level 5), but the distance-sliced maps in Fig. \ref{fig:Gmap} contain more details. At $d_c\,{=}\,1.05$\,kpc
and 1.53\,kpc, some clumpy regions with large EW are consistent with nearby molecular clouds, such as Phoenix ($\ell\,{\sim}\,30^{\circ}
-40^{\circ}$) and Cygnus complex ($\ell\,{\sim}\,80^{\circ}$) in the middle plane and Ophiuchus ($\ell\,{\sim}\,0^{\circ}$ and
$b\,{\sim}\,15^{\circ}$) at high latitude, although the resolution of the present DIB EW map is pretty low ($1.8^{\circ}$).
Nevertheless, some clouds seem to disappear in the DIB EW map, like the Cepheus and Polaris Flare ($\ell\,{\sim}\,110^{\circ}
-120^{\circ}$), which can be clearly seen in the \gaia Total Galactic Extinction map (see Fig. 24 in \citealt{Delchambre2022}) and 
the Planck dust map (see Fig. 3 in \citealt{Planck2016dust}). Cepheus and Polaris Flare are nearby clouds ($<$400\,pc) with strong
CO emission, but no strong DIB signals were detected in this region in either this FPR or DR3, in spite of many RVS observations there.
With the increase of distance, at 3.23\,kpc, the high-EW regions are linked up into a bright stripe along the galactic plane.
%--------------------------------------------------------------------------------------------------------------------------------

At low galactic latitudes, $\lambda$862.1 was detected in most longitudinal directions, but a gap can be found at $\ell\,{=}\,-120^{\circ}$,
where the CO emission is also weak (see \citealt{Dame2001}). It seems that $\lambda$864.8 is more concentrated around the regions
with larger EW, which should be a bias due to the difficulty in measuring weak $\lambda$864.8 with the present spectral S/N level. 
On the other hand, $\lambda$862.1 and $\lambda$864.8 seem to occupy similar latitude ranges, that they are mainly distributed within
$\pm30^{\circ}$ and this range decreases with distance. In principle, the DIB signals detected in nearby voxels should also be seen 
in distant voxels along the same direction as EW is an integrated variable of the abundance of DIB carriers. But in fact, 
the number of DIB detections about $\pm30^{\circ}$ of latitude decreases with the distance for the control sample shown in 
Fig. \ref{fig:Gmap}. The reason could be that some of the DIB fittings in distant voxels were filtered out by the criteria 
of the control sample due to the low quality of their ISM spectra. Furthermore, the measured DIB strength in distant voxels would
be a biased mean, as faint stars among or behind dense clouds are hard to be observed by \gaia. Therefore, it is possible to see the
decrease of DIB EW with distance even if the voxels all contain reliable DIB detections. This observational bias could also happen
in nearby regions behind dense clouds. 
%--------------------------------------------------------------------------------------------------------------------------------
% The longitude distribution is consecutive for $\lambda$862.1 at low latitudes, 

% DIB signals above $\pm30^{\circ}$ are mainly within 1--2\,kpc and 
% disappear in more distant voxels where the DIB detections could be filtered out by the criteria because of the decreasing amount 
% of RVS spectra and consequently the lower quality of measurements.

\begin{figure}
\centering
\includegraphics[width=8cm]{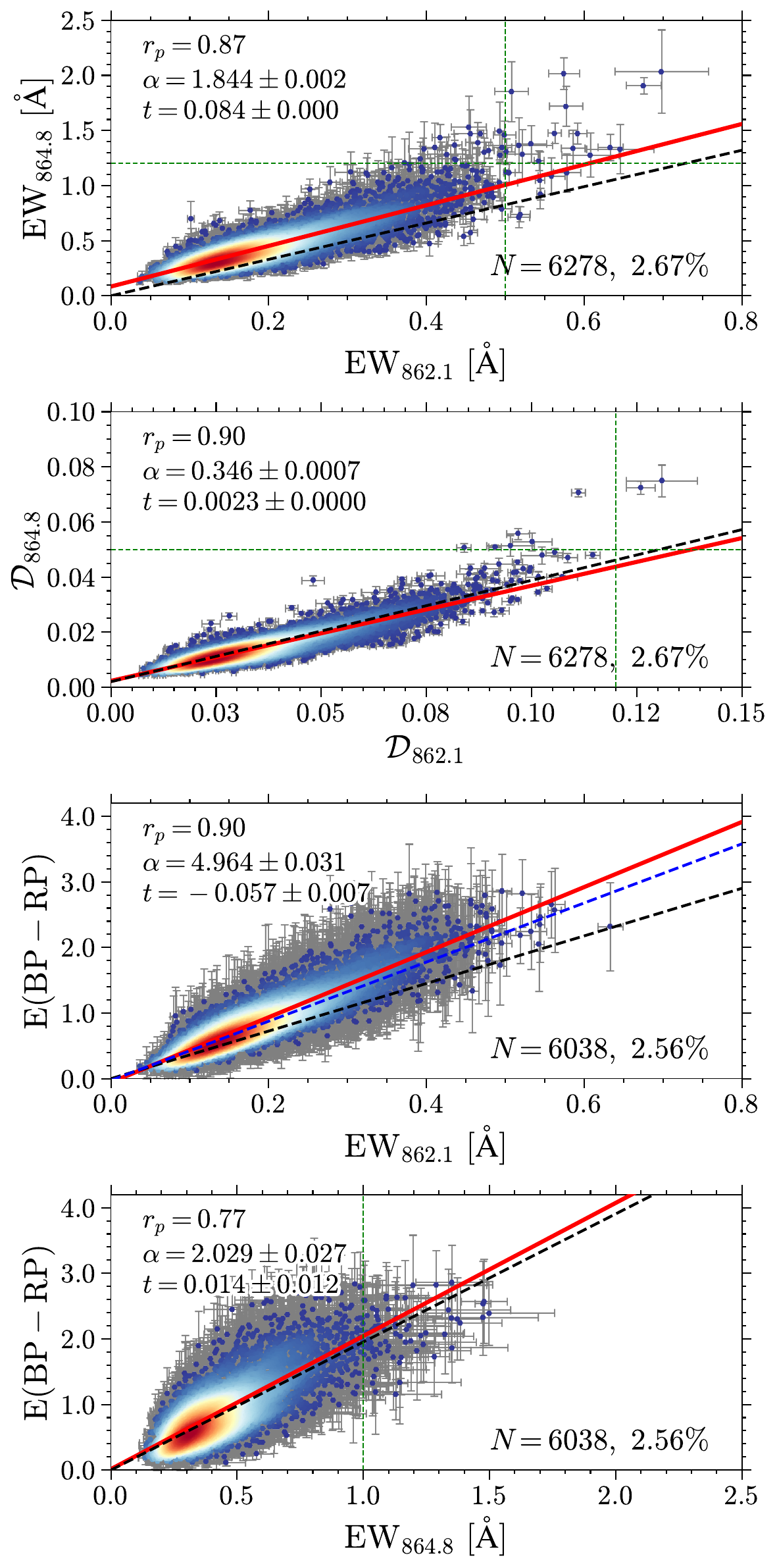}
\caption{Diverse correlations between $\lambda$862.1, $\lambda$864.8, and mean $\rm E(BP-RP)$ from {\gspphot} \citep{Andrae2022}
in each voxel. The data points are coloured by their number densities estimated by a Gaussian KDE. The red lines are the linear 
fit to the data points. The dashed green lines indicate the upper limits on the variables for the linear fits. And the fitted 
slope ($\alpha$) and intercept ($t$) are marked in each panel, together with the Pearson correlation coefficient ($\rm r_p$) and 
the number of DIB detections ($N$). Dashed black lines are previous results from \citetalias{Zhao2022}, and the dashed blue 
line is from \citetalias{PVP}. }  
\label{fig:intensity}
\end{figure}
%--------------------------------------------------------------------------------------------------------------------------------

\subsection{Correlations between DIBs and dust reddening} \label{subsect:dib-dust}

The linear correlation between DIB strength and dust reddening is a general property for many strong DIBs \citep[e.g.][]{Friedman2011,Lan2015}
and could be treated as 
a validation test of the DIB measurement. For this comparison, we only select DIB detections with $\rm QF_{862.1}\,{=}\,0$ and 
$\rm QF_{864.8}\,{=}\,0$, resulting in 6278 cases (2.67\% of total) to increase the reliability and to focus on the Galactic middle 
plane (see QF distribution in Fig. \ref{fig:qf-distrib}) where the interstellar materials are generally well mixed with each other, 
so a tight linear correlation can be expected between $\lambda$862.1, $\lambda$864.8, and dust. There are 3\,677\,773 (60\%) RVS
objects that have $\EBPRP$ from {\gspphot} \citep{Andrae2022}. The mean $\EBPRP$ in a voxel was calculated with all the 
stars with $\EBPRP$ in that voxel. Thus, the number of used stars in a voxel for mean $\EBPRP$ and DIB fitting could be different.
There are 6038 (2.56\%) voxels containing at least ten stars having $\EBPRP$ and $\rm QF\,{=}\,0$ for the fittings of both 
$\lambda$862.1 and $\lambda$864.8, which were used to compare DIB EW and $\EBPRP$ (see the third and fourth panels from top to 
bottom in Fig. \ref{fig:intensity}). On the other hand, for the comparison of EW and depth between $\lambda$862.1 and $\lambda$864.8
(see the first and second panels from top to bottom in Fig. \ref{fig:intensity}), we only required $\rm QF\,{=}\,0$ and got 6278 
(2.67\%) voxels. The requirement of the highest level of QF is the main constraint to the sample size. The comparison including
other QFs and discussions will be made in a follow-up work. The linear fit to each correlation was achieved by an ordinary 
least squares regression using the Python package {\it statsmodels}, with some upper limits on the variables (indicated in Fig. 
\ref{fig:intensity}) to exclude the nonlinear realms. The fitted slope ($\alpha$), intercept ($t$), and Pearson coefficient 
($\rm r_p$) are marked in each panel in Fig. \ref{fig:intensity} as well.
%--------------------------------------------------------------------------------------------------------------------------------
% Mean $\EBPRP$ was calculated in each {\dibspec} voxel and 6038 (2.56\%)
% of them with $\rm QF\,{=}\,0$ contain at least ten objects having $\EBPRP$, which will be used to compare EW and $\EBPRP$. 
% The correlations between $\rm EW_{862.1}$, $\rm EW_{864.8}$, and $\EBPRP$, as well as between $\mathcal{D}_{862.1}$ and $\mathcal{D}_{864.8}$,
% are shown in Fig. \ref{fig:intensity}. 

Tight linearity can be found between $\rm EW_{862.1}$ and $\rm EW_{864.8}$, between $\mathcal{D}_{862.1}$ and $\mathcal{D}_{864.8}$, 
and between $\rm EW_{862.1}$ and $\EBPRP$, and a weaker one for $\rm EW_{864.8}$ and $\EBPRP$. In the correlation of EW for the 
two DIBs, some cases present greater $\rm EW_{864.8}$ than expected when $\rm EW_{862.1}\,{\gtrsim}\,0.5$\,{\AA}. Similar deviation
from the linearity could also be found for their depth. This deviation could indicate the departure of the carriers of the two
DIBs (if we assume they have different origins) as the trend is gradual and continuous with EW and the ISM spectra of these cases
show prominent DIB features. No apparent deviation is seen between EW and $\EBPRP$, but the linear fit needs to be limited to
$\rm EW_{864.8}\,{<}\,1$\,{\AA} to get a small intercept between $\rm EW_{864.8}$ and $\EBPRP$. 
%--------------------------------------------------------------------------------------------------------------------------------
% correlation

Linear fits done in \citetalias{PVP} and \citetalias{Zhao2022} are also shown for comparison. We note that \citetalias{Zhao2022} 
fixed the intercept as zero (except $\mathcal{D}_{862.1}$ and $\mathcal{D}_{864.8}$) and applied 2$\sigma$-clipping for the linear 
fits due to its small sample size (1103 DIB detections). The $\EBPRP/{\rm EW_{862.1}}$ ratios from FPR and \citetalias{PVP} are 
consistent with each other with a difference of 9.20\%, but the degree of dispersion is much lower in FPR because of stronger 
quality control and the higher S/N in general after stacking RVS spectra. As a reference, the difference of $\EBV/{\rm 
EW_{862.1}}$ between literature studies varies from 4\% to 41\% with a mean of 20\%\footnote{The differences between seven
studies are listed in Table 3 in \citet{PVP} except \citet{Wallerstein2007}.} The comparison between the DIB results in FPR and 
\citetalias{Zhao2022} is more meaningful because they both use BNM to build the stellar templates and fit DIB profiles in stacked 
RVS spectra with the same model, only with differences in the sample size and stacking strategy. By an internal comparison, a 
consistent tendency can be found for each correlation between this FPR and \citetalias{Zhao2022}, with a similar degree of dispersion 
of the scatter plots, but the DIB results in FPR contain much stronger DIB signals. For example, $\rm EW_{862.1}$ is mainly within 
0.15\,{\AA} in \citetalias{Zhao2022} and 0.4\,{\AA} in this FPR. Both the slope and intercept are highly consistent with each other 
for $\mathcal{D}_{862.1}$ and $\mathcal{D}_{864.8}$ in this FPR and \citetalias{Zhao2022}, with a difference of only 6.50\%. The 
difference in the correlation between $\rm EW_{862.1}$ and $\rm EW_{864.8}$ is slightly larger ($\sim 10\%$), and the fit in FPR 
has a positive and non-negligible intercept. The cause is unclear. Strict QF control in FPR, which leads to the lack of $\lambda$864.8 
with $\rm EW_{864.8}\,{<}\,0.1$\,{\AA}, could have an impact but cannot fully explain the offset as the tight linear correlation 
keeps until $\rm EW_{862.1}\,{\sim}\,0.5$\,{\AA}. Including all the other QFs (1--5) could reduce the intercept from 0.084
to 0.076, but the degree of dispersion will significantly increase. The biggest difference between FPR and \citetalias{Zhao2022} 
occurs for $\EBPRP/{\rm EW_{862.1}}$ (26.93\%), which is caused by the sigma-clipping in \citetalias{Zhao2022}. We refit 
$\EBPRP/{\rm EW_{862.1}}$ using the data in \citetalias{Zhao2022} without sigma-clipping and get a ratio of $4.468\pm0.081$. The 
difference then becomes much smaller (9.50\%). Additionally, the 2$\sigma$-clipping caused the linear fit in \citetalias{Zhao2022} 
to be dominated by DIBs with $\rm EW_{862.1}$ between 0.04 and 0.1\,{\AA}, where much fewer cases were seen in this FPR after QF 
control. Thus, the $\EBPRP/{\rm EW_{862.1}}$ ratio would also vary in different $\rm EW_{862.1}$ ranges. The sigma-clipping, on 
the other hand, has a very light effect on the correlation between $\EBPRP$ and $\rm EW_{864.8}$. The difference is only 3.77\%, 
although in \citetalias{Zhao2022} the intercept was fixed as zero and in FPR we got a positive one of 0.014. Besides the sample 
size and sigma-clipping, many other factors can affect the fitting results as well, such as the QF control and the source of dust 
reddening. More detailed discussions are beyond the scope of this work. And the deviation and variation in the correlation between 
different ISM, like DIB and dust, need to be understood by considering the interstellar environment as well.
%--------------------------------------------------------------------------------------------------------------------------------

\begin{table*}
\centering
\caption{DIB parameters of $\lambda$862.1 and filed information of 12 voxels within the Local Bubble (see Sect. \ref{localbubble}).}
\label{tab:local-bubble}
\begin{tabular}{crrrcccccc}
\hline\hline 
Number{\tablefootmark{a}} & healpix & lc & bc & dc & p08620 & p18620 & p28620 & ew8620 & flags8620  \\ 
       &         & (deg) & (deg) & (kpc) &  & (nm) & (nm) & (nm) &            \\ [0.05ex]
\hline        
1  & 6212        &  304.28       & 36.91        & 0.165 & 0.0082 & 862.20 & 0.19 &        0.0038 &  1 \\
2  & 6348        &  284.14       & 47.74        & 0.165 & 0.0075 & 862.28 & 0.25 &        0.0047 &  1 \\
3  & 7685        &    4.39       & 19.37        & 0.065 & 0.0087 & 862.33 & 0.22 &        0.0049 &  1 \\
4  & 4793        &   59.97       & --33.73      & 0.165 & 0.0081 & 862.25 & 0.34 &        0.0068 &  2 \\
5  & 2312        &   40.31       & 43.28        & 0.165 & 0.0055 & 862.24 & 0.24 &  0.0033 &  1 \\
6  & 8047        &   51.13       & 16.46        & 0.165 & 0.0112 & 862.35 & 0.13 &        0.0036 &  2 \\
7  & 9695        &  272.04       & 29.70        & 0.235 & 0.0086 & 862.28 & 0.18 &        0.0039 &  1 \\
8  & 10008   &  257.39   & 11.41        & 0.235 & 0.0069 & 862.27 & 0.25 &       0.0043 &  1 \\
9  & 5853        &  188.17       & --18.54      & 0.165 & 0.0087 & 862.35 & 0.24 &        0.0052 &  1 \\
10 & 5235        &  226.30       & --8.49       & 0.235 & 0.0099 & 862.37 & 0.24 &        0.0059 &  1 \\
11 & 10041   &  249.64   & 12.91        & 0.235 & 0.0115 & 862.32 & 0.23 &       0.0067 &  0 \\
12 & 7347        &    7.37       & 13.09        & 0.165 & 0.0151 & 862.25 & 0.22 &        0.0082 &  0 \\ [0.05ex]
\hline
\end{tabular}
\tablefoot{ \\
\tablefoottext{a}{The number of voxels matches those marked in Fig. \ref{fig:LB}. The rest column names correspond to those
in Table \ref{tab:catalog-params}.} }
\end{table*}
%--------------------------------------------------------------------------------------------------------------------------------
%--------------------------------------------------------------------------------------------------------------------------------

\begin{figure*}[!htbp]
  \centering
  \includegraphics[width=14cm]{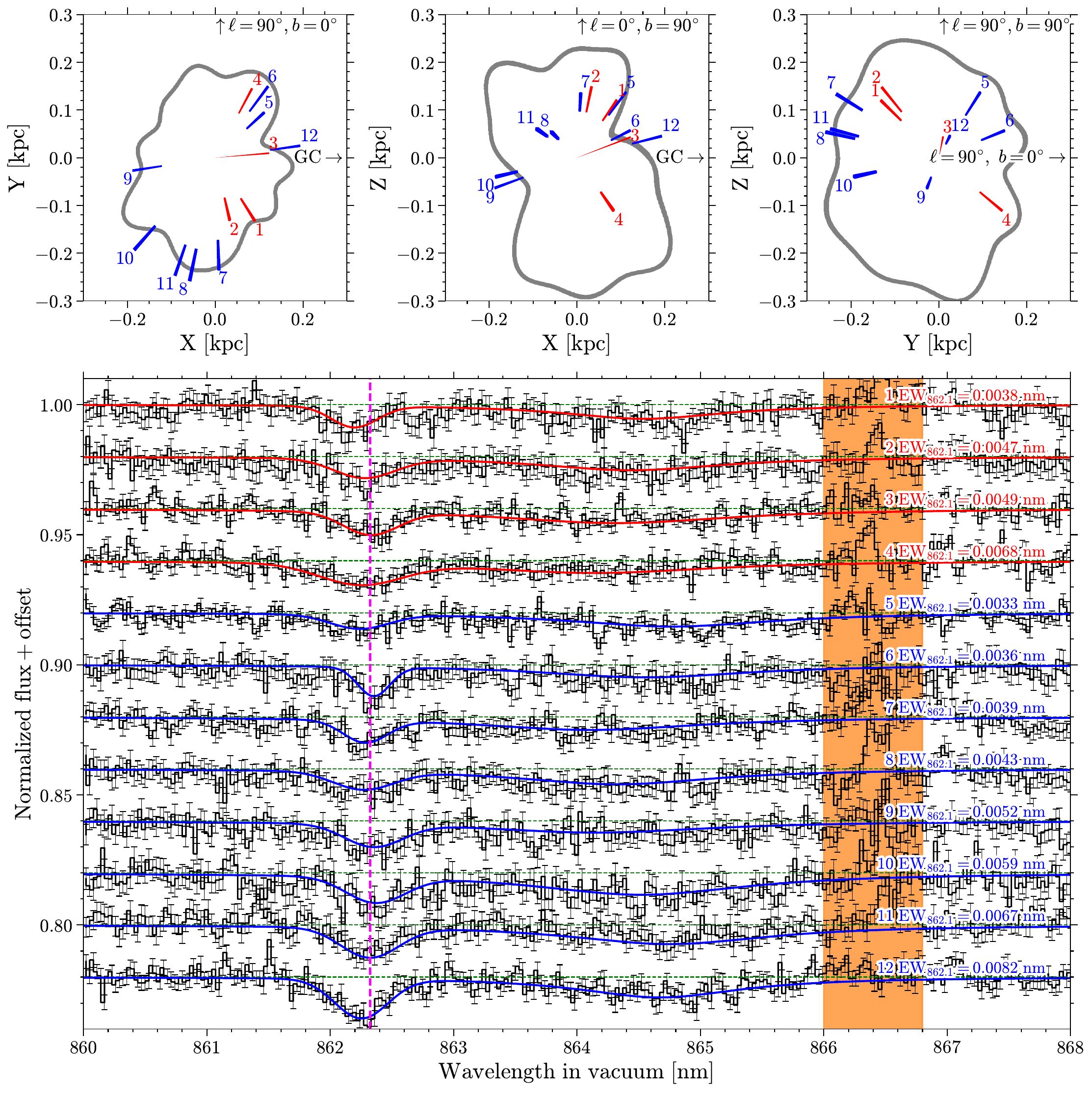}
  \caption{{\it Upper panels:} Spatial distribution of 12 reliable DIB detections by visual inspection. Their voxels are projected
  into the Galactic (XY), meridian (XZ), and rotational (YZ) planes. The reds are inside the Local Bubble and the blues are crossed
  with the surface of the Local Bubble. The grey marks the surface of the Local Bubble determined by \citet{Pelgrims2020}. 
  {\it Lower panel:} Black lines are stacked ISM spectra, and the red/blue lines are the DIB fitting results of the corresponding
  DIB profiles. The orange indicates the masked spectral region between 866 and 866.8\,nm in the fitting. The vertical dashed
  magenta line indicates the rest-frame wavelength of DIB\,$\lambda$862.1 of 862.323\,nm determined in \citet{PVP}.  }
  \label{fig:LB}
\end{figure*}
%--------------------------------------------------------------------------------------------------------------------------------

\section{The Local Bubble} \label{localbubble}

It is widely known that the Local Bubble has a much lower density than the average of the ISM in the solar neighbourhood 
because of its harsh environment (high temperature and low density; \citealt{Welsh2010}; \citealt{Lallement2014}), but \citet{Farhang2019} 
reported the detections of DIBs $\lambda$578.0 and $\lambda$579.7 in the Local Bubble in the spectra of 359 early-type stars and 
mapped the 3D density distribution of their carriers. \citetalias{PVP} also found some relatively strong signals of $\lambda$862.1 
that are very close to the Sun and generally suggested that $\lambda$862.1 could also be detected in the Local Bubble. However, by 
reanalysing the public RVS spectra (about one million) in DR3, \citet{Saydjari2022} claimed no detection of $\lambda$862.1 with 
high confidence levels in their analysis within the Local Bubble. As \citetalias{PVP} only presented the median $\rm EW_{862.1}$ 
distribution in the Galactic plane and \citet{Saydjari2022} only analysed a small part of the RVS spectra, we plan to make a 
thorough investigation to see if we can reliably find $\lambda$862.1 in the Local Bubble. 
%--------------------------------------------------------------------------------------------------------------------------------
% almost free of interstellar atoms and molecules

As a preliminary investigation, we focus in this work only on 145 DIB detections with a high level of S/N and QF, that is $\rm 
S/N\,{\geqslant}\,300$, $\rm QF_{862.1}\,{\leqslant}\,2$, $\rm QF_{864.8}\,{\leqslant}\,2$, and $d_c\,{<}\,300$\,pc. Then their 
ISM spectra were further visually inspected. As DIB signals within the Local Bubble would be very weak, a highly reliable DIB
detection needs to satisfy that the flux uncertainty of its ISM spectrum within the DIB profile is smaller than the depth of each
wavelength bin. Finally, we found four DIB detections whose voxels are mostly inside the Local Bubble and 12 detections whose 
voxels are crossed with the surface of the Local Bubble determined by \citet{Pelgrims2020} using the 3D dust map of \citet{Lallement2019}. 
The DIB parameters of $\lambda$862.1 and field information of these 12 voxels are listed in Table \ref{tab:local-bubble}. And 
the projection of these 12 voxels and the surface of the Local Bubble in the Galactic (XY), meridian (XZ), and rotational 
(YZ) planes are shown in the upper panels in Fig. \ref{fig:LB}, and their ISM spectra and DIB fittings are presented in the lower
panel. The profile of the DIB\,$\lambda$862.1 is conspicuous in the noisy spectra, with all $\rm QF_{862.1}\,{\leqslant}\,2$, 
although the DIB fitting is not good, especially for $\lambda$864.8 (whose profile is too shallow for the S/N of the spectra). 
The Local Bubble is known to contain assembling molecular clouds on its wall so that the dust abundance is significantly different 
inside and on the surface of the Local Bubble.  Besides the selection bias (weaker signals cannot be detected by present data), 
the DIB profile would be broadened by the heavy noise, or the continuum placement was affected by the poorly fitted $\lambda$864.8 
profile, which leads to an overestimation of EW. The four internal voxels mostly inside the Local Bubble, containing 
apparent DIB profiles, seem to indicate the possible detection of $\lambda$862.1 in the Local Bubble, but further investigation
is necessary, especially taking into account all available ISM spectra.
%--------------------------------------------------------------------------------------------------------------------------------
% Therefore, we can conclude the existence of $\lambda$862.1 in the Local Bubble.
% A more systematic study of the DIBs in the LB is necessary, especially taking into account all available ISM spectra.

\section{Caveats and known issues} \label{sect:caveats}

We list the caveats and the known issues of the first results of {\dibspec} as presented in this FPR. Future developments for DR4 
aim to tackle and possibly remove these issues. 
%--------------------------------------------------------------------------------------------------------------------------------

\begin{enumerate}
    \item The inverse of parallax ($1/\varpi$) of the background stars was directly used in {\dibspec} for stacking spectra in 
    different voxels. But $1/\varpi$ would not be the true distance of the stars in distant zones (several kpc, \citealt{Bailer-Jones2015}). 
    The detected DIB signal accounts for an integration of the DIB carriers between the background stars and us. If the distance 
    between the DIB carriers and the background stars is much larger than the distance uncertainty of $1/\varpi$, the DIB measurement 
    would be safe. Otherwise, additional uncertainty of the DIB detection will be introduced. Distant voxels or voxels with few 
    targets will get heavier influence.
    % the distance uncertainty would significantly exceed the distance bin size for very distant zones!
    \item About 5\% (12\,692) of stacked ISM spectra have zero flux uncertainty at each wavelength bin. Some of them are due to the
    zero flux error in observed RVS spectra used for stacking. But the cause for others is still unknown. The flux uncertainties
    of these spectra were fixed as 0.01 during the DIB fitting.
    \item The flux uncertainty at each wavelength bin was taken as the standard error in the mean (SEM) from the individual RVS
    spectra (see Sect. \ref{subsect:stack-fit}). Nevertheless, as the median of individual fluxes was taken for stacking, the
    flux uncertainty should be $1.253 \times {\rm SEM}$. Presently used values in the spectra table underestimate the uncertainties
    of the stacked ISM spectra.
    \item Negative $\dibdepth$ (`p08620' and `p08648' in Table \ref{tab:catalog-params}) can be found for $\lambda$862.1 
    (166) and $\lambda$864.8 (198), although the prior of $\dibdepth$ is always positive. These cases are all badly fitted due to 
    the low-S/N spectra and/or the weak DIB signals.
    \item A scaling factor is suggested to be used to correct the lower and upper confidence levels of EW for $\lambda$862.1 and 
    $\lambda$864.8 (`ew8620\_lower', `ew8620\_upper', `ew8648\_lower', `ew8648\_upper' in Table \ref{tab:catalog-params}), 
    respectively (see Sect. \ref{subsect:dib-ew} for details).
    \item There are a set of detections reported in the output table `interstellar\_medium\_params' with inconsistencies in their 
    EW and the lower or upper confidence levels of EW, even after the correction by the scaling factors (see Sect. 
    \ref{subsect:dib-ew}). Specifically, there are 421 cases with $\rm EW_{862.1}\,{<}\,EW_{862.1,lower}$, 351 cases with $\rm EW_{862.1}\,{<}\,EW_{862.1,upper}$, 
    2791 cases with $\rm EW_{864.8}<EW_{864.8,lower}$, and 440 cases with $\rm EW_{864.8}<EW_{864.8,upper}$. 
    The cause of this inconsistency could be some problems in recording the lower and upper confidence levels when producing
    the DIB results. We note that this inconsistency does not mean a bad DIB fitting, but the EW confidence levels would be problematic 
    for such cases.
    \item When we compare the target and reference spectra to build the stellar template, we have to reduce the weights of the 
    \ion{Ca}{ii} lines to have a better model of weaker lines such as \ion{Fe}{i}. Thus, we cannot model \ion{Ca}{ii} lines very well 
    and mask the specific region (866.0--866.8\,nm) during the fitting. Such a region falls within the profile wing of DIB\,$\lambda$864.8,
    introducing an uncertainty on the determination of the precise boundary of the DIB. We will try other data-driven methods in the 
    future to model all the stellar lines without downweighting the \ion{Ca}{ii} lines.
    % We checked that the recorded EW is the same as the one calculated by $\dibdepth$ and $\dibwidth$ for both $\lambda$862.1 and 
    % $\lambda$864.8. 
    % Thus, the inconsistency is caused by the problems in recording the lower/upper values of EW, that the drawing 
    % from the MCMC posterior samplings to estimate the EW distribution was not well done.  
\end{enumerate}
%--------------------------------------------------------------------------------------------------------------------------------

\section{Summary and conclusions} \label{sect:conclusion}

% We summarize here the detection and measurement of two DIBs at 862.1\,nm ($\lambda$862.1) and 864.8\,nm ($\lambda$864.8) in volumetric pixels (voxels) from stacked ISM spectra derived from DR3 RVS spectra.
% The results, produced by the {\dibspec} module, are published as part of \gaia Focused Product Release (FPR), and the two tables, one for
% DIB parameters and the other for stacked ISM spectra, can be accessed in the \gaia Archive\footnote{\url{https://gea.esac.esa.int/archive/}}.
We summarize here the processing and validation of two published tables produced by the {\dibspec} module in the \gaia FPR, 
one for fitted DIB parameters (Table \ref{tab:catalog-params}, `interstellar\_medium\_params') and the other for stacked ISM spectra 
(Table \ref{tab:catalog-spectra}, `interstellar\_medium\_spectra'). {\dibspec} derived ISM spectra for 5\,983\,289 RVS objects
with $|b|\,{<}\,65^{\circ}$ as targets using the other 160\,392 RVS spectra ($|b|\,{\geqslant}\,65^{\circ}$) as references and
the BNM. The individual ISM spectra were stacked to increase the spectral S/N in defined 3D voxels with 
a resolution of $1.8^{\circ}$ (level 5 HEALPix binning) on the celestial sphere and 0.07--1\,kpc in distance, and on average 20 
spectra in each voxel. {\dibspec} fitted and measured the two DIBs in 235\,428 voxels, with a DIB model applying a Gaussian
profile for $\lambda$862.1 and a Lorentzian profile for $\lambda$864.8. A median FWHM was determined as $0.52\pm0.05$\,nm for
$\lambda$862.1 and $1.91\pm0.44$\,nm for $\lambda$864.8, which can be referred to as a typical value considering a large space coverage.
Users are encouraged to use QF to control the quality of the DIB fittings for specific investigations and the related discussions 
in Sect. \ref{sect:catalog} are also useful.
%--------------------------------------------------------------------------------------------------------------------------------

Taking advantage of the stacking procedure, {\dibspec} extends the DIB detection in distance compared to the DIB catalogue in
\gaia DR3 \citep{PVP}. The DIB strength of $\lambda$862.1 is highly consistent between DR3 and this FPR (a mean difference of 
0.002\,{\AA} with a standard deviation of 0.025\,{\AA}), and some systematic differences along with $\rm EW_{862.1}$ would be 
caused by the selection bias between the two samples. We provide several other validation tests as well: The DIB EW map in a 
Galactic view at different distances shows the integration of DIB carriers along the sightlines, revealing some clumpy dense 
regions corresponding to notable molecular clouds, such as Phoenix, Cygnus, and Ophiuchus, while Cepheus and Polaris Flare were 
not seen in either of the DIB EW maps. Based on a high-quality subsample with $\rm QF\,{=}\,0$ for both $\lambda$862.1 and $\lambda$864.8 
(6278 detections, 2.67\% of total), linear correlations between $\lambda$862.1, $\lambda$864.8, and dust reddening ($\EBPRP$) 
were found to be consistent with those in \citet{PVP} and \citet{Zhao2022}, with the smallest difference of 3.77\% and biggest one
of 26.93\%. We also found some detections of $\lambda$862.1 inside and around the surface of the Local Bubble with prominent 
profiles in the derived ISM spectra.
%--------------------------------------------------------------------------------------------------------------------------------

The DIB work in this FPR is not only complementary to \gaia\ DR3, but a pathfinder for future releases. The acquired experience 
and caveats will benefit the development of the {\dibspec} module for future \gaia\ releases. DIB results in this FPR have already 
shown the power of using numerous RVS spectra to map both the intermediate and strong DIB\,$\lambda$862.1 and the broad 
and shallow DIB\,$\lambda$864.8 in the solar neighbourhood and
reaching over 4\,kpc. In particular, we note that broad DIBs such as $\lambda$864.8 were never measured in a sample of hundreds of thousands of spectra before. 
%--------------------------------------------------------------------------------------------------------------------------------

% We present here the pipeline of the \gaia FPR product {\dibspec} where we stack spectra in 3D voxels at a HEALPix level of 5 and a side length of $\rm 1.8^{o}$. The output of {\dibspec} is an archive table with the DIB parameters of $\lambda$862.1 and $\lambda$864.8  where their profile was fitted by a Gaussian and Lorentzian function, and a spectra table containing the stacked ISM spectra in each voxel. Compared to the \gaia DR3 product of individual DIB measurements (\citealt{PVP}), {\dibspec} has the advantage due to the stacking procedure to  construct a full sky map of  both DIBS ($\lambda$862.1 and $\lambda$864.8). In addition, due to the higher SNR of {\dibspec} with respect to \gaia DR3, it is possible to trace more distant regions such as the Scutum-Centaurus arm or the Perseus-Outer Arm. We provide several validations tests such as the comparison with \gaia DR3, the equivalent width map in a galactic view and the correlation between the DIBs and the dust reddening and show evidence of detections of DIBs in the Local Bubble.
%--------------------------------------------------------------------------------------------------------------------------------

\begin{acknowledgements}
  This work has made use of data from the European Space Agency (ESA) mission \gaia (https://www.cosmos.esa.int/\gaia), processed 
  by the \gaia Data Processing and Analysis Consortium (DPAC, https://www.cosmos.esa.int/web/\gaia/dpac/consortium). Funding for 
  the DPAC has been provided by national institutions, in particular, the institutions participating in the \gaia Multilateral Agreement.
  % T. Z. acknowledges financial support of the Slovenian Research Agency (research core funding No. P1-0188) and the European Space 
  % Agency (Prodex Experiment Arrangement No. C4000127986).  Part of the calculations have been performed with the high-performance 
  % computing facility SIGAMM, hosted by the Observatoire de la Côte d'Azur. The GSP-spec group acknowledges financial supports from 
  % the french space agency (CNES), Agence National de la Recherche (ANR 14-CE33-014-01) and  Programmes Nationaux de Physique Stellaire 
  % \& Cosmologie et Galaxies (PNPS \& PNCG) of CNRS/INSU. H.Z. is funded by the China Scholarship Council (No.201806040200). YF acknowledges 
  % the BELgian federal Science Policy Office (BELSPO) through various PROgramme de D\'eveloppement d'Exp\'eriences scientifiques (PRODEX) grants.
\end{acknowledgements}

\bibliographystyle{aa}
%\vspace*{-0.4cm}
\bibliography{dibs}

% \onecolumn
\appendix

\section{Error analysis of the  BNM} \label{app:ref-check}

\begin{figure*}
  \centering
  \includegraphics[width=14cm]{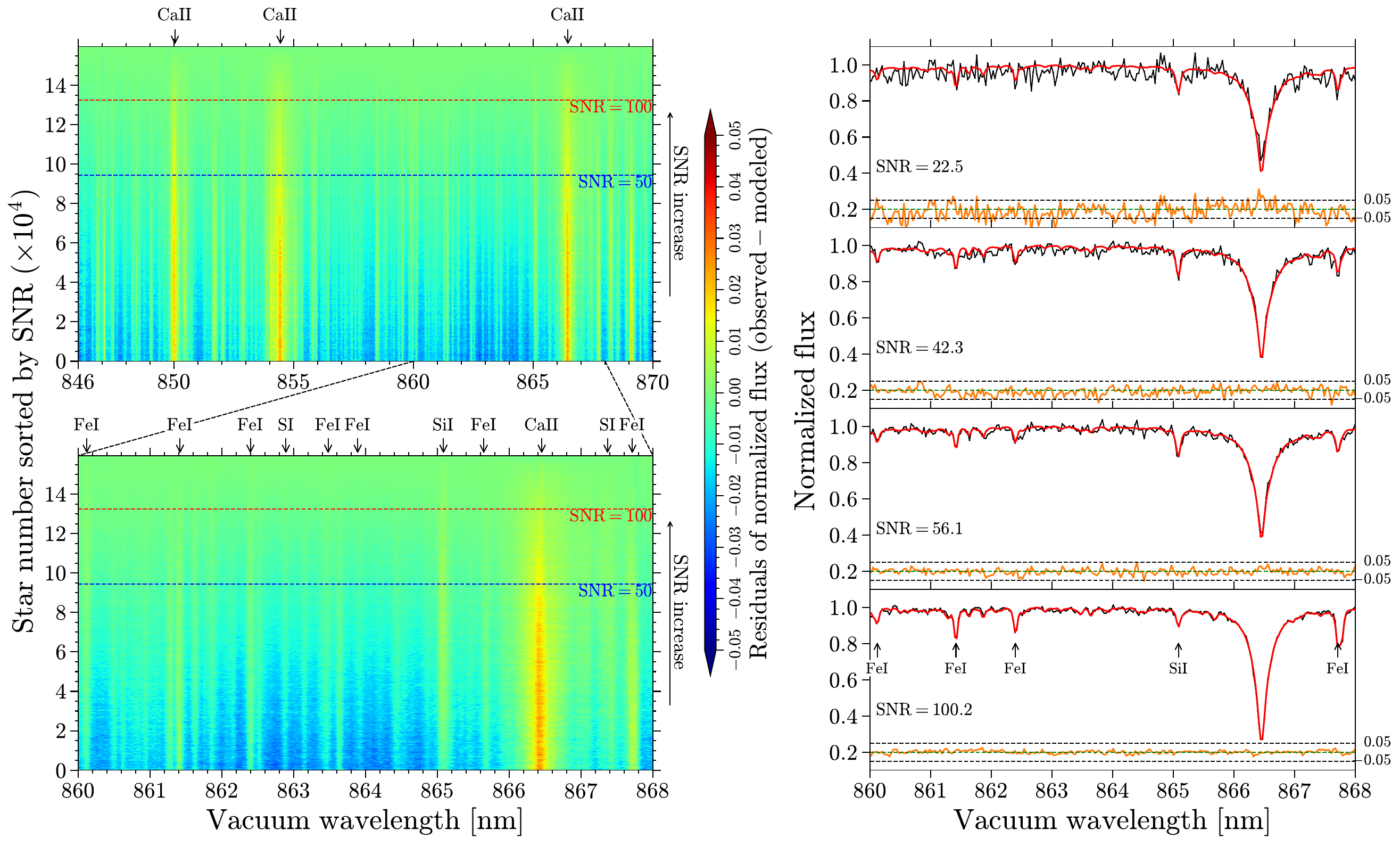}
  \caption{{\it Left panel:} Distribution of flux residuals ($\rm observed-modelled$) with spectral wavelength for 159\,591 
  reference spectra. Each row presents one spectrum and the spectra are sorted by their S/N. The dashed blue and red lines 
  indicate the position of S/N equalling 50 and 100, respectively. Some typical stellar lines within the RVS spectral region 
  determined by \citet{Contursi2021} are indicated as well. The lower panel is a zoom-in plot of the upper one to show the 
  distribution of the residuals in the DIB window (860--868\,nm). {\it Right panel:} Four examples of the reference spectra 
  (black lines with observed flux errors) and their derived stellar templates (red lines). The orange lines are the flux 
  residuals ($\rm observed-modelled$) with dashed black lines indicating $\pm$5\% of the continuum.}
  \label{fig:ref-check}
\end{figure*}

In this section, we estimate the magnitudes of the errors introduced by the BNM, which was used in {\dibspec} to build the stellar 
templates for target spectra (Sect. \ref{subsect:ism-spec}). For each spectrum in the reference sample (160\,392, see Sect. 
\ref{subsect:ref-spec}), we apply the BNM to generate its stellar template using all the other reference spectra. In total, 
159\,591 (99.5\%) spectra have enough best neighbours ($>$10) to build their stellar templates. The distribution of the flux 
residuals between the observed RVS spectra and the generated stellar templates ($\rm observed-modelled$) is shown on the left-hand 
side in Fig. \ref{fig:ref-check}, and four generated stellar templates are shown as examples in the right panels together with 
the corresponding observed RVS spectra. 
%--------------------------------------------------------------------------------------------------------------------------------

The flux residuals in the spectra present some patterns, especially in low-S/N spectra, instead of being uniform or noise-dominated. 
Positive residuals (red region) can be found near the \ion{Ca}{ii} triplets because of their reduced weights and some other stellar 
lines like \ion{Si}{i} at 8658.4\,{\AA}, \ion{Fe}{i} at 8623.97\,{\AA} and 8677.13\,{\AA}, indicating an overestimation of the depth 
of these stellar lines. On the other hand, negative residuals (blue region) mainly appear in spectra with $\rm S/N\,{<}\,50$, which 
could be due to a failed modelling of the lines by BNM in low-S/N spectra or to the improper normalization. The positions of some 
strong stellar lines are indicated in Fig. \ref{fig:ref-check}. These lines are determined by \citet{Contursi2021} for the RVS 
spectra. %And a full list can be accessed online\footnote{\url{http://cdsarc.u-strasbg.fr/viz-bin/cat/J/A+A/654/A130}}.
%--------------------------------------------------------------------------------------------------------------------------------

\begin{figure}
  \centering
  \includegraphics[width=7cm]{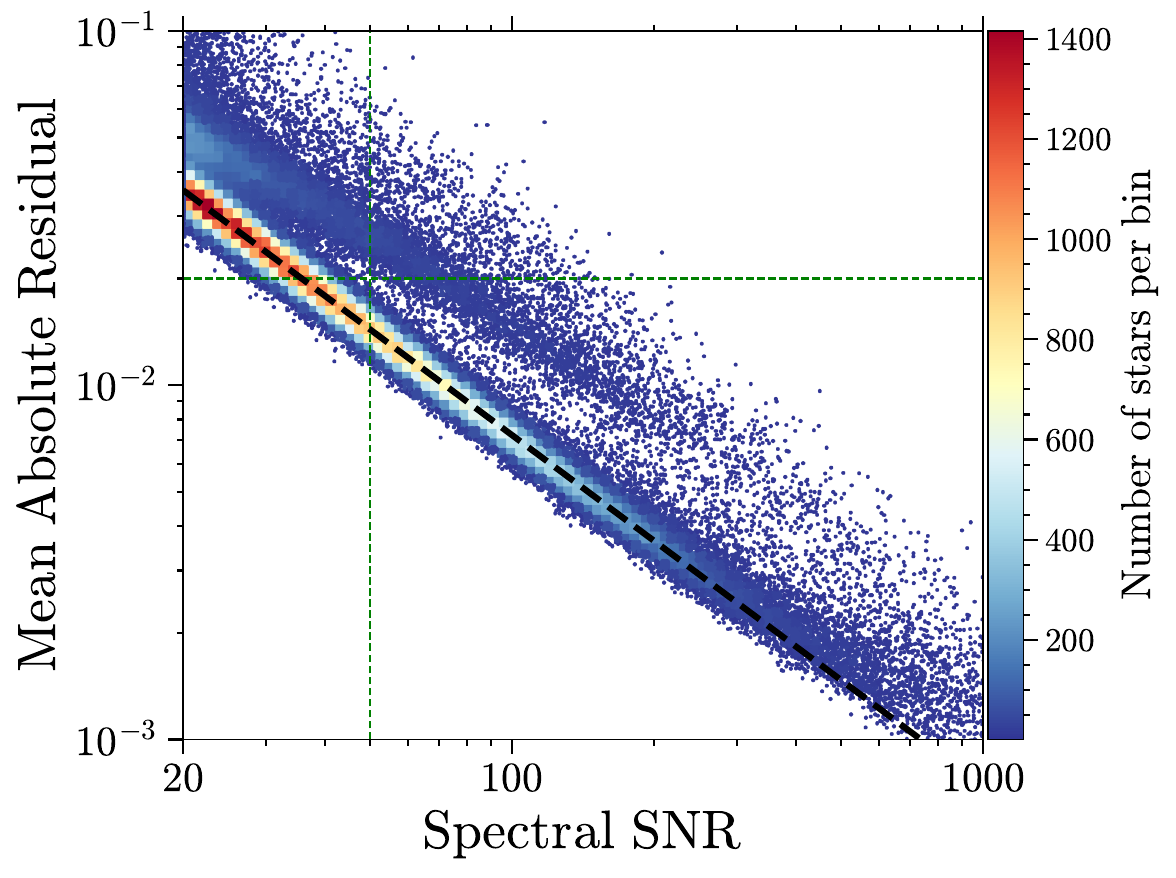}
  \caption{Variation of the mean absolute residual (MAR) between observed and modelled RVS spectra, calculated within the DIB window 
  (861.2--866.0\,nm), with the spectral S/N. The dashed green lines indicate $\rm S/N=50$ and $\rm MAR=0.02$, 
  respectively. The dashed black line is fitted to MAR and S/N on a logarithmic scale.}
  \label{fig:ref-check-snr}
\end{figure}

To quantitatively characterize the uncertainty introduced by BNM into the ISM spectra and consequently to the DIB measurement, we 
calculate the mean absolute residuals ($\rm MAR=mean(|observed-modelled|)$) of normalized flux between 861.2 and 866.0\,nm (region 
of the two DIBs in priors). The relationship between MAR and spectral S/N is shown in Fig. \ref{fig:ref-check-snr}. MAR has a 
strong dependence on spectral S/N. We applied a linear fit to them in logarithmic scale with $3\sigma$ clipping, and got a relation 
of $\rm log_{10}(MAR)\,{=}\,{-}0.16\,{-}\,0.99\,{\times}\,log_{10}(S/N)$ (the dashed black line in Fig. \ref{fig:ref-check-snr}) 
that can be used to describe the detections in the main branch, that is MAR linearly increases with $\rm 1/(S/N)$ for most of the 
stars, although MAR would be slightly larger than expected for very large S/N. Other 11.1\% of stars in a secondary branch have 
larger MAR, which could be caused by the bad normalization of RVS spectra (most with $\rm S/N\,{<}\,50$) and/or the low number 
density in the vicinity of the queried stars. The reason for the dependence of MAR on the spectral S/N could be that high-S/N 
spectra are less affected by the random noise and have best neighbours with higher S/N, as BNM rejects the reference spectra with 
morphological differences larger than $\rm 3/(S/N)$ of the queried spectrum. 

From our test, we conclude that the uncertainty introduced by BNM strongly depends on the S/N of the queried spectra. When $\rm 
S/N\,{>}\,50$, MAR, the average magnitude of the flux residuals in the DIB vicinity (861.2--866.0\,nm), is smaller than 0.02 (2\% of 
the continuum) for 96.8\% of spectra in the reference sample, and smaller than 0.01 for 61.7\% of spectra. We expect BNM to have a similar 
performance on the RVS target sample because the target and reference samples have similar S/N distribution (see the right panel 
in Fig. \ref{fig:sp}).
%--------------------------------------------------------------------------------------------------------------------------------

% \clearpage

%--------------------------------------------------------------------------------------------------------------------------------

%--------------------------------------------------------------------------------------------------------------------------------

% \begin{figure*}
%   \centering
%   \includegraphics[width=14cm]{figures/ref-check-params.pdf}
%   \caption{{\it (a):} Variation of the mean absolute residual (MAR) between observed and modeled RVS spectra, calculated within 
%   the DIB window (861.2--866.0\,nm), with the spectral SNR. The dashed green lines indicate $\rm SNR=50$ and $\rm MAR=0.02$, 
%   respectively. The dashed black line is fitted to MAR and SNR in logarithmic scale. {\it (b)--(d):} Distribution of MAR in the 
%   stellar atmospheric parameters planes ($\teff-\logg$, $\teff-\meta$, and $\logg-\meta$). The colors represent the average MAR 
%   in bins of $\Delta \teff=40$\,K, $\Delta \logg=0.05$\,dex, and $\Delta \meta=0.04$\,dex.}
%   \label{fig:ref-check-params}
% \end{figure*}
%--------------------------------------------------------------------------------------------------------------------------------

\clearpage

\section{Corner plots of the DIB fittings} \label{app:corner}

\begin{figure}
  \centering
  \includegraphics[width=8cm]{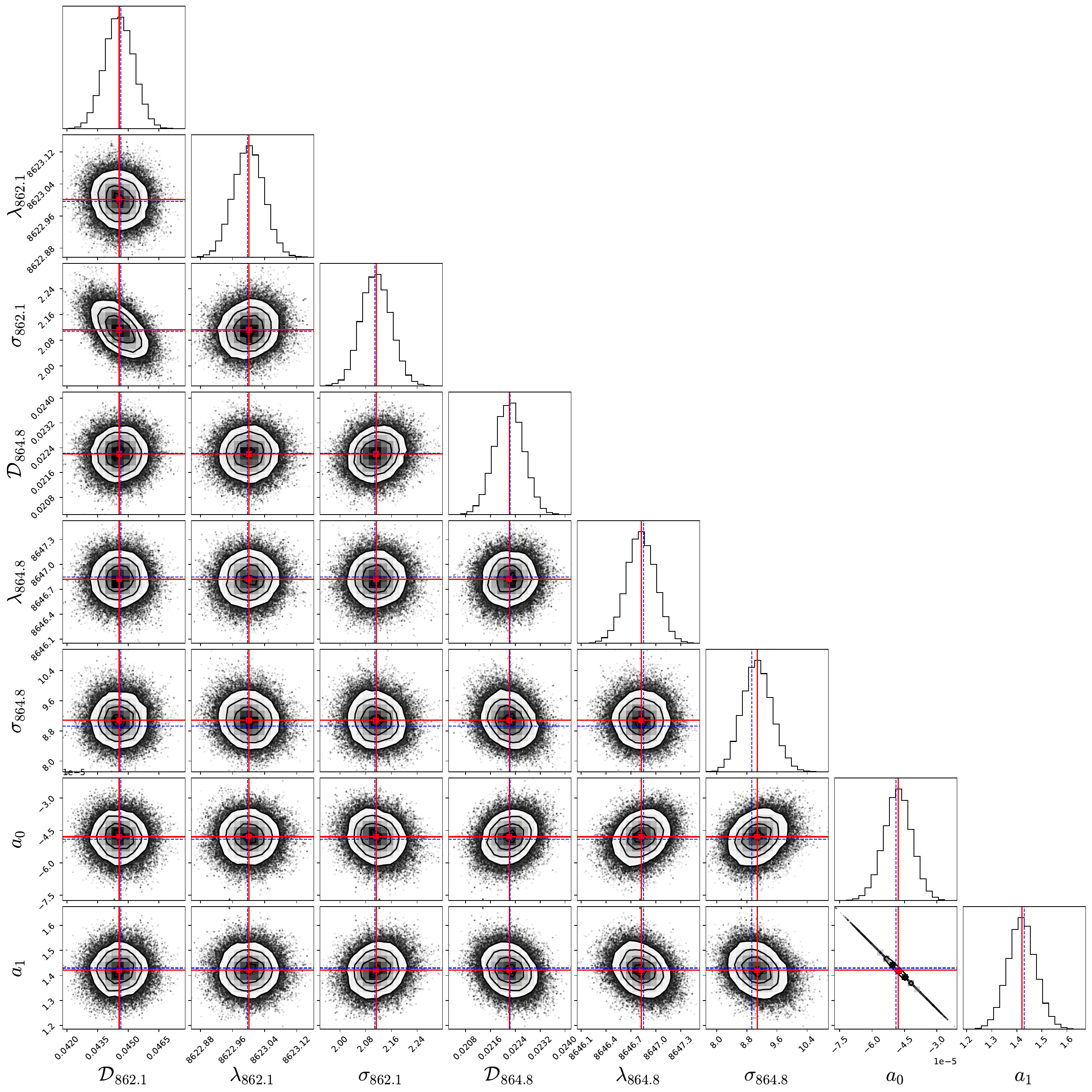}
  \caption{Corner plot of the DIB fitting in the voxel with $I_{\rm pix}=10450$ and $d_c=1.05$\,kpc (the first panel in Fig.
  \ref{fig:stack-fit} from top to bottom). The histograms and scatter plots show the one- and two-dimensional projections of the
  posterior distributions of the fitted parameters. The red squares and lines indicate the best-fit estimates for each parameter
  in the reproduced fitting. And the dashed blue lines mark the fitted parameters in the output table of {\dibspec}.}
  \label{fig:corner0}
\end{figure}

\begin{figure}
  \centering
  \includegraphics[width=8cm]{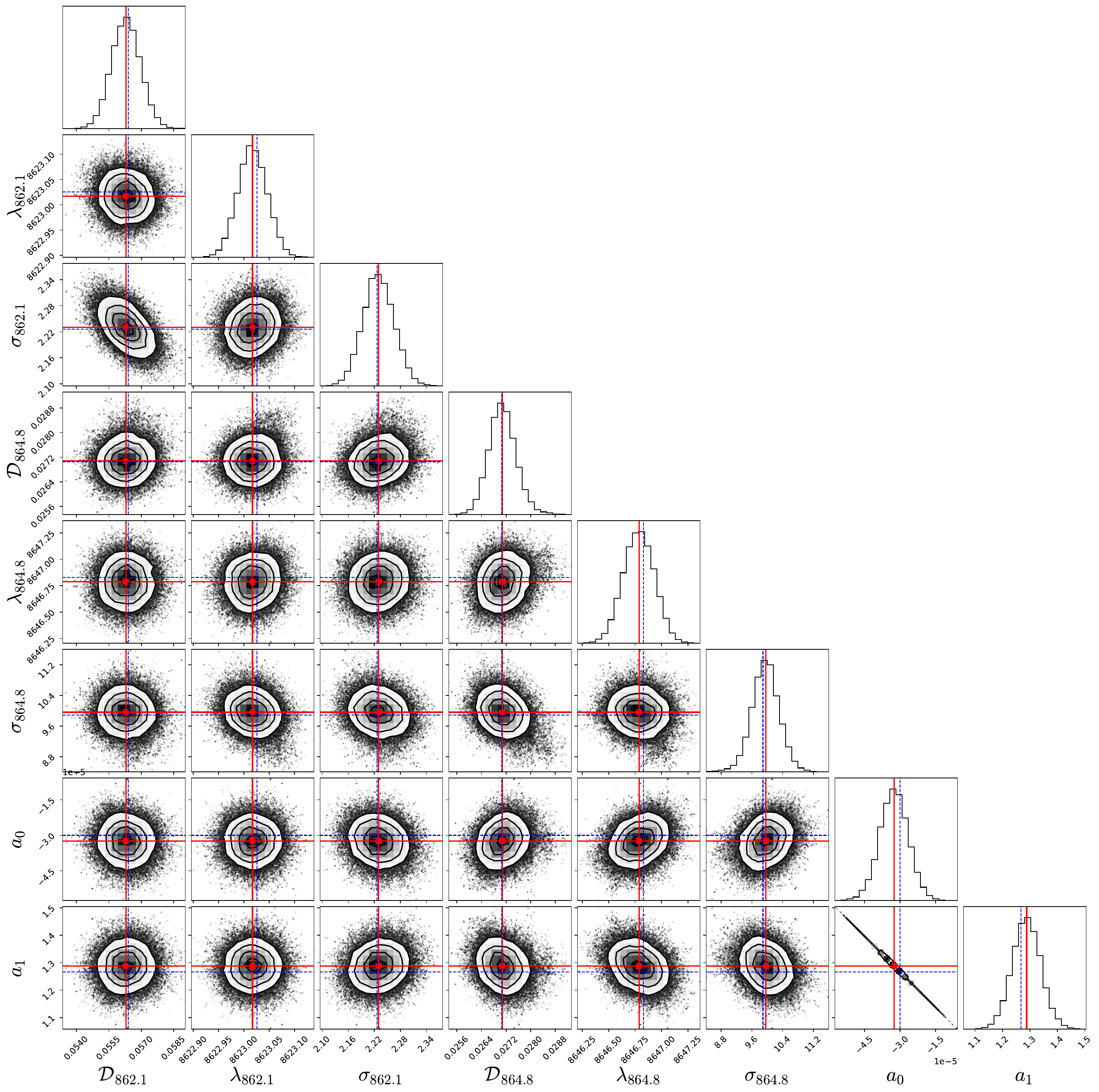}
  \caption{Same as Fig. \ref{fig:corner0}, but for the voxel with $I_{\rm pix}=10450$ and $d_c=1.27$\,kpc (the second panel in 
  Fig. \ref{fig:stack-fit} from top to bottom)}
  \label{fig:corner1}
\end{figure}

\begin{figure}
  \centering
  \includegraphics[width=8cm]{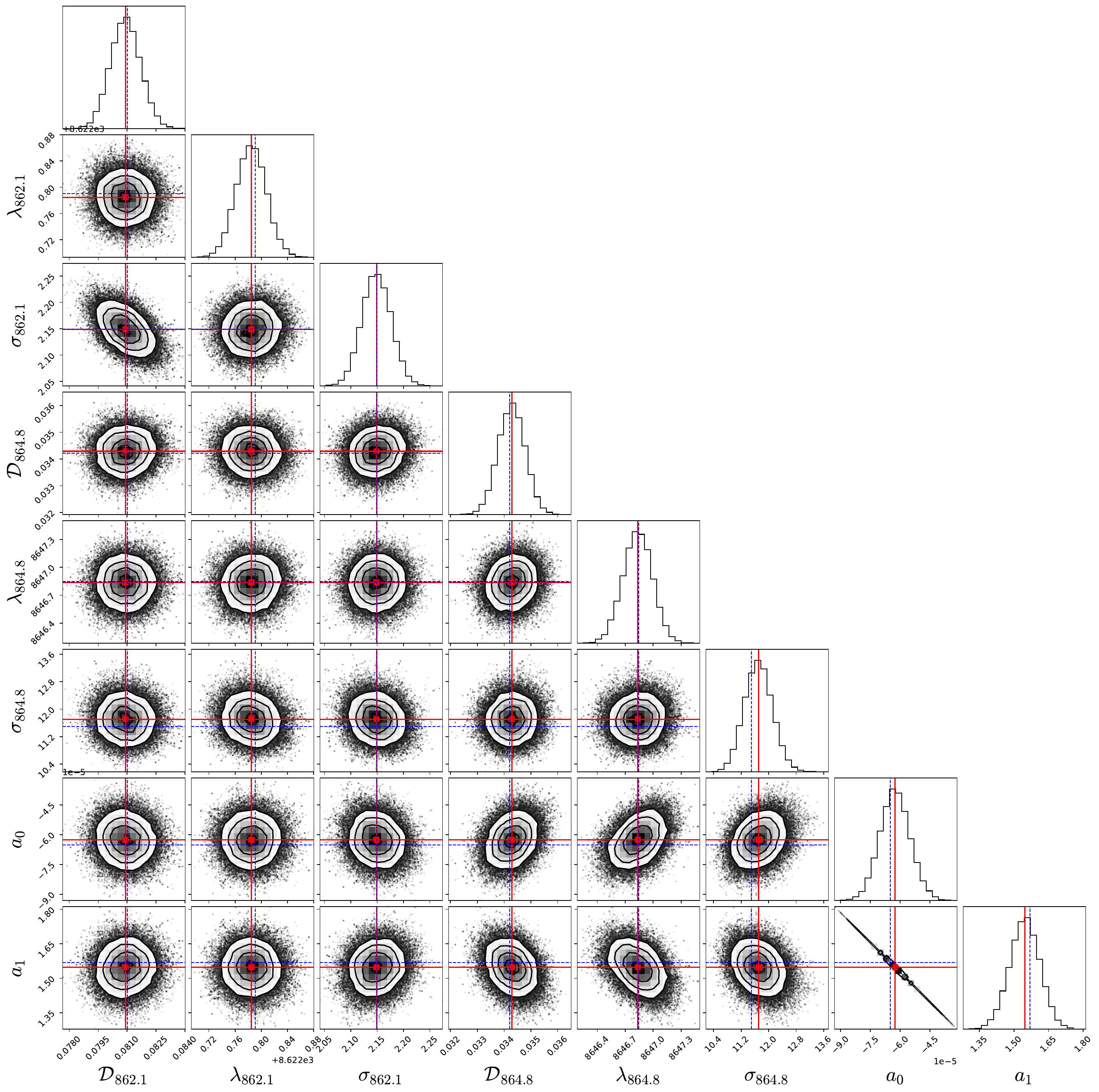}
  \caption{Same as Fig. \ref{fig:corner0}, but for the voxel with $I_{\rm pix}=10450$ and $d_c=2.12$\,kpc (the third panel in 
  Fig. \ref{fig:stack-fit} from top to bottom)}
  \label{fig:corner2}
\end{figure}

\begin{figure}
  \centering
  \includegraphics[width=8cm]{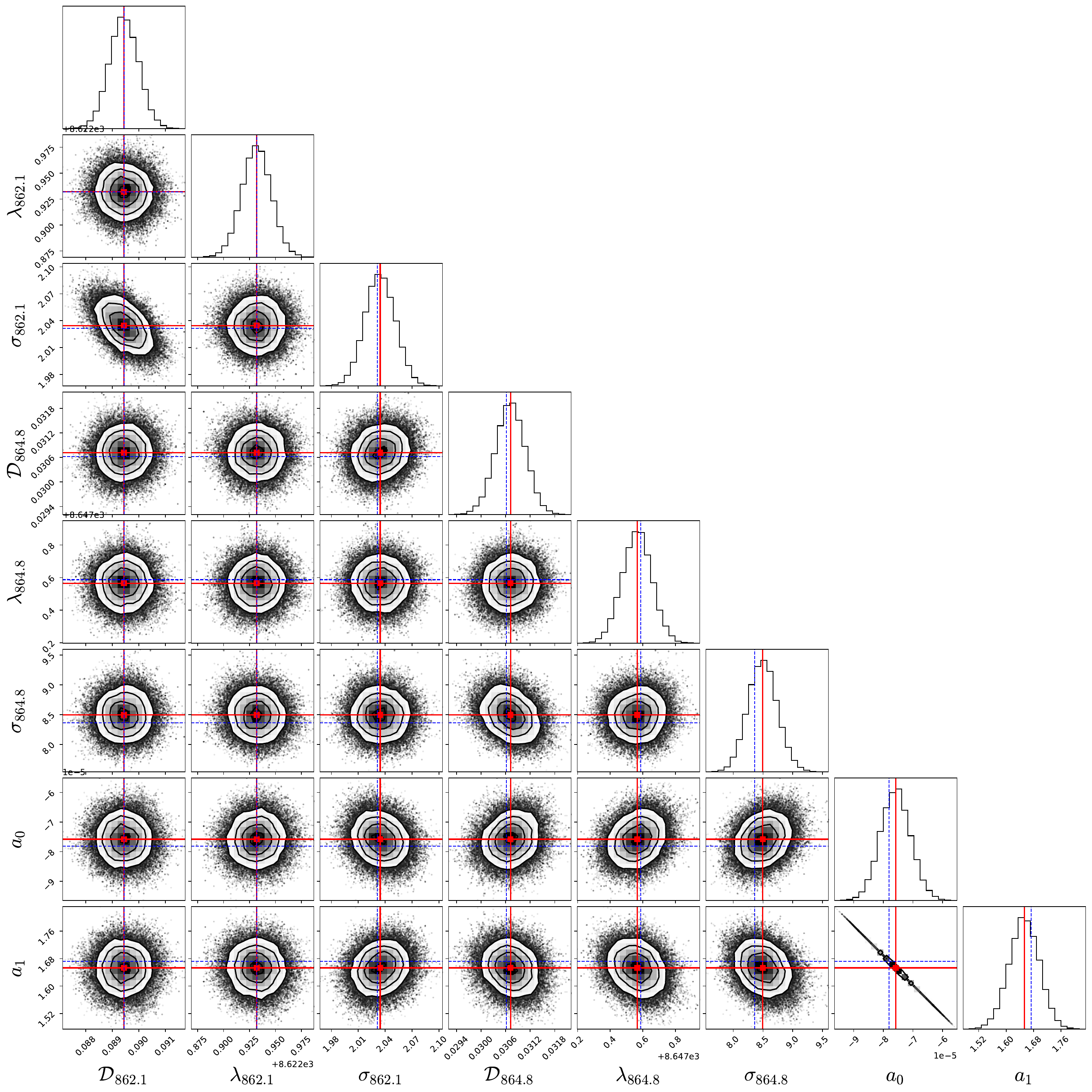}
  \caption{Same as Fig. \ref{fig:corner0}, but for the voxel with $I_{\rm pix}=10450$ and $d_c=2.40$\,kpc (the forth panel in 
  Fig. \ref{fig:stack-fit} from top to bottom)}
  \label{fig:corner3}
\end{figure}

\begin{figure}
  \centering
  \includegraphics[width=8cm]{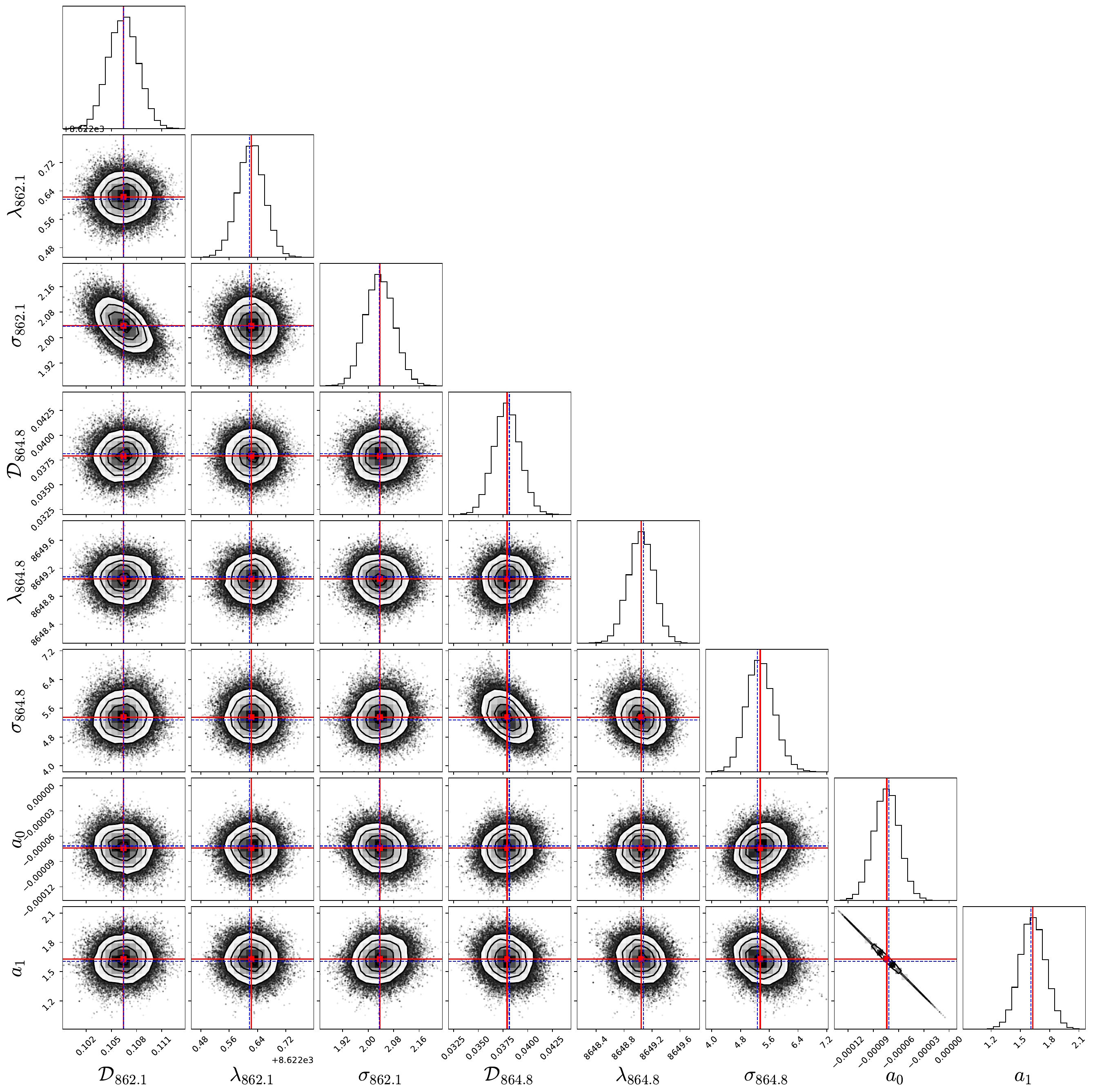}
  \caption{Same as Fig. \ref{fig:corner0}, but for the voxel with $I_{\rm pix}=10450$ and $d_c=3.23$\,kpc (the fifth panel in 
  Fig. \ref{fig:stack-fit} from top to bottom)}
  \label{fig:corner4}
\end{figure}

Figures \ref{fig:corner0}--\ref{fig:corner4} show the corner plots of the DIB fitting in the voxels according to the examples
shown in Fig. \ref{fig:stack-fit}. The histograms and scatter plots show the one- and two-dimensional projections of the posterior 
distributions of the fitted parameters, with red squares and lines indicating the best estimates. Because the intermediate 
quantities of {\dibspec} were dropped, the posterior distributions of the parameters were drawn by refitting the stacked ISM
spectra in the same way as {\dibspec}, some tiny differences can thus be found between the best estimates (red) and the fitting
results in the output table of {\dibspec} (blue). 
%--------------------------------------------------------------------------------------------------------------------------------

%--------------------------------------------------------------------------------------------------------------------------------

%--------------------------------------------------------------------------------------------------------------------------------

%--------------------------------------------------------------------------------------------------------------------------------

%--------------------------------------------------------------------------------------------------------------------------------

%--------------------------------------------------------------------------------------------------------------------------------

\clearpage

\section{Python script for converting the spectra table} \label{app:python-code}

The python script below shows a simple method to convert the spectra table to a fits file, in which each row stands for one stacked
ISM spectra.
%--------------------------------------------------------------------------------------------------------------------------------

\begin{lstlisting}[language=Python]
import numpy as np
import pandas as pd
from tqdm import tqdm
from astropy.io import fits

dibs = pd.read_csv(somepath+'parameter_table.csv')
tabe = pd.read_csv(somepath+'spectral_table.csv')

spec_lc = tabe.lc.values
spec_bc = tabe.bc.values
spec_dc = tabe.dc.values
cata_lc = dibs.lc.values
cata_bc = dibs.bc.values
cata_dc = dibs.dc.values
wave = np.unique(tabe['lambda'].values)
flux = np.zeros((dibs.shape[0],wave.shape[0]),dtype='float')
ferr = np.zeros_like(flux)
for i in tqdm(range(dibs.shape[0])):
    t = (spec_lc == cata_lc[i]) & (spec_bc == cata_bc[i]) & (spec_dc == cata_dc[i])
    flux[i] = data['flux'][t]
    ferr[i] = data['flux_uncertainty'][t]
    
hdr0 = fits.Header() 
hdr0['COMMENT1'] = 'stacked ISM spectra'
hdr0['EXTNAME'] = ('flux', 'normalized flux') 
hdr1 = fits.Header()
hdr1['EXTNAME'] = ('ferr', 'flux uncertainty')
hdr2 = fits.Header() 
hdr2['EXTNAME'] = ('wave', 'wavelength bins in vacuum')
hdu0 = fits.PrimaryHDU(flux, header=hdr0)   
hdu1 = fits.ImageHDU(ferr, header=hdr1)
hdu2 = fits.ImageHDU(wave, header=hdr2)
hdu  = fits.HDUList([hdu0, hdu1, hdu2])
hdu.writeto(somepath+'DIBSpec_ISM_spectra.fits',overwrite=True)
\end{lstlisting}

\clearpage

\end{document}